\documentclass[useAMS,usenatbib,a4paper]{mnras}
\usepackage{color,soul}
\usepackage[dvipsnames]{xcolor}
\usepackage{footmisc}
\usepackage{hyperref}
\usepackage{amsmath}
\usepackage{amssymb}
\usepackage{amstext}

\usepackage{mathtools}
\usepackage{aas_macros}
\usepackage{graphicx}

\graphicspath{ {./source_files/} }

\usepackage{natbib}

\title[Discovery of New Dipper Stars with K2]{Discovery of New Dipper Stars with K2: A Window into the Inner Disk Region of T Tauri Stars}

\author[Hedges et al]{Christina Hedges$^{1}$\thanks{E-mail: chedges@ast.cam.ac.uk},
  Simon Hodgkin$^{1}$\thanks{E-mail: sth@ast.cam.ac.uk} \&
  Grant Kennedy$^{2}$
  \\
  $^{1}$Institute of Astronomy, University of Cambridge, Madingley Road, Cambridge, CB3 0HA, UK \\
  $^{2}$Department of Physics, University of Warwick, Gibbet Hill Road, Coventry, CV4 7AL, UK \\
}
\begin{document}
  \pdfpagewidth 8.27in 
  \pdfpageheight 11.65in
\label{firstpage}
\pagerange{\pageref{firstpage}--\pageref{lastpage}}
\maketitle

\begin{abstract}
In recent years a new class of Young Stellar Object has been defined, referred to as dippers, where large transient drops in flux are observed. These dips are too large to be attributed to stellar variability, last from hours to days and can reduce the flux of a star by 10-50\%. This variability has been attributed to occultations by warps or accretion columns near the inner edge of circumstellar disks. Here we present 95 dippers in the Upper Scorpius association and $\rho$ Ophiuchus cloud complex found in K2 Campaign 2 data using supervised machine learning with a Random Forest classifier. We also present 30 YSOs that exhibit brightening events on the order of days, known as bursters. Not all dippers and bursters are known members, but all exhibit infrared excesses and are consistent with belonging to either of the two young star forming regions. We find 21.0 $\pm$ 5.5\% of stars with disks are dippers for both regions combined. Our entire dipper sample consists only of late-type (KM) stars, but we show that biases limit dipper discovery for earlier spectral types. Using the dipper properties as a proxy, we find that the temperature at the inner disk edge is consistent with interferometric results for similar and earlier type stars.

\end{abstract}
\section{Introduction}

\begin{figure*}
  \centering
  \includegraphics[width=\textwidth]{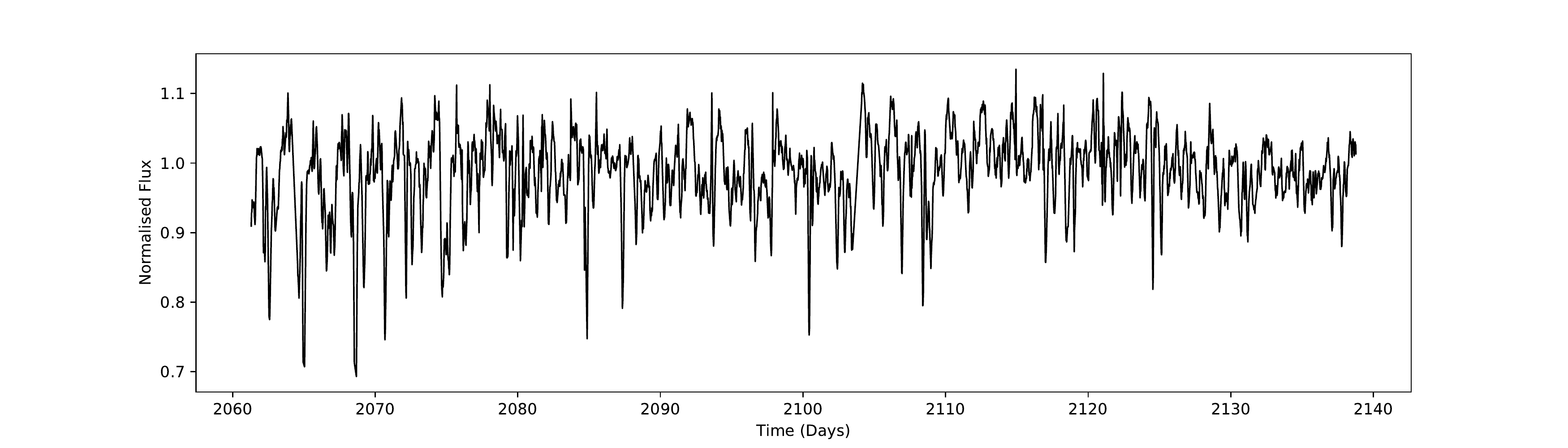}
  \caption{An example of a star in K2 C02 exhibiting the dipper phenomenon. Notable features include the magnitude of the dips, the periodicity of the events and the changing dip depth over time. Some dippers are aperiodic where either there is no clear period or the events stop occurring.}
  \label{dipperexample}
\end{figure*}

Young stellar objects (YSOs) are known to exhibit a wide range of variability based on many astrophysical processes such as stellar spots, accretion bursts, hot spots and occultations by circumstellar disks (see \cite{Herbst2012} for a discussion of variable YSOs). Of these YSOs a new class has emerged over the past few years, known as `dippers'. (e.g. \cite{Alencar2010,Morales-Calderon2011,Cody2014,Ansdell2016}). These are characterised as classical T Tauri stars (CTTS) with largely flat lightcurves and sharp, short duration dips. An example of a dipper is shown in Figure~\ref{dipperexample}.

The dipper phenomenon is characterised by the depth of the occultations (10-50\%), the reoccurrence of similar events on short time scales (from hours to days), irregularity in dip depth and changes in dip shape as a function of time. The dip depth can vary greatly in the light curves of individual objects, suggesting the material blocking the stellar light is depleted and replenished regularly, (see Figure~\ref{dipperexample}).

The stochastic and highly variable nature of dippers is attributed to occulting dusty material at the inner edge of the circumstellar disk \citep{Bouvier1999,Bodman2016}. Extinction differences between the optical and IR have been observed during dips, also suggesting that the dips are caused by dust (see \cite{Bouvier2003,Alencar2010,Mcginnis2015}).

Several works have suggested that the photometric variability of young T Tauri stars can be attributed to warps in the disk at the inner edge and/or accretion streams. \cite{Bouvier2003} suggest this warp would be caused by interactions between an inclined stellar magnetic field (with respect to the stellar rotation axis) and the circumstellar disk. \cite{Bodman2016} suggest the occulting material originates from accretion streams. \cite{Mcginnis2015} show the extinction is large and consistent between the optical and IR during dips. \cite{Ansdell2015} also suggest vortices introduced by Rossby Wave instabilities or inhomogeneities in the disk as mechanisms for creating the dipper phenomenon.

The occurrence rate of dippers ($\sim$20\%) is attributed to an inclination effect by \cite{Mcginnis2015} and \cite{Bodman2016}, whereby the occultations are only observable at inclinations of $\sim$70$^{\circ}$. However, \cite{Ansdell2016} note that three of the young dipper sample in Upper Sco and $\rho$ Oph with resolvable disks in archival sub-mm data are not at the same inclinations, and are not edge-on. They show on the contrary that one is close to face-on. They suggest that the observed dipper phenomenon is not due to a viewing angle observational bias alone. However, it is possible that these cases have inner disks which are inclined to the observer at a preferential angle for viewing the occultations of material at the inner edge. Several young stars have been observed with inner disks that are inclined with respect to their outer disks (for example \cite{Loomis2017,Thalmann2015,Marino2015}) and so such behaviour may be expected in young stars.

Dippers have been found to only occur for K and M spectra type stars \citep{Stauffer2015,Ansdell2016}. These small, cool stars have a sublimation radius that is close to the stellar surface, allowing dusty material to persist close to the host star \citep{Bodman2016}. This spatial scale is inaccessible to interferometric observations. Dippers therefore offer an alternative way to explore the inner regions of these young stars. Observing dipper stars may lead to new insights in the structure and evolution of inner disk regions.

Dippers have been observed photometrically with Spitzer, CoRoT and K2, which have provided light curves of more than 40 days of continuous monitoring with a cadence of up to 30 minutes. \cite{Cody2014} present a study of the NGC 2264 cluster using CoRoT and Spitzer, observing both the optical and infrared simultaneously. They measure the number of dippers as a fraction of the disk bearing population, naming this metric the ``dipper fraction". For NGC 2264 the dipper fraction is 21.6\%$\pm$3.7\%.

\cite{Ansdell2015} present 10 new confirmed dipper targets discovered in $\rho$ Oph and Upper Sco with a range of periods and occultation depths. K2 light curves were initially searched by eye resulting in a sample of 100 candidates, which showed evidence of aperiodic or semi-periodic dimming. They then passed their 100 light curves through a high-pass filter with a cut-on frequency of 1 day$^{-1}$ to suppress periodic variation  (e.g. from stellar rotation) and highlight quasi-periodic and aperiodic dimming events. Several metrics were applied to the light curves, resulting in their final sample of 10 dippers. 15 additional dipper candidates were also presented. These were later analysed in \cite{Bodman2016}. The metrics presented in \cite{Ansdell2015} are discussed further in Section \ref{ansdellcomp}.

In this paper we present a Machine Learning (ML) approach to finding dippers in the same K2 Campaign 2 dataset (hereafter C02) used by \cite{Ansdell2015}. Our approach is to use the K2 lightcurves alone to discover and accurately classify dippers. \cite{Bloom2011} have argued that ML approaches have a number of advantages over human-based classification, including: (1) speed and throughput, (2) repeatability of the exercise (after tuning or retraining), and (3) reproducibility. These advantages combine to enable possible biases in the ML approach to be tested, and mitigated against. In contrast, it can be hard to persuade a human to classify 13000 light curves by eye for a second time with slightly different selection criteria.

The goals of this paper is to define a reasonably unbiased sample of a class of objects which remains rather poorly defined. ML approaches have been shown to be powerful tools for the classification of objects which appear to be self-similar and distinct from other variable sources \citep[e.g.][]{Richards2012}. We can test the completeness of the machine using a confusion matrix (see Section \ref{iterations}), and find which features are highly important for distinguishing this particular class. Additionally, one advantage of an ML approach is that the classification is probabilistic, enabling statistical analysis of derived samples.

\cite{Ansdell2015} note their survey is not complete, presenting an additional 15 dipper candidates awaiting confirmation. By building a larger sample of dippers we are able to investigate population statistics of this unusual object. A larger sample also helps to define shared properties and provide further insights into the physical mechanisms which give rise to the observed variability.

In this paper we revisit the K2 C02 observations of all targets, including known (and unknown) members of Upper Sco and $\rho$ Oph. In Section \ref{observations} we discuss the input K2 data used in this paper and summarise the two regions. In Section \ref{machine} we discuss the details of the Machine Learning algorithm we have used. In Sections \ref{sample} and \ref{characteristics} the sample of dippers found by the machine learning algorithm and their characteristics are discussed. In Section \ref{bursters} we discuss the bursters found by this work. We summarise our findings and inferences from this dipper sample in Section \ref{summary}.

\section{K2 Observations}
\label{observations}

The K2 mission is the follow-up to the original \emph{Kepler} mission, launched by NASA in 2009, which failed in 2013 due to the loss of a second reaction wheel. Raw K2 photometry is known to be up to a factor of 3-4 times less precise than seen in the original \emph{Kepler} mission. K2 data is split into 80 day Campaigns, distributed around the ecliptic plane. In this work we use the data from Campaign 2, which observed Upper Scorpius and $\rho$ Ophiuchus, (see Section \ref{clusters}). The data is taken in the long cadence mode with a time sampling of 29.4 minutes.

The data from K2 C02 have been processed and made publicly available by several groups. \cite{Vanderburg2014}, hereafter VJ14, employ a Self-Flat-Fielding technique to remove the noise from the roll of the spacecraft. \cite{Aigrain2016} employ Gaussian Process regression using the 2D position of each star to model the instrumental systematics and astrophysical variability in each light curve. \cite{Luger2016} use a pixel level decorrelation to remove systematics as well as Gaussian Processes to find the astrophysical variability for each source. All of these pipelines improve on the raw K2 aperture photometry and claim to reach precision of within a factor of two of the original mission.

Dippers are by their nature highly variable targets. Their strong astrophysical variability may well be misclassified as systematic error by many reduction pipelines. Notably some pipelines sigma clip their final data product, which could potentially remove the dipper signal. We find that the reduction from VJ14 did not heavily penalise dippers by sigma clipping. Additionally it is robust against neighbouring contaminants which are expected in young star forming regions such as $\rho$ Oph. We employ the VJ14 reduction in our work, which can be accessed through the \emph{Mikulski Archive for Space Telescopes} (MAST) under the acronym K2SFF.

There are still systematics in the VJ14 K2 light curves, most notably offsets between the first half and second half of the campaign. This artefact creates a step function in the data. In order to correct for the low-order trends we remove a 30-day Gaussian smoothing kernel from the data. (Similarly \cite{Cody2014} remove a 20-day smoothing kernel from their CoRoT light curves.) This does not affect the short duration dippers (timescales $\leq$ few days) but does not completely remove the step-function artefact either. A more thorough correction could be attempted.

K2 pixels are approximately 4 arcseconds in scale. It is possible for sources to be close enough for light curves to be contaminated by a neighbouring source. Using the input RA, Dec and median K2 magnitude values for K2 C02 (which can be accessed through MAST) the catalogue searched for any targets within 20 arcseconds (5 pixels) of their nearest neighbour. In these cases the brightest source is accepted and light curve from the dimmest source is removed from further reduction. 228 contaminating sources were removed from the sample.

\subsection{Upper Scorpius and $\rho$ Ophiuchus}
\label{clusters}

Upper Scorpius and $\rho$ Ophiuchus were observed in K2 C02 for 80 days, starting on the 23$^{\rm rd}$ August 2014. The Upper Sco association is at 145$\pm$2 pc \citep{dezeeuw}, with the younger $\rho$ Oph cluster in front at 120$\pm_{4.2}^{4.5}$ pc \citep{Loinard2008}. $\rho$ Oph is a smaller cloud complex which is highly extincted, causing many stars to be extremely reddened. Upper Sco has been found to have an age of 11 $\pm$ 2 Myrs \citep{Pecaut2011} while $\rho$ Oph is a younger association with an age of 0.1-1 Myrs  \citep{Andrews2007,wilking,Luhman1999}. However, as noted in \cite{wilking}, the two cloud complexes can be hard to distinguish; there are several members on the edges of $\rho$ Oph with ages more consistent with Upper Sco.

\begin{figure}
  \centering
  \includegraphics[width=\columnwidth]{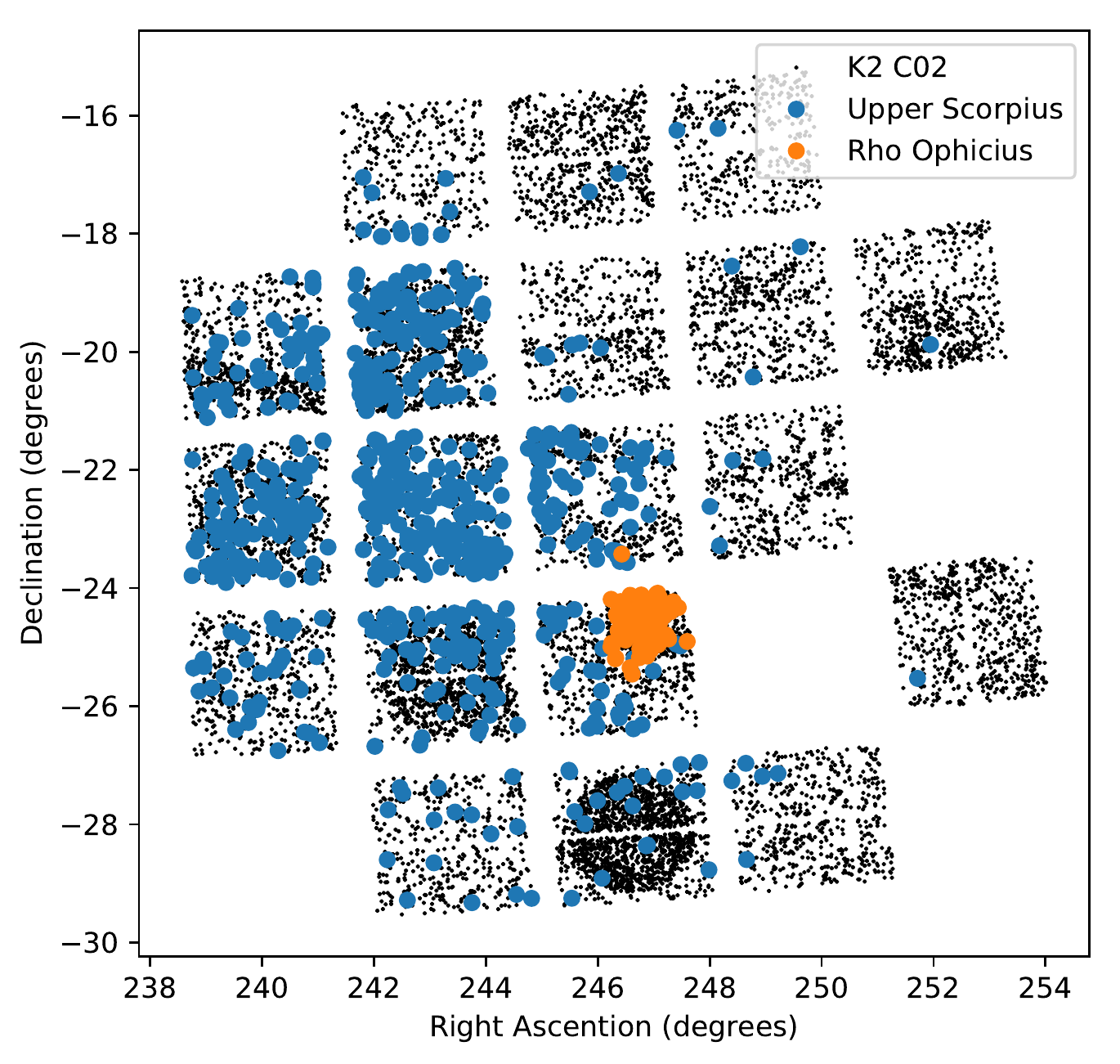}
  \caption{K2 C02 field. $\rho$ Ophiuchus members have been highlighted in red and are much more localised than the Upper Scorpius members, highlighted in blue. Here only members that are also observed in K2 C02 are shown.}
  \label{field}
\end{figure}

Figure~\ref{field} shows the K2 C02 field (centered on RA =246.12$^\circ$, Dec=-22.44$^\circ$) where 13,399 stars were observed in long cadence mode. (Note that two modules, 3 and 7, are no longer functional and appear as gaps in the field.) Targets have been taken from the K2 target list, available at MAST.

Members of $\rho$ Oph and Upper Sco are highlighted in Figure~\ref{field}. Membership lists from \cite{parks,wilking,ratzka} were compiled to create a catalogue of 2153 $\rho$ Oph members. Membership lists from \cite{luhman,Rizzuto2015,lodieu2013,dezeeuw,slesnick2008} were compiled to create a catalogue of 1349 members for Upper Sco. For each of the light curves in K2 C02 (having had any neighbouring sources removed) a membership was assigned by cross matching with each member list using a radius of 1 arcsecond. Any matches were flagged with the region name, any non matches were flagged with an 'Unknown' membership. Where membership was ambiguous, Upper Scorpius was chosen as the parent region. In the K2 C02 sample we find 708 stars to be members of Upper Sco and 199 to be members of $\rho$ Oph.

This membership list is incomplete, as we show below. There are young stars in K2 C02 with IR excesses consistent with being a member of either Upper Sco or $\rho$ Oph, some of which we identify to be dippers. With the anticipated release of Gaia parallaxes for these targets in 2018, the two star forming regions will be more easily distinguished and membership lists will be more robust.

\section{Building a Machine Learning Algorithm}
\label{machine}

Machine Learning (ML) has been used to classify the light curves of variable stars from a variety of surveys (see e.g. \cite{Dubath2011}, \cite{Richards2011}), up to and including K2 (\cite{Armstrong2015,Armstrong2016}). These previous studies have shown how to construct and test frameworks for the reliable (and testable) classification of light curve data. Our aim was to build upon these methods, and extend them to handle the new class of dipper. In particular we felt that we could add to the sample of objects discovered by \cite{Ansdell2015}. The sample of dippers used to train the machine is shown in Figure~\ref{knowndip} and given in Table \ref{dippertab}.

In this paper we have used a Random Forest (RF) algorithm to classify the K2 light curves, implemented with Python's \emph{scikit-learn} package \cite{scikit-learn}. \cite{Richards2011} tested a variety of tree-based Machine Learning classifiers on noisy time-series data, and demonstrated that RF classifiers were superior and fast.

\subsection{The Training Set}
\label{trainingset}

RF is a classification method that relies on many decision trees (which are individually weak classifiers) grouped together to create one strong classifier. The algorithm is trained on data with known classifications. The RF classifier built from the training set is then applied to the remaining K2 C02 dataset, where the classifications are unknown. The decision at each node in the decision tree for a given lightcurve depends on the numeric value of a single feature (e.g. median light curve level, number of dips, median dip depth and so on). The ML algorithm finds the thresholds at which these numeric features are cut in order to separate the training data into the known (``true") classifications.

\begin{figure}
  \centering
  \includegraphics[width=\columnwidth]{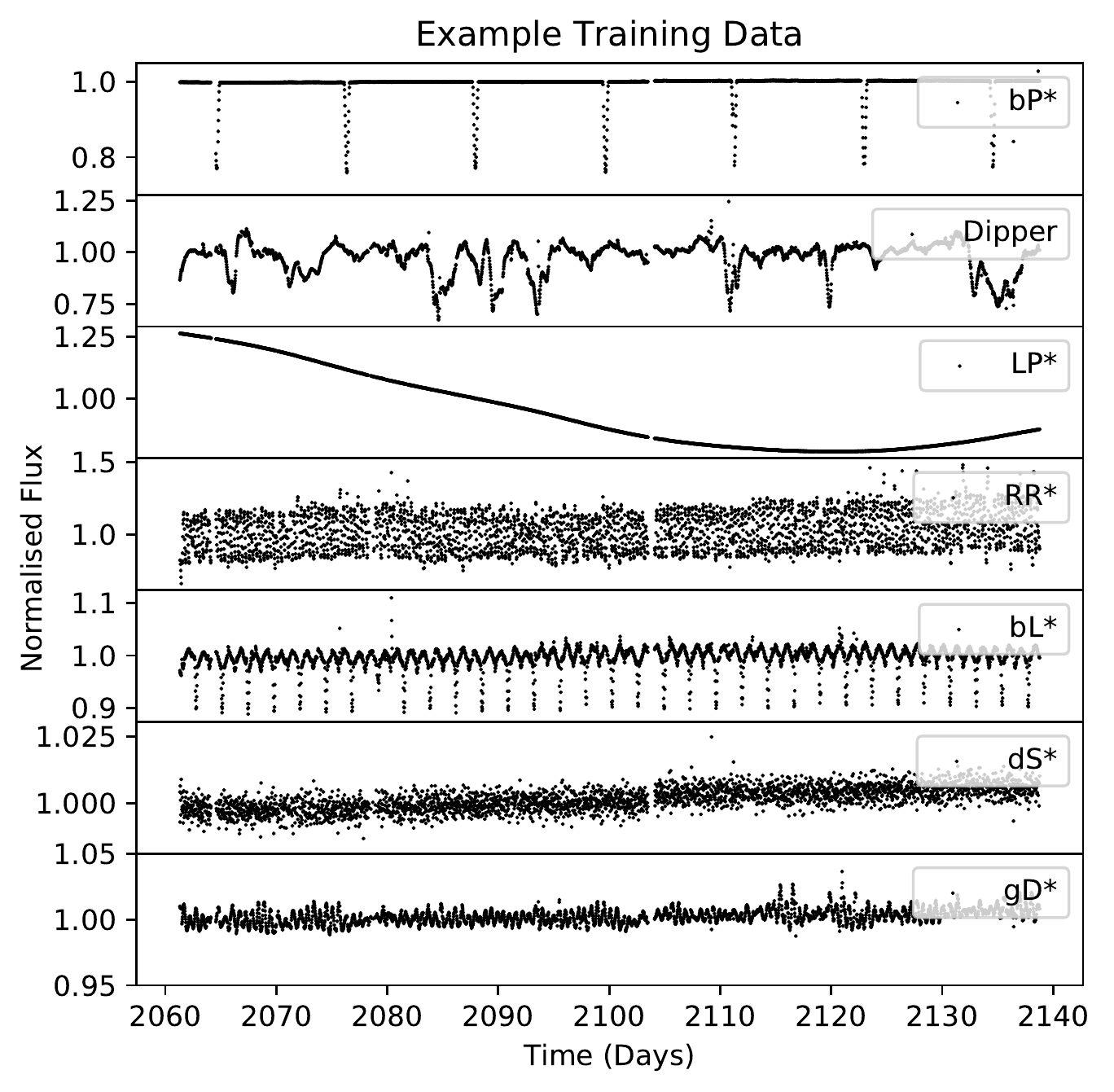}
  \caption{Example light curves of each training class for the machine. From top to bottom the classifications are Beta Persei Eclipsing Binary, Dipper, Long Period Variable, RR Lyrae, Beta Lyrae Eclipsing Binary, Delta Scuti and Gamma Doradus.}
  \label{examples}
\end{figure}

RF is a supervised classification method which requires a sample of objects with known classifications, referred to as the training set. The number of each object is given in Table~\ref{knowntab}. Examples of each object are given in Figure~\ref{examples}. To create the training set, we identified a set of K2 C02 targets which are already classified in the literature, and are clear examples of those classes that we want the machine to find. Our training set was built from four primary references, detailed below.

\begin{enumerate}
\item the K2 Variability Catalogue II \citep[K2VarCatII, ][]{Armstrong2016} (A2016), based on K2 data from campaigns 0--4, and containing 1137 classified light curves, discounting noise and periodic stars (which are mostly spot modulated). They include 7 distinct classes in their final results table: RR Lyrae, Delta Scuti, Gamma Doradus, Eclipsing Binary (Type A), Eclipsing Binary (Type B), Other Periodic and Noise.
\item the Machine-learned ASAS Classification Catalog  \citep[MACC, ][]{Richards2012}). The catalogue has been trained on 27 different variability classes, and contains 74 targets that have a classification probability of more than 60\% and are in K2 C02 sample of targets.
\item stars from the Simbad database \citep{Wegner2000} that have been classified in the literature.
\item A sample of 22 dippers in \cite{Ansdell2015} and \cite{Bodman2016}. There are 25 in total, but 3 were rejected as having low signal-to-noise light curves on inspection in the VJ14 reduction.
\end{enumerate}

\begin{table}
\centering
\caption{Initial number of objects in each training class, built from MACC, K2VarCatII, Simbad and \protect\cite{Ansdell2015}}
\label{knowntab}
\begin{tabular}{ll}
\hline
\textbf{Classification}	&	\textbf{Number}	\\ \hline
Dipper & 22 (Increasing to 40)\\ \hline
Beta Persei Type Binaries& 69\\ \hline
Beta Lyrae Type Binaries& 55\\ \hline
Long Period Variable & 45\\ \hline
RR Lyrae & 74\\ \hline
Delta Scuti & 121\\ \hline
Gamma Doradus & 32\\ \hline
\end{tabular}
\end{table}

In this paper, we have focussed on distinguishing dippers from other classes of variable. We choose not to run a more comprehensive classification exercise across larger numbers of classes and sub-classes, as in other works \citep[e.g.][]{Richards2012, Dubath2011}. However, we find the approach of classifying many types of variable produces better results than classifying dippers and 'non-dippers' only, improving our correct classification probability from 40\% to 86\%. We selected 7 classes to train the RF against:  Two classes of eclipsing binary: Beta Lyrae type (or EA in K2VarCat) and Beta Persei type (or EB in K2VarCat) We also use classifications for Long Period Variable, Gamma Doradus, RR Lyrae, Delta Scuti and Dipper. A class for 'Noise' or 'Quiet' was also trialled in early iterations. It was found while this helped the overall accuracy of the machine for other variables, it did not improve the quality of dipper classification.

The initial numbers of training targets we assigned for K2 C02 are listed in Table~\ref{knowntab}. Classes were used where at least 30 exemplars could be identified. This was found to be large enough to find meaningful results for dippers.

\subsection{Defining the Features}
\label{features}
In order to distinguish classes a set of numeric features (derived from each lightcurve) are needed. We designed features which would highlight key differences between dippers and the classes in our training set. These included measures of asymmetry, periodicity and ensemble properties of variability within the light curve, as detailed below.

It is important to choose a set of features which are representative and are, as far as possible, independent. An initial large set of features of more than 60 unique features was defined. We identified the highest ranking features using the {\em scikit-learn} package 'extra-tree-classifier' (see Section \ref{importance}). We then removed any features that were strongly correlated and allowed the feature with the highest importance score to survive. The surviving 25 features are discussed below.

We have chosen to construct features based solely on the K2 photometry, rather than auxiliary data from other surveys. Our reasons for ignoring other sources of information beyond K2 (specifically photometric colours) are discussed in Section~\ref{colours}.

\subsubsection{Literature Features}

We identified three light curve features in the literature that are particularly well suited to the classification of dippers and irregular variables, via flux asymmetries. These are summarised in Table~\ref{Richardstab}, and described briefly here. The {\it M}-statistic, described in \cite{Cody2014},

\begin{equation}
\label{Mstateqn}
M=(<d_{10\%}>-d_{med})/\sigma_d
\end{equation}

where $<d_{10\%}>$ is the mean of all magnitudes at the top and bottom decile of the light curve, $d_{med}$ is the median of all the data and $\sigma_d$ is the RMS of the light curve. In this paper we use magnitudes for all features rather than normalised flux, adapting the literature metrics where necessary.

We also adapt two features that are loosely based on features from \cite{Richards2011}. The first, \emph{mag$\_$mid20}, is the difference between the median of the lower 40th percentile and the median of the upper 40th percentile divided by \emph{mag$\_$95}, (which is the median magnitude in the top 5$^{th}$ percentile less the median flux in the bottom 5$^{th}$ percentile). The second, \emph{perc$\_$amp}, is the maximum magnitude in the light curve divided by the median magnitude.

\begin{table}
  \centering
  \caption{RF features adapted from \protect\cite{Richards2011} and \protect\cite{Cody2014}}
  \label{Richardstab}
  \begin{tabular}{lp{5cm}r}
    \hline
    \textbf{Feature}	&	\textbf{Description}	\\ \hline
    mag$\_$mid20	&	The difference between the median of the lower 40th percentile and the median of the upper 40th percentile of magnitude divided by mag$\_$95 (see text for details)\\ \hline
    perc$\_$amp	& The maximum magnitude divided by the median magnitude value. \\ \hline
    {\it M}	& The {\it M}-statistic defined in \cite{Cody2014}, see Equation~\ref{Mstateqn}  \\ \hline
  \end{tabular}
\end{table}

\subsubsection{Statistical Features}

Table \ref{Christinatab} shows simple features which are commonly used to describe photometric variability. Features such as \emph{jack1} are particularly suited to identifying stars which changed behaviour during the observing campaign. This is a useful discriminant between dippers and eclipsing binaries, and other regular variables.

\begin{table}
  \centering
  \caption{Additional light curve features used in the random forest algorithm designed to highlight dippers.}
  \label{Christinatab}
  \begin{tabular}{lp{5cm}}
    \hline
    \textbf{Feature}	&	\textbf{Description}	\\ \hline
    \emph{smooth$\_$std}	&	The standard deviation in magnitudes of the smoothed data with a kernel size of 20 days.	\\ \hline
    \emph{std}	&	The standard deviation of the light curve in magnitudes.	\\ \hline
    \emph{med$\_$abs$\_$dev}	&	The median absolute deviation of the photometry about the \emph{median: $median (|X_i-median(X_i))|$}	\\ \hline
    \emph{jack1}	&	The ratio of the standard deviation of the first half of the data to the standard deviation of the second half of the data.	\\ \hline
    \emph{diff}	&	The maximum of the magnitude less the minimum of the magnitude.	\\ \hline
    \emph{dtav}	&	The difference between the mean of the magnitudes and the median of the magnitudes.	\\ \hline
    \emph{period} 	&	The period of the light curve found by a Lomb-Scargle periodogram \citep{Lomb,Scargle}.	\\ \hline
  \end{tabular}
\end{table}

\subsubsection{Gaussian Mixture Models}
\label{groups}

\begin{figure*}
  \centering
  \includegraphics[width=\textwidth]{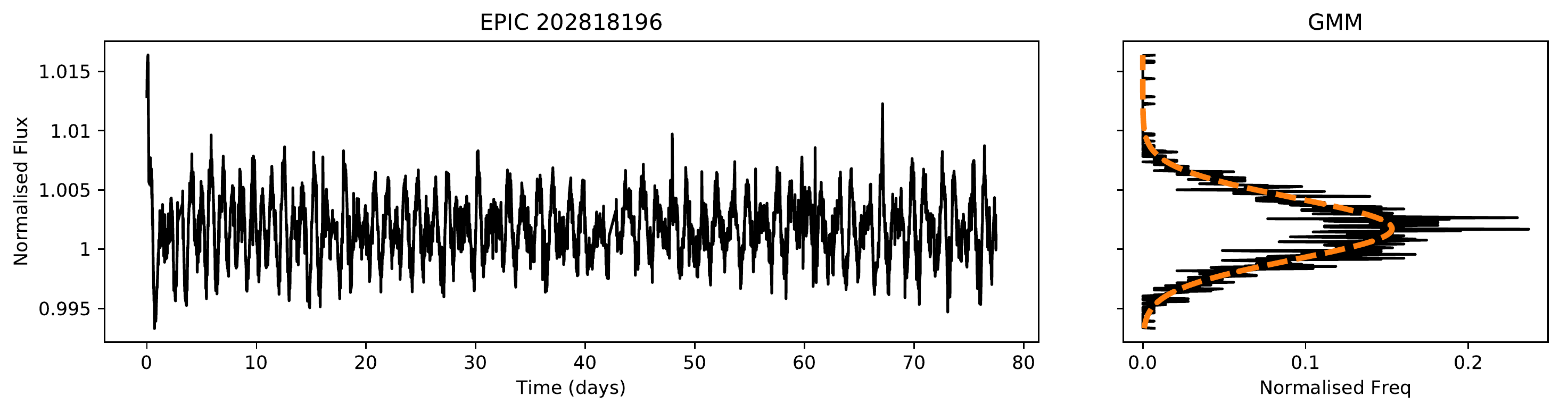}
  \includegraphics[width=\textwidth]{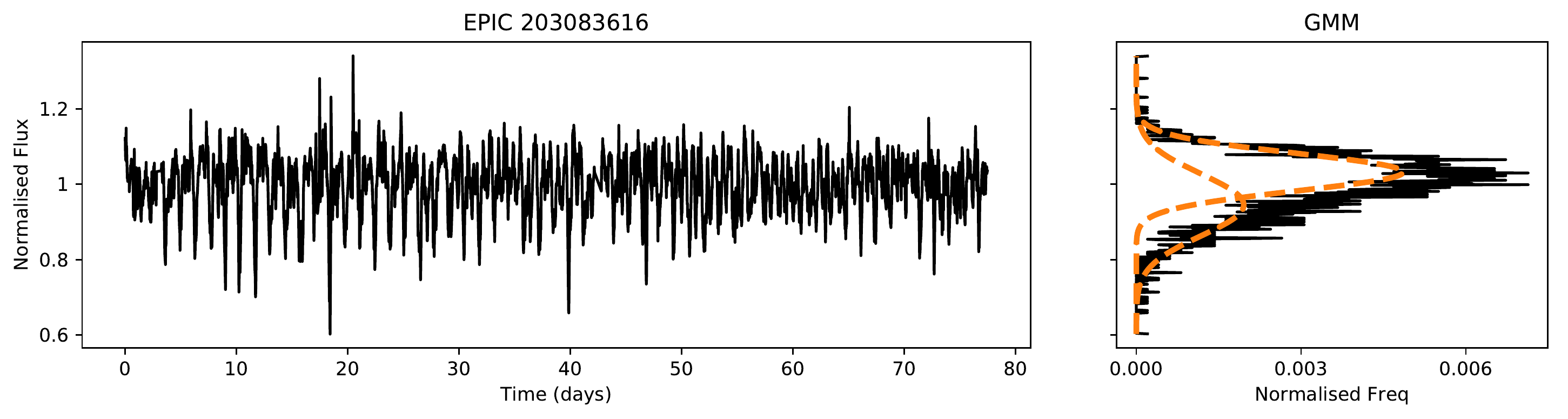}
  \caption{Example of Gaussian Mixture Models in 1D for a single light curve of a quiet star (top) and a dipper star (bottom). Both light curves have been corrected by removing a 30-day Gaussian smoothed version of the light curve. In the dipper case we see a main population from the star and a second broad population due to the dips.}
  \label{GMM}
\end{figure*}

We used 1D Gaussian Mixture Models (GMMs) in our feature set to split each light curve into different Gaussian populations. A GMM assumes all data points are drawn from an underlying, finite number of unknown Gaussian populations. Using scikit-learn's Gaussian Mixture Model package we calculated the optimal number of Gaussians to describe each light curve.

For a star showing low activity levels or noise a single Gaussian is sufficient to describe the distribution of observed fluxes, as shown in Figure~\ref{GMM}. A target with dips is also shown in Figure~\ref{GMM}, where a second Gaussian component is required to describe the flux deviations. We used the inbuilt Bayesian Inference Criteria function (BIC) to select the optimum number of Gaussians. We fixed the maximum number of Gaussians required for any light curve to be 6. In practice high numbers of Gaussians are only reached when there is large, long term variability in a light curve which has been poorly detrended (e.g. when there is a sharp discontinuity); only 2.6\% of targets reached the maximum of 6 Gaussians.

This separation of the light curve into different components enabled us to attempt to distinguish {\it intrinsic} stellar variability (e.g. photometric modulation arising from spots), from variability arising from some other process, e.g. dips or bursts. We used the median (\emph{GMMmed}) and the standard deviation (\emph{GMMstd}) of the highest weighted gaussian as a measure of the baseline and amplitude of the intrinsic stellar variable flux. We then perform a search for any dips that lie more than 3$\times$\emph{GMMstd} away from \emph{GMMmed}. Dips must consist of three or more consequtive points (all of which are 3$\times$\emph{GMMstd} away from \emph{GMMmed}). By using at least three connected points this reduces the sensitivity to small spikes due to cosmic ray hits and thruster firings. This also limits our sensitivity, any dippers that varied on time scales of less than than 90 minutes are undetectable (as K2's cadence is approximately 30 minutes).

By finding dips with respect to \emph{GMMmed} we created an additional set of features based on clusters of points, given in Table \ref{Grouptab}. These include the number of events (\emph{ngroups}), the median depth of the events (\emph{medd}) and the median duration of the events (\emph{medl}).

\begin{table}
  \centering
  \caption{RF features derived from the Gaussian Mixture Model.}
  \label{GMMtab}
  \begin{tabular}{lp{5cm}}
    \hline
    \textbf{Feature}	&	\textbf{Description}	\\ \hline
    \emph{diff2}	&	The difference between the median of the highest weighted gaussian from the GMM to the median of the second highest weighted gaussian from the GMM.	\\ \hline
    \emph{GMMstd}	&	The standard deviation of the highest weighted Gaussian from the GMM	\\ \hline
    \emph{std\_gauss}	&	The standard deviation of the entire set of Gaussian Mixture Models	\\ \hline
  \end{tabular}
\end{table}

\begin{table}
  \centering
  \caption{Features based on groups of points used in the random forest algorithm.}
  \label{Grouptab}
  \begin{tabular}{|l|p{5cm}|}
    \hline
    \textbf{Feature}	&	\textbf{Description}	\\ \hline
    \emph{ngroups}	&	Number of dipping events, defined as a cluster of at least 3 points outside 3$\times$GMMstd.		\\ \hline
    \emph{medd}	&	Median value of the depth of the dips.	\\ \hline
    \emph{medd2}	&	Median value of the depth of the 5 deepest dips.	\\ \hline
    \emph{medl}	&	Median duration of the dips.	\\ \hline
    \emph{dh}	&	Maximum depth of events less the minimum depth.	\\ \hline
    \emph{dl}	&	Maximum duration of dips less the minimum duration.	\\ \hline
    \emph{grad$\_$groups}	& The number of groups of 3 points at least 3$\sigma$ from the mean of the gradient of the light curve.		\\ \hline
  \end{tabular}
\end{table}

\subsubsection{Wavelet Analysis}
\label{Wavelets}
Wavelet analysis \citep{Torrence98apractical} is a method of describing periodicity in a time series while preserving information about the localisation of the periodicity (see \cite{Bravo2014} for a review of the method for finding rotation periods in Kepler data.) The procedure is similar to applying a windowed Fourier Transform. In wavelet analysis a wavelet function is moved along the light curve in time. The correlation between the lightcurve and the wavelet is measured at each point. The wavelet can be scaled as a function of time and the analysis repeated. This creates power as a function of wavelet time scale and position in time along the light curve, which can be used to find localised periodicity. In this case we us a 6th order Morlet function (used in \cite{Torrence98apractical}) which is a sine wave modified by a Gaussian.

Wavelet analysis is often used where the variability of a source is changing. For dippers this is particularly relevant. An object with one mode of variability (such as a pulsating star or an eclipsing binary) has a simple wavelet power spectrum that does not vary significantly with time. However, a dipper will have complex variability, changing power wherever there are changes in dip depth.

\begin{figure*}
  \centering
  \includegraphics[width=\textwidth]{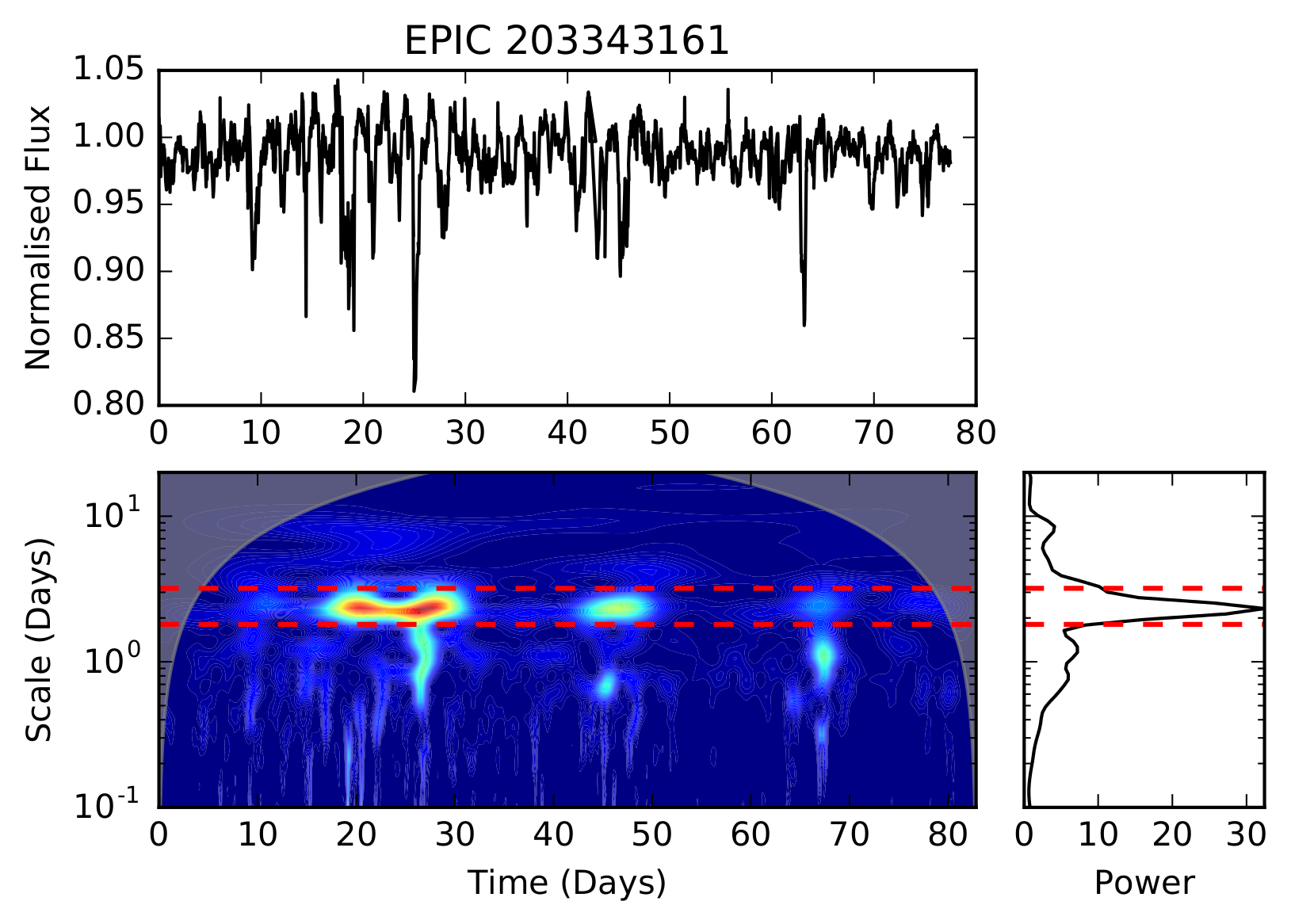}
  \caption{Wavelet power spectrum of dipper target. Top: Original light curve. Bottom left: The wavelet power as a function of time and wavelet scale. Scale refers to the width of the wavelet function in days.  Bottom right: The summed power across time. Although the periodicity changes in power as the dips decrease in depth they reappear at the same scale in days. The red line drawn over the power spectrum indicates the region selected for determining the \emph{wa\_period, s\_per} and \emph{std\_flat} features in Table \ref{Wavelettab}.}
  \label{wavelets}
\end{figure*}

An example of a wavelet power spectrum is given in Figure~\ref{wavelets}. The high power features correspond to the deepest dips in the light curve. In Figure~\ref{wavelets} although the dips change and almost disappear, they are always on the same time-scale of 2.3 days. In Figure~\ref{wavelets} the dipper is aperiodic, as the wavelet is only highly correlated at 20-30 days and 40-50 days at a wavelet time-scale of 2.3 days.

The use of wavelet analysis here allows an important feature to be measured; the change in aperiodic features over time. This can either be by changes in amplitude or by the time-scale of the wavelet. To measure this we take the region of the wavelet power spectrum that contains the highest power. This region is obtained by fitting a Gaussian to the total power (summed over all time points) and taking the one sigma region around the peak power. (The width of this peak is a feature in the machine, referred to as \emph{s\_per}.) This region is highlighted in Figure~\ref{wavelets}. This region is then averaged in the wavelet time-scale dimension to produce power as a function of time. We use the standard deviation of this slice as a feature in our machine, referred to as \emph{std\_flat} in Table \ref{Wavelettab}. This feature is one of the top 10 most important features as it helps to distinguish dippers (which behave irregularly in terms of dip depth and dip period) from beta Lyrae type variables (which have a large M statistic but are regular).

Periodic features found using this method are given in Table \ref{Wavelettab}. We additionally compute the standard deviation of the power of the Lomb-Scargle periodogram as a further feature based on the periodicity of the data. Unlike Lomb-Scargle periodograms, Wavelet Analysis requires evenly spaced data and that all time stamps have a corresponding magnitude value. For any gaps in the data we fill in points with a linear interpolation. These gaps occur infrequently throughout the data and usually last only a few hours.

\begin{figure*}
  \begin{center}
    \includegraphics[width=\textwidth]{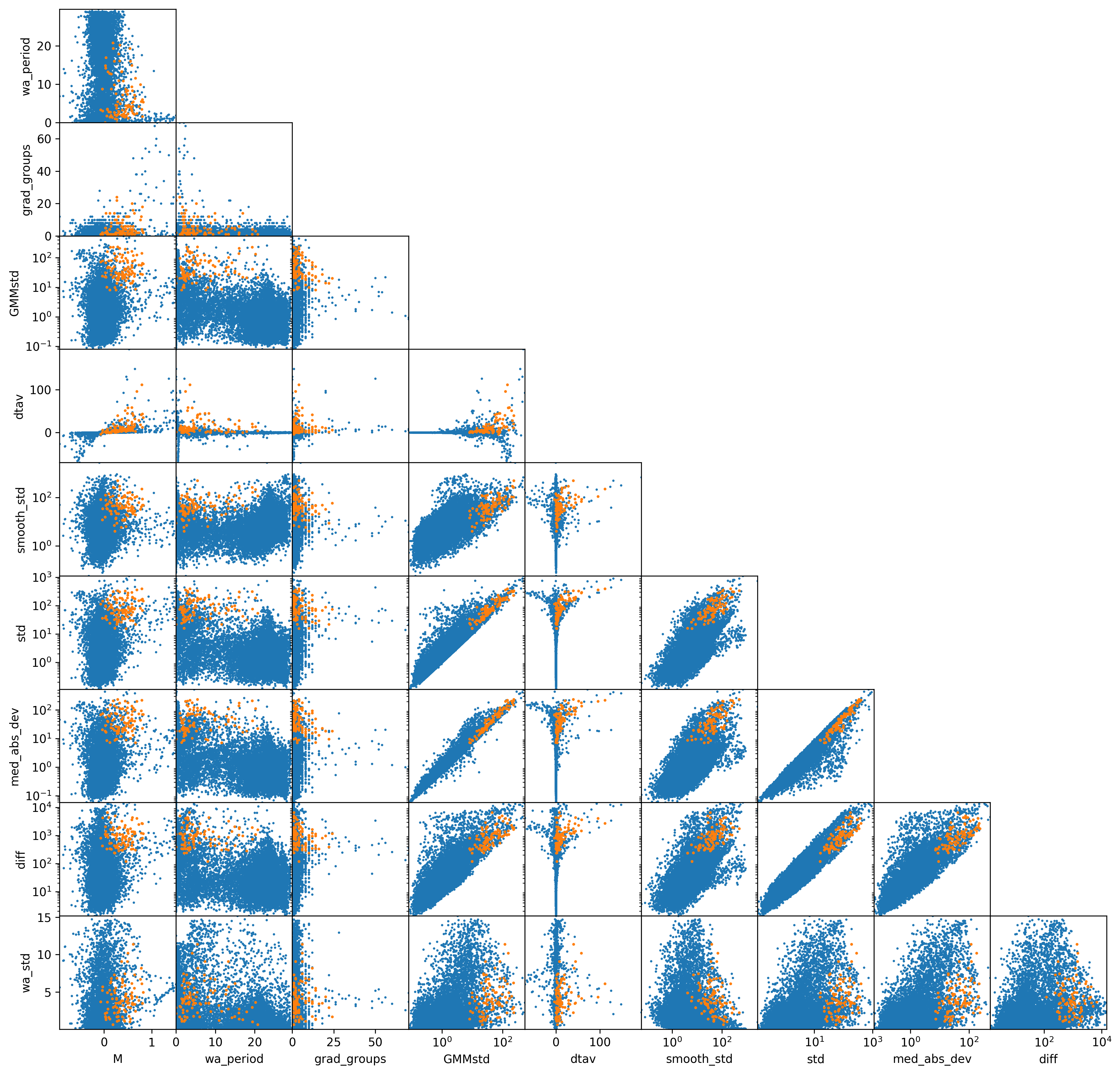}
    \caption{Correlation between the top 10 features of the ML algorithm as ranked by importance. Blue points show all 13000 light curves used in this work, orange points show the 95 dipper candidates that were confirmed by eye as discussed in Section \ref{sample}. Dippers are mixed with the population in most features and are not uniquely identifiable in any one feature. This shows the need for a ML approach.}
    \label{fig:corner}
  \end{center}
\end{figure*}

\begin{table}
  \centering
  \caption{Features based on wavelet analysis and the Lomb-Scargle periodogram used in the random forest algorithm.}
  \label{Wavelettab}
  \begin{tabular}{|l|p{5cm}|}
    \hline
    \textbf{Feature}	&	\textbf{Description}	\\ \hline
    \emph{wa$\_$period}	&	Period from wavelet analysis	\\ \hline
    \emph{s$\_$per}	&	Width of the gaussian fit to the period peak in wavelet analysis.	\\ \hline
    \emph{std$\_$flat} 	&	Standard deviation of the flattened wavelet analysis	\\ \hline
    \emph{wa$\_$std} 	&	Standard deviation of the power spectrum from wavelet analysis	\\ \hline
    \emph{pgram$\_$std} 	&	Standard deviation of the Lomb-Scargle periodogram	\\ \hline
  \end{tabular}
\end{table}

\subsection{Feature Importance}
\label{importance}
\begin{figure}
  \centering
  \includegraphics[width=1.0\linewidth]{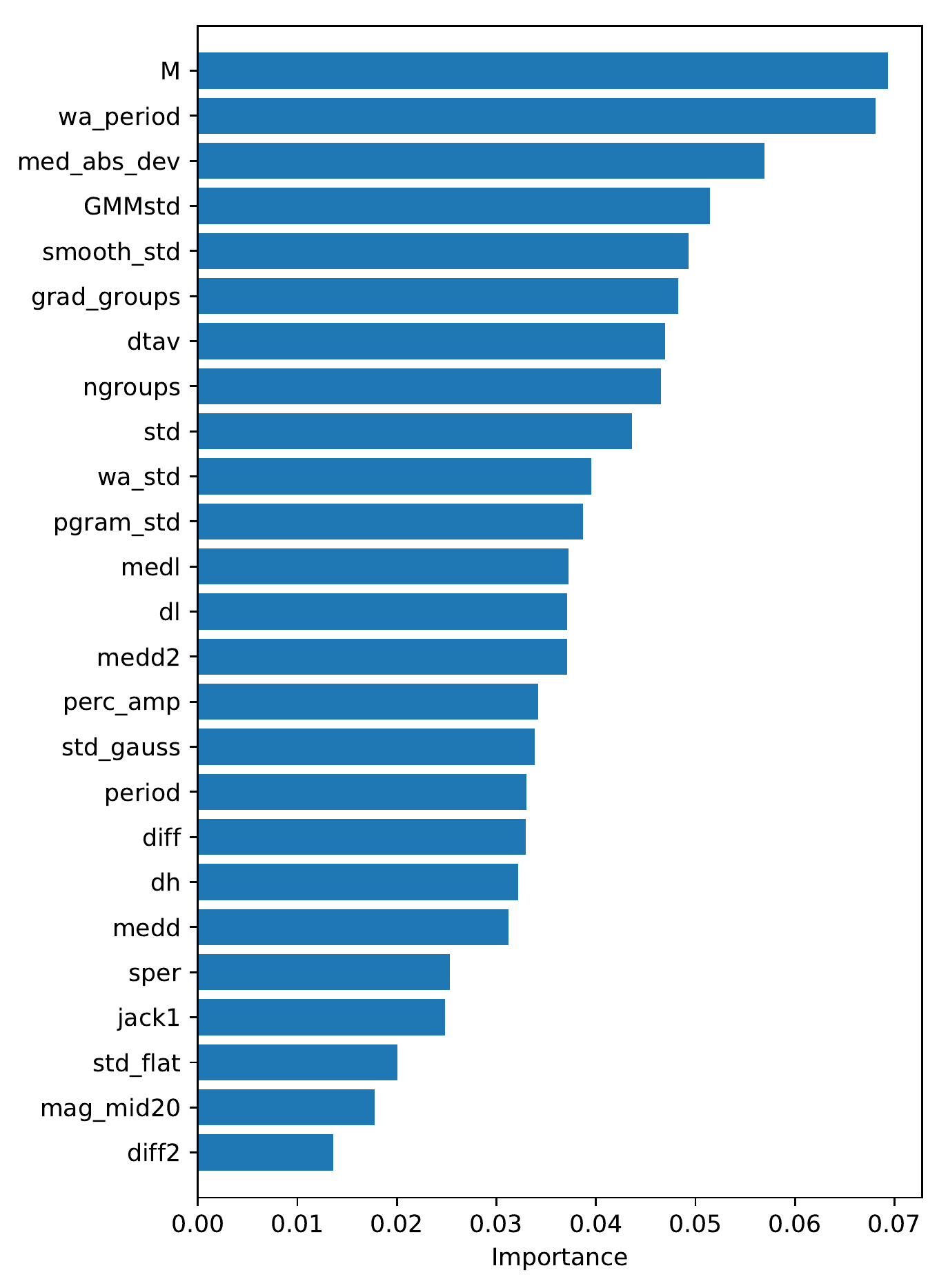}
  \caption{Importance score for the classifier when using only two classifications, ('Dipper' and 'Non-Dipper'.) We see that the most important features for this class of object are period, \emph{M-statistic}, the number of groups of turning points and the standard deviation. Note that this is the relative importance of each feature, with all features summing to one.}
  \label{fig:importance}
\end{figure}

Using scikit-learn's 'extra-tree-classifier' function it is possible to rank each feature by weighted importance to the classification, which sums to 1. The importance scores for each of the features used in this work are shown in Figure~\ref{fig:importance}. Periodic features and \emph{M-statistic} rank highly, showing they are important to the classification.

Figure~\ref{fig:corner} shows how the top 10 high importance features correlate, with the dipper sample found by this work highlighted. The dipper sample is not clear or unique in any of these features, showing the need for a large number of features and an ML algorithm to distinguish between classifications.

\begin{figure}
  \begin{center}
    \includegraphics[width=1\columnwidth]{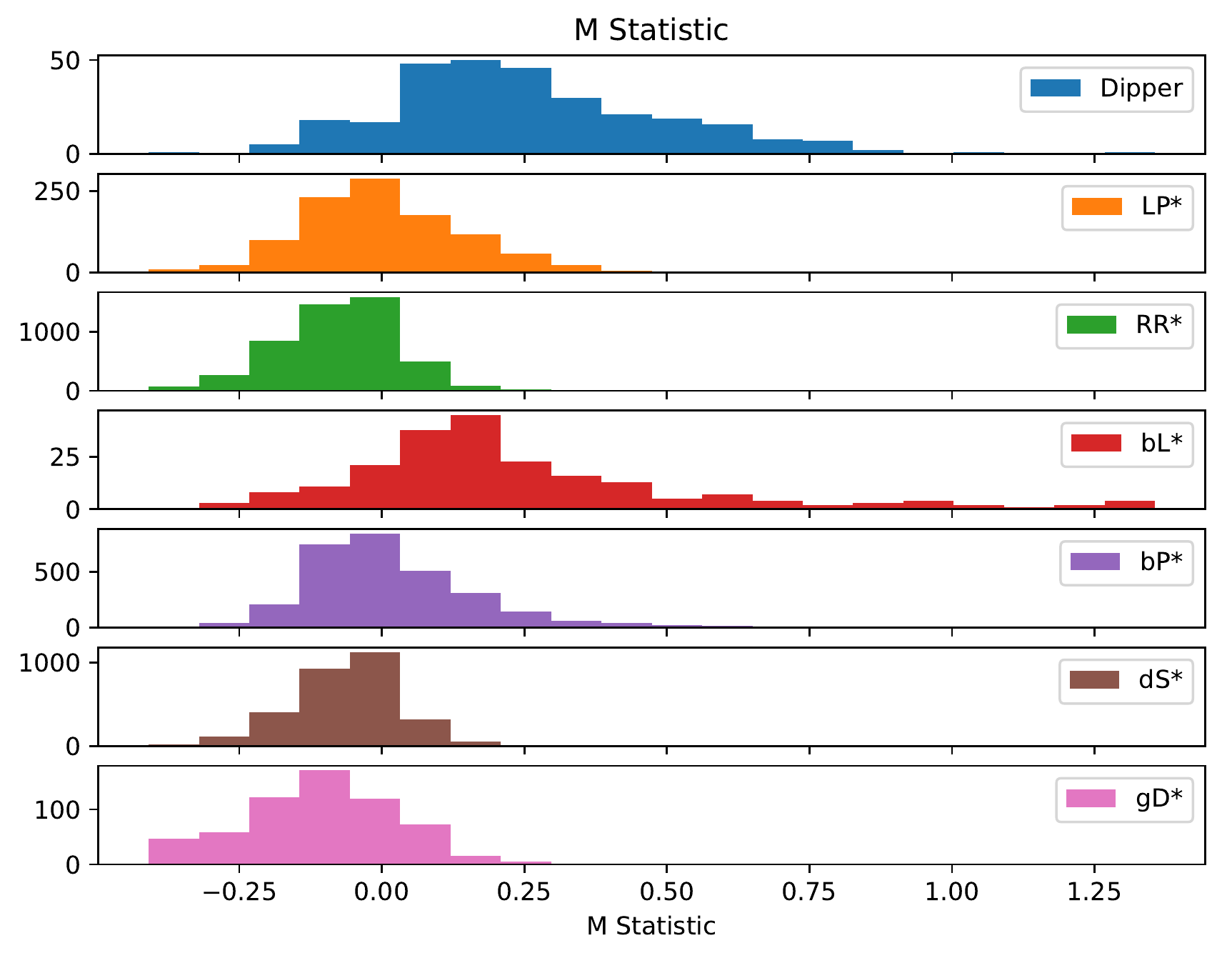}
    \caption{Distribution of each class of object in \emph{M-statistic}, having run the ML algorithm. The classes are (from top to bottom) Dipper, Long Period Variable, RR Lyrae, Beta Lyrae Eclipsing Binary, Beta Persei Eclipsing Binary, Delta Scuti and Gamma Doradus. Note that using just the feature \emph{M-statistic} is not enough to distinguish dippers from Beta Lyrae Eclipsing Binaries. Additional features from the machine are needed.}
    \label{fig:1dfeature}
  \end{center}
\end{figure}

For example, Figure~\ref{fig:1dfeature} shows the frequency of each classification across the value of the \emph{M-statistic}, the most important feature. While dippers are skewed towards a high value of the \emph{M-statistic}, the behaviour is mimicked in Beta Lyrae type eclipsing binaries (bL*) and Gamma Doradus stars (gD*). This is expected, as \emph{M-statistic} measures the asymmetry of the light curve. Eclipsing binaries will show frequent, deep eclipses and some pulsators will also be skewed. More features are needed than a simple \emph{M-statistic} to distinguish these classes. For example, features such as \emph{std\_flat} are helpful in distinguishing dippers from Beta Lyrae eclipsing binaries.

\subsection{Testing and Tuning the Performance of the Random Forest Classifier}
\label{Performance}

In order to test the efficiency of the RF classifier, a `test' portion of the training sample is kept to one side. The true classifications of this set are already known. The RF classifier is then trained on the remaining `training' data set. The trained RF classifier is then run on the remaining `test' set to find the efficiency of the machine. The true classifications are compared to the predicted classifications, producing a confusion matrix. We use a 20\%/80\% test-train split. There are a relatively small number of exemplars in each classification, shown in Table \ref{knowntab}, which can produce some scatter in the results. To avoid errors from the small training sample we run the classifier 200 times and record the result for each iteration.

\subsubsection{Iterative Training Set Selection}
\label{iterations}
\begin{figure}
  \centering
  \includegraphics[width=\columnwidth]{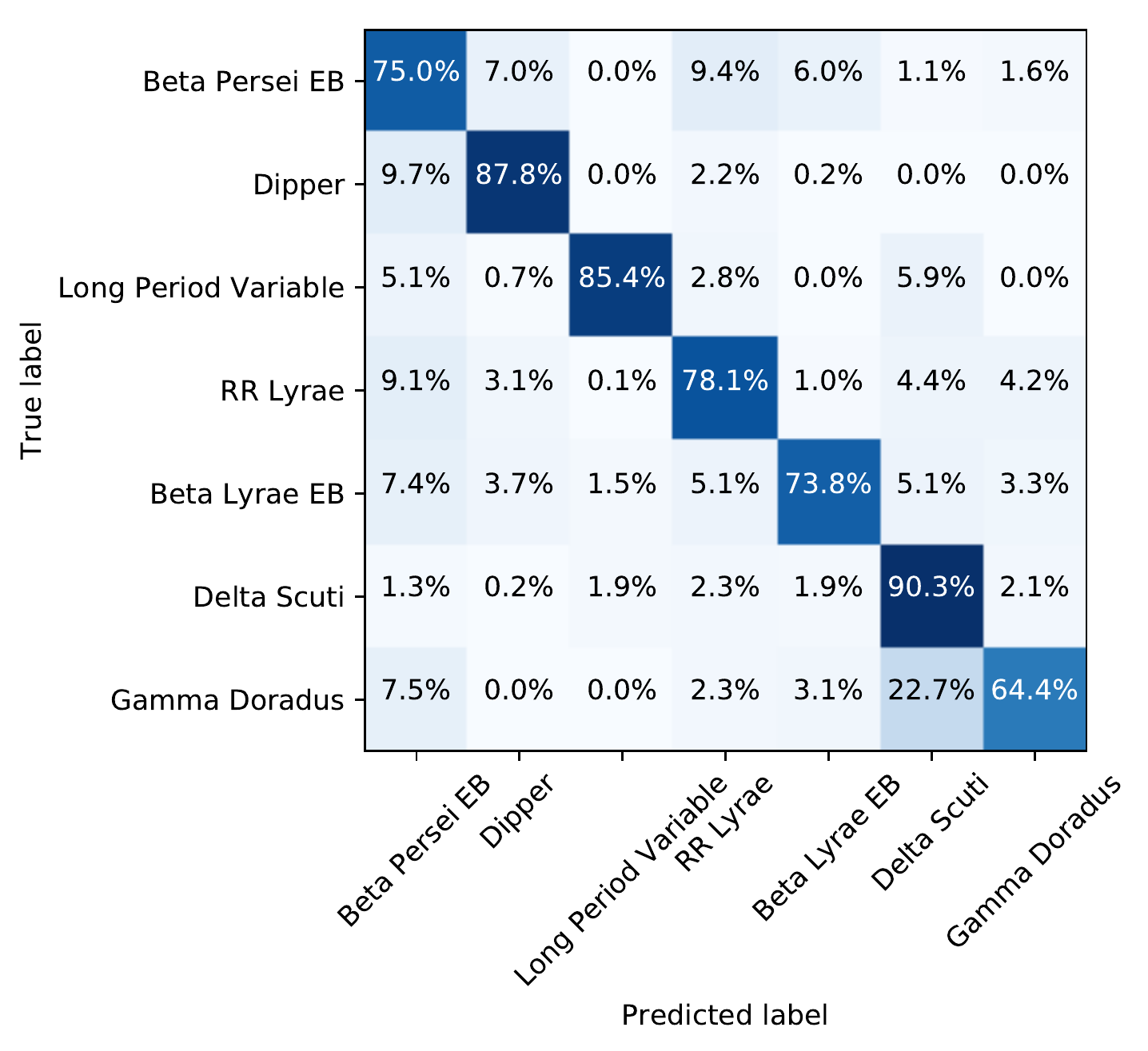}
  \caption{Confusion matrix for the final classifier after 200 iterations using an 20\%-80\% test-train split. $\sim$86\% of dippers were correctly classified, with most of the incorrect dipper classifications lying in Beta Persei or Beta Lyrae type binaries. The matrix is highly diagonalised showing that the method works well for classifying variables.}
  \label{confusion}
\end{figure}

Once the RF classifier has been built it is then possible to use metrics such as the confusion matrix, (see Figure~\ref{confusion}), to evaluate the quality of the machine. After the first iteration of the machine the classifier did not perform adequately with our small initial sample of dippers. Using only the literature sample of 22 dippers in K2 C02 provided a correct classification of the training set of 68\% for dippers. This small sample caused approximately 32\% of dipper candidates to be misclassified, mostly as Beta Lyrae or Beta Persei type eclipsing binaries. To improve on this we manually vetted the results of the machine learning algorithm in this first iteration with the 22 literature dippers as a training sample, and identified 18 targets that were clear examples of dippers. These new candidate dippers stars were added into the training set to give 40 training targets. After retraining the classifier this increased our correct classification percentage from 68\% to 86\%. This classifier was then applied again to the K2 C02 data to find our final sample. Figure~\ref{confusion} shows the confusion matrix after the machine had been trained on 40 dippers. The matrix is highly diagonalised showing that the classifier performs well and correctly classifies each variable type in the majority of cases.

\subsubsection{Exclusion of IR Colours}
\label{colours}
When infrared colours from the 2MASS and WISE surveys \citep{twomass,WISE} are included as features, the classification efficiency for dippers is increased to 95\%, a significant improvement from 86\% without colours.

We elected not to use photometric colour information for two reasons. Firstly there are many stars in the K2 C02 field with incomplete colour information, which complicates their inclusion in the ML algorithm. Secondly, adding colour information would add weight to targets with obvious disks. To ensure this classifier works for stars of any age (and will still perform well for stars with little to no IR excess) colour has been omitted as a feature. This ensures we are not preferentially selecting young stars as dippers, and should not miss dipping candidates that are diskless, thus further testing the hypothesis is related to the inner regions of the circumstellar disk.

\section{Dipper Candidates}
\label{sample}

Once the classifier had been trained on the targets found in Section \ref{trainingset} the same algorithm was used on unclassified data. This algorithm was run on 13,399 unique lightcurves with no information on membership. This gave each lightcurve a classification probability of being a dipper. The exercise was repeated 200 times (where the machine was retrained each time with a randomised training set of 80\% of the dippers), and the probabilities for each lightcurve were recorded. The machine returns 290 candidate dippers. We took targets that had a probability assigned by the ML algorithm of 40\% or more averaged over all 200 iterations, resulting in 198 candidates. Targets fainter than 18th magnitude were not included in this analysis; on inspection they have signal to noise ratios that are too low to reliably classify. We remove targets that are 18th magnitude or fainter, leaving 193 candidates.

The 193 lightcurves were examined by eye to remove any cases where strong systematics may have contributed to a misclassification. Of this list 69 were found to be other variable types that had been misclassified, including stellar pulsators and eclipsing binaries. Misclassifications were identified most commonly periodic variable type stars with additional systematics that mimicked dipper behaviour. (For example a discontinuity in the lightcurve caused an apparent change in the variability of the star.) The set was further cleaned to remove any dipper candidates with small flux variations that could reasonably be due to spots on the surface of the star; candidates with flux dips less than 10\% or less than three dips were removed from consideration. The number of targets with each classification after the procedures described above is given in Table \ref{Resultstab1}.

\begin{table}
  \centering
  \caption{Results of the machine learning algorithm, having been cleaned by eye. Here a fringe case has been rejected for being close to star spot behaviour. A second stage of eyeballing was undertaken for targets that were classified with a low dipper probability. This increased the number of dipper targets slightly and the number of burster targets significantly.}
  \label{Resultstab1}
  \begin{tabular}{|l|p{3cm}|}
    \hline
    \textbf{Type}	&	\textbf{Number of Targets}	\\ \hline
    Dipper	&	91 (rising to 95) \\ \hline
    Burster	&	10 (rising to 30) \\ \hline
    Fringe/Spots 	&	23	\\ \hline
    Misclassification 	&	69\\ \hline
  \end{tabular}
\end{table}

A second subset of 204 lightcurves were classified as a dipper with a probability of more that 20\% but less than 40\% across the 200 iterations. These were also eyeballed to find any false negatives. Out of these 204 objects, 4 candidates were manually relabelled as dippers.

In these two phases we also identified a set of 30 lightcurves by eye, which belong to a distinct class that we have labelled as bursters (see Section \ref{bursters}). There are cases where the classification is uncertain, for example when the target is mostly symmetric about the median flux. In such cases the target was put in the burster class to ensure a clean dipper sample. Candidates were labelled as bursters when they were visually either mostly symmetric around the mean flux or mostly exhibiting flares with little to no dipping. Some dipper candidates have been removed by visually reclassifying these objects (e.g. see EPIC203870058 in the Appendix, which both dips and bursts.) 

All bursters found in this work are shown in the Appendix. As the machine was not trained on this particular class we can not be certain of completeness. We expect there to be other burster examples that have not been found here. As discussed in Section \ref{bursters}, the sample of Bursters found in this work is too small to warrant a new training classification in the machine learning process.

After eyeballing, we were left with a final set of 95 dippers, including the 40 objects used to train the machine. Our final sample is given in the appendix of this work. Of this 95, 22 belong to $\rho$ Oph, 43 belong to Upper Sco and 31 have an unknown membership.

\subsection{Verifying the Machine With Colour}

\begin{figure}
  \centering
  \includegraphics[width=0.5\textwidth]{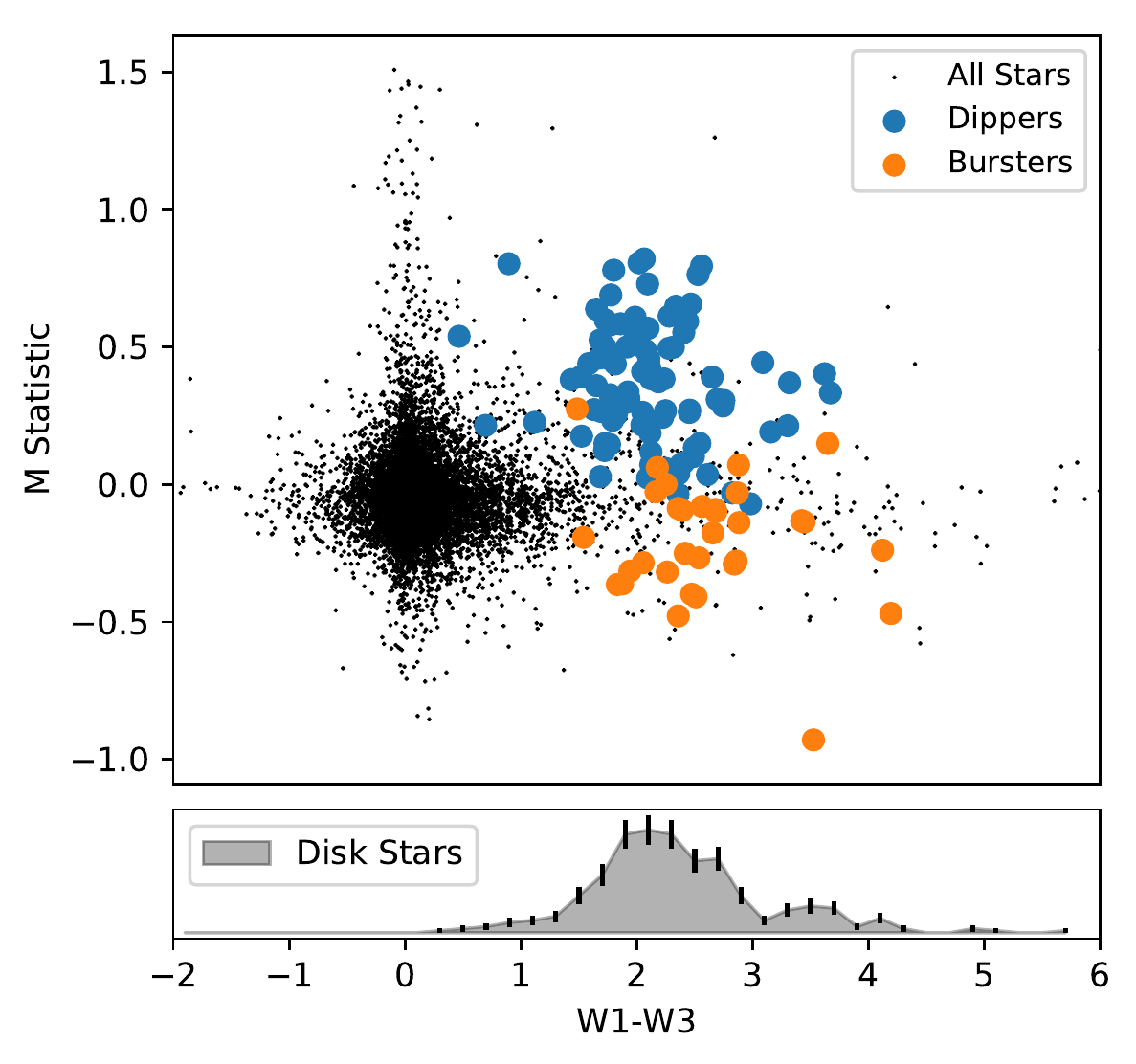}
  \caption{\emph{M-statistic}, defined in Eq \ref{Mstateqn}, as a function of W1-W3 IR excess for K2 C02 found using a random forest algorithm. The algorithm was trained only on light curve parameters and shows dippers in K2 C02 occur in systems with an infrared excess indicative of hosting a disk. Upper panel shows Dipper and Burster population as a function of colour and M statistic. Lower panel shows the distribution of disk bearing stars as a function of colour based on $\alpha>-1.6$ where $\alpha=d\log\lambda F_\lambda/d\log\lambda$. All dippers and bursters are consistent with disk bearing stars as discussed in Section~\ref{disks}}
  \label{Mstat}
\end{figure}

By omitting colour information we can independently check the performance of the ML algorithm. As dippers are expected to host protoplanetary disks, we would expect most, if not all the dippers to show an IR excess and be members of either young star forming region (distinguished in colour space from the background population). Figure~\ref{Mstat} shows the dipper sample as a function of IR excess and \emph{M-statistic}. Despite not training on colour information we see that the dipper and burster populations are clearly biased towards a high infrared excess. We find all are consistent with being young, disk bearing stars, providing yet more evidence that the dipper and burster phenomenon are related to disk material. Any targets where there is a high IR excess and high \emph{M-statistic} were inspected by eye and found to be either eclipsing binaries or contain discontinuities due to the VJ14 data reduction process. Some targets also show a large increase in flux due to the transit of Mars through the K2 campaign, which can also cause a high \emph{M-statistic}.

\section{Dipper Characteristics}
\label{characteristics}

With this new large sample of 95 dippers we can now revisit some of the basic properties of the class. We pay particular attention to the features derived from the K2 photometry, which were used to classify the dippers, combined with infrared photometry from 2MASS \citep{twomass} and WISE \citep{WISE}. We also investigate differences between dippers in Upper Scorpius and $\rho$ Ophiuchus.

\subsection{Dip Parameters}

Parameters such as the dip depth and duration are useful for characterising the behaviour of individual dippers. If the theory of the inner edge of the disk obscuring the star is correct (such as those from \cite{Bouvier1999,Alencar2010,Mcginnis2015,Bodman2016}), the depth of the dip should relate to the cross-sectional area of material obscuring the star. Additionally, the period and duration of the dips should relate to the orbital distance from the star. To measure these parameters we first find all dips using the methods discussed in Section \ref{groups}.

We measure the depth and duration of dips using a more detailed procedure than the one used in to build features (see Section~\ref{features}). This more detailed approach takes too long to run on all 13,399 targets in K2 C02 to generate features, but more accurately describes the dip parameters. This procedure is performed on a lightly smoothed version of the data; a boxcar is used at a 25\% of the best fit lomb-scargle period to remove any short term variability unrelated to the dipping.

First we calculate a level that approximates the stellar baseline. We create a heavily smoothed version of the lightcurve using a Gaussian smoothing kernel at twice the period of the dipping events (up to a maximum of 20 days). (We use the Lomb-Scargle period found above). This ensures sharp features and dips are removed.  This smoothed lightcurve is offset to the median level of the stellar baseline as assessed by the GMM (see Section \ref{groups}). We find all dips that have minima 3 \emph{GMMstd} fainter than this continuum level.

\begin{figure*}
  \includegraphics[width=1.\textwidth]{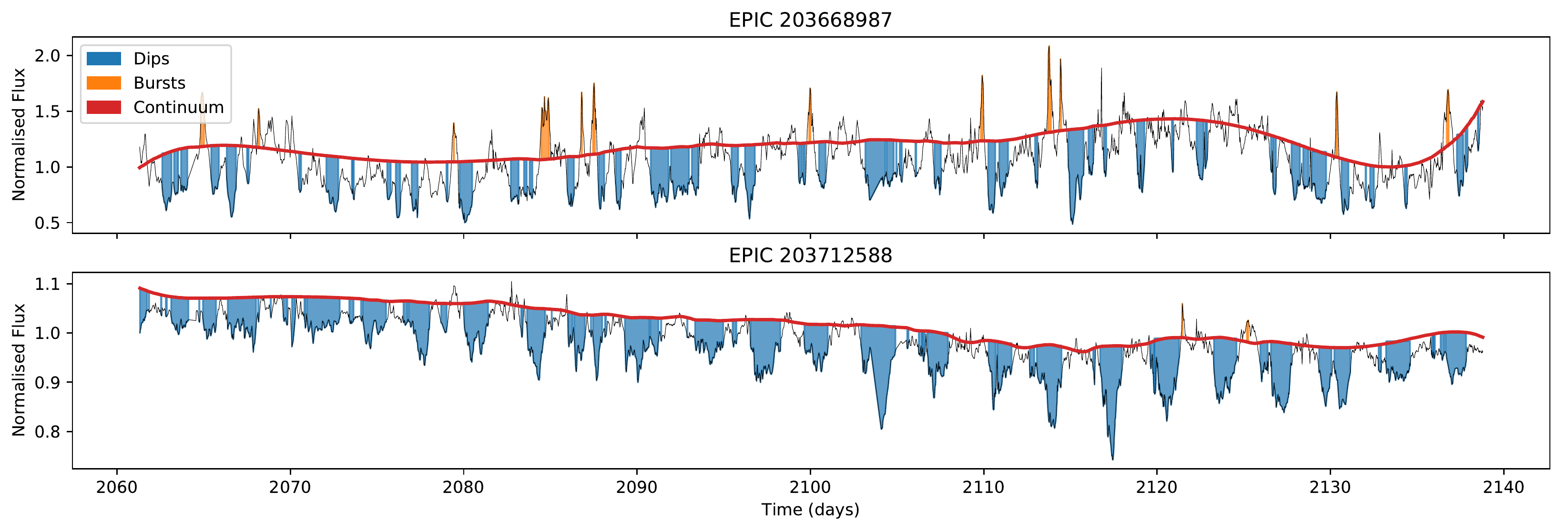}
  \caption{Example of the output of the dip depth and duration finding algorithm. Here the red line shows the continuum level, which is a simple 20 day smooth of the data after outliers have been removed. This is also offset to the continuum level as assessed by the GMM. Blue areas indicate features that have been labelled as dips, with the width and height of the area indicating the duration and depth of the dip. Orange areas have been labelled as bursts.}
  \label{fig:walkexample}
\end{figure*}

We measure the depth and duration of each event by finding the minimum of the dip and 'walking' up either side until either the continuum is met or there is a turning point (assessed by the gradient of the data). This creates an allowed region (shown in blue in Figure~\ref{fig:walkexample}). The duration of the dip is the width of the final allowed region. The dip depth is the average continuum level in that region less the minimum flux found in the dip.

The same procedure is used to find bursts by finding all events that are 3 \emph{GMMstd} above the continuum. These are shown in orange in Figure~\ref{fig:walkexample}.

This approach is still simplistic. Bursts and dips are sometimes confused, particularly in cases where there is a strong long term trend in the data. Some dips are also clearly the convolution of two or more events, which are harder to separate. While this method is not ideal, it provides statistics on the dipper population as a whole.

\subsection{Comparison with Ansdell et al. 2015}
\label{ansdellcomp}
In \cite{Ansdell2015} dippers were defined by first smoothing with a high pass filter at 1 day and then measuring the following: the number of dips (N$_{Dip}$), the average dip depth of the three deepest dips as measured from the normalised light curve (D$_{Dip}$), the ratio of the dip depth to the standard deviation of the data (R$_{Dip}$) and the rotational period of the star P$_{Rot}$.

\begin{figure*}
  \centering
  \includegraphics[width=\textwidth]{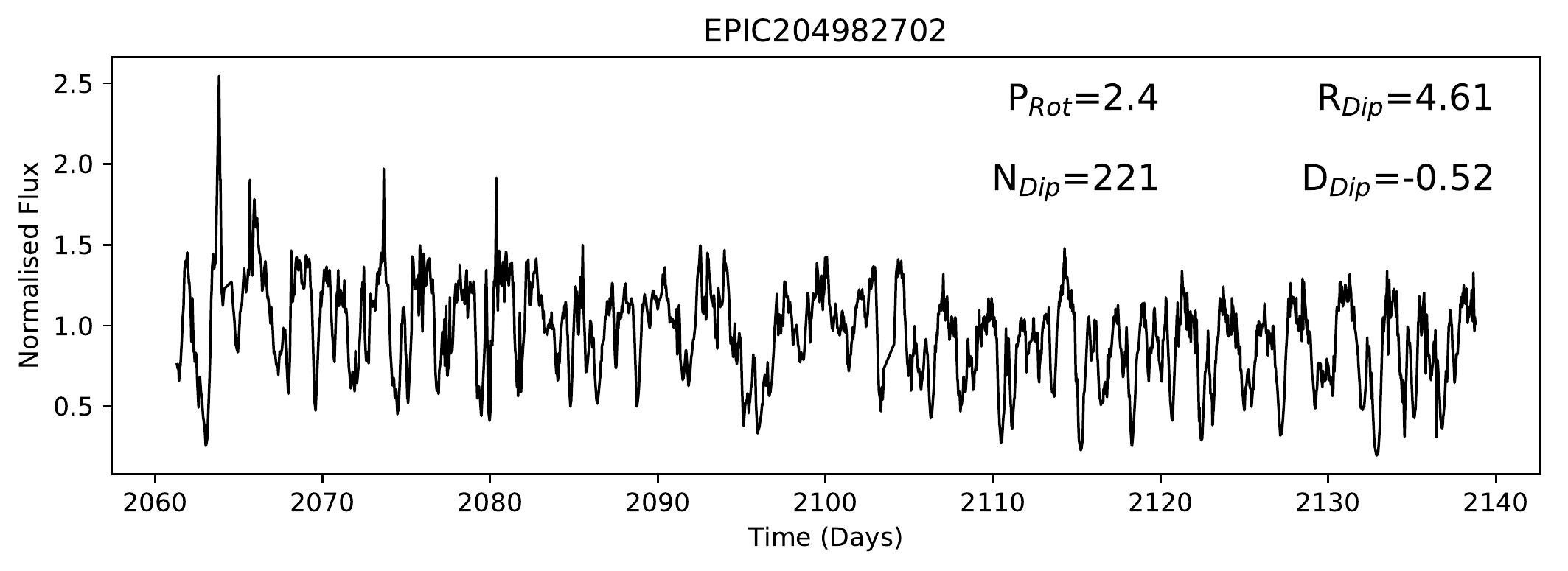}
  \includegraphics[width=\textwidth]{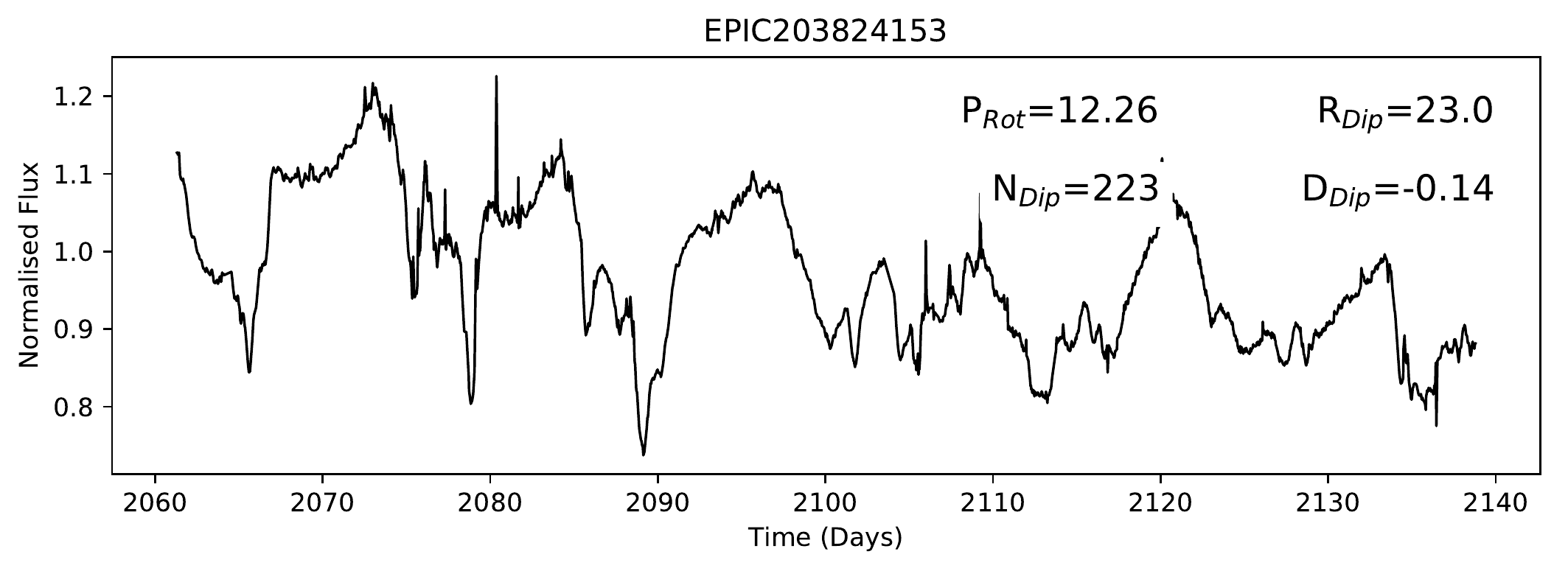}
  \caption{Example of two dipper targets that did not meet the criteria set out in \protect\cite{Ansdell2015}. \emph{Top}: The R$_{Dip}$ parameter here is slightly less than 5 as the dips are narrow and frequent. \emph{Bottom}: The period of the dipping is approximately 12 days which is longer than the 10 day cut.}
  \label{ansdellexample}
\end{figure*}

A dipper was then defined as a target with N$_{Dip}>$5, D$_{Dip}>$0.07, R$_{Dip}>$5 and P$_{Rot}$ to be less than 10 days. We find when we compare our sample of 95 dippers to this metric 35\% do not qualify. This is in most cases due to our dippers being aperiodic with no clear value for P$_{Rot}$. We also found that the R$_{Dip}$ criteria clipped out a small number of dippers that were symmetric (i.e. bursted as well as dipped). Two examples of dippers found by this work that do not qualify based on this metric are shown in Figure~\ref{ansdellexample}. We suggest that these criteria may be too strict and cut out valid dipper candidates.

\begin{figure}
  \centering
  \includegraphics[width=1.\columnwidth]{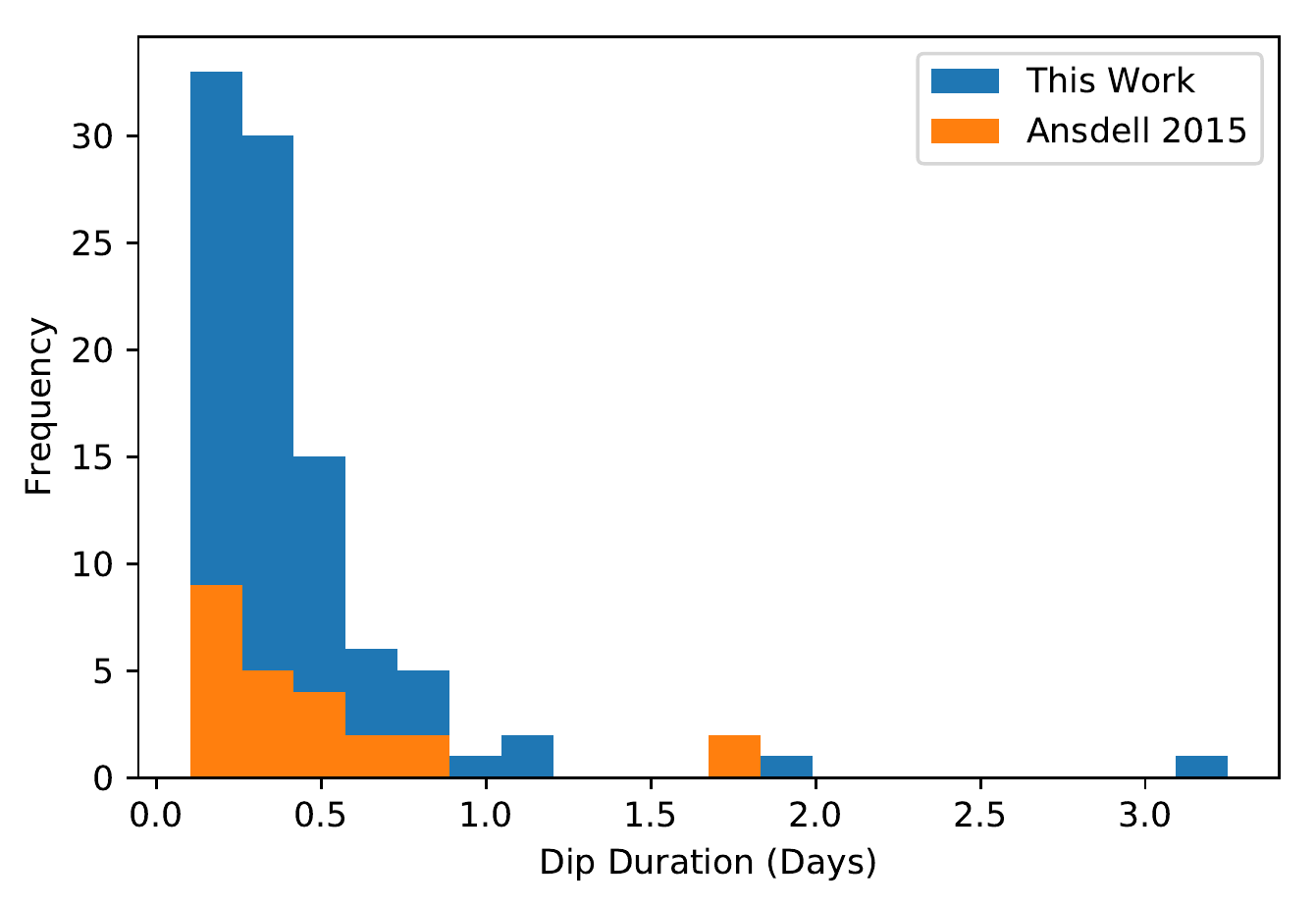}
  \includegraphics[width=1.\columnwidth]{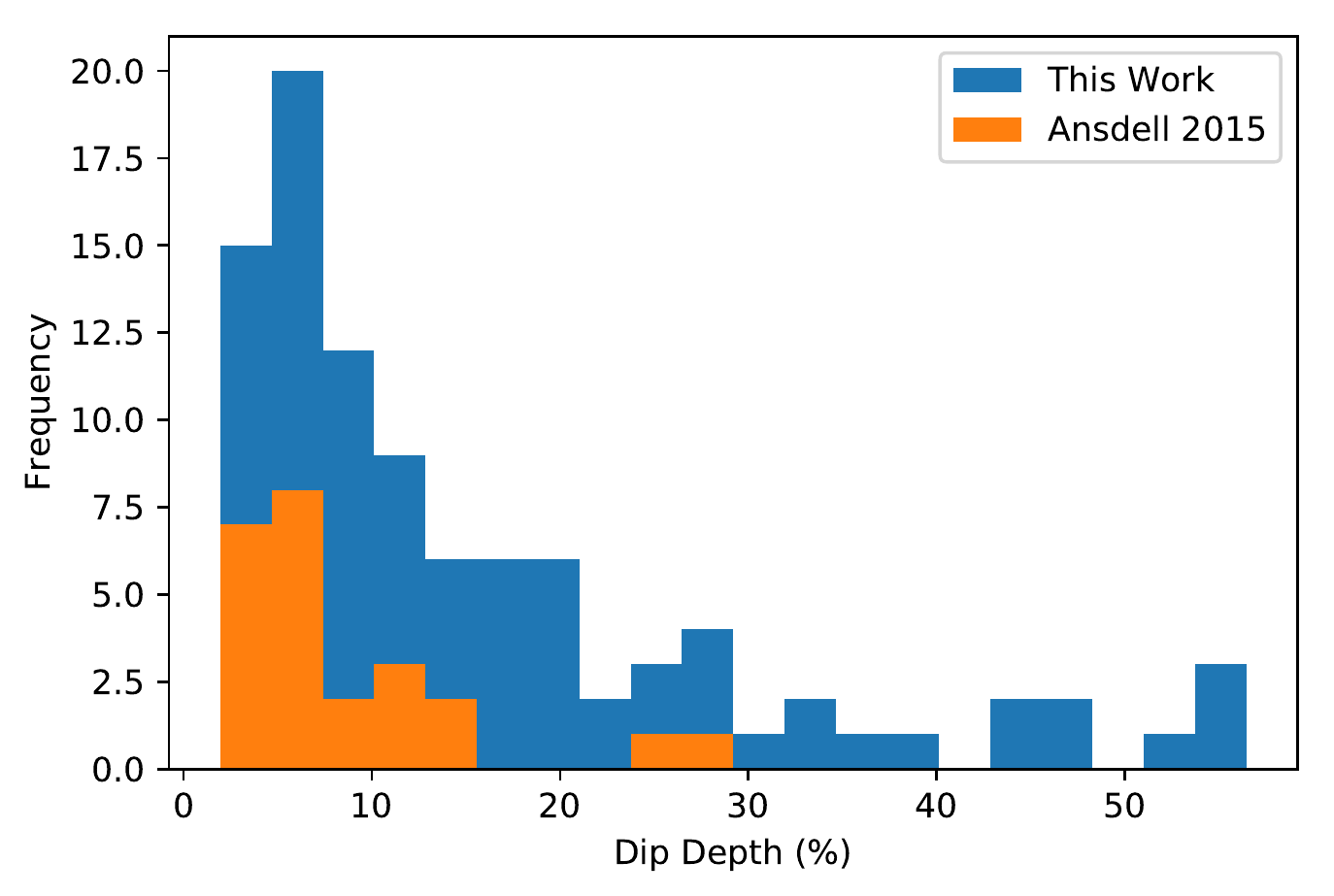}
 \caption{Distributions of median dip depth and median dip durations for all 95 dipper targets. Blue: Dipping events found in this work using machine learning. Orange: 24 dippers found in \protect\cite{Ansdell2016} also used in this work. We see the two distributions are similar, suggesting there are no additional biases for or against these two parameters when machine learning is used.}
  \label{fig:completeness}
\end{figure}

Figure~\ref{fig:completeness} shows the distributions of median depth and median duration calculated for our sample of 95 dippers and calculated for the 24 dippers found in \cite{Ansdell2015}. They show similar distributions, suggesting the surveys have similar sensitivities.

In general it is easier to find deeper dips as they are more easily distinguished from stellar noise. As we discount any targets with dips small enough to be explained with star spots, we may therefore be under-representing the dippers at the smallest dip depths. Similarly, as the cadence of the data is 30 minutes, we do not have the time resolution needed to observe shorter duration dips than 1.5 hours. We may be under-representing the sample of 'short period' dippers.

\begin{figure}
  \centering
  \includegraphics[width=\columnwidth]{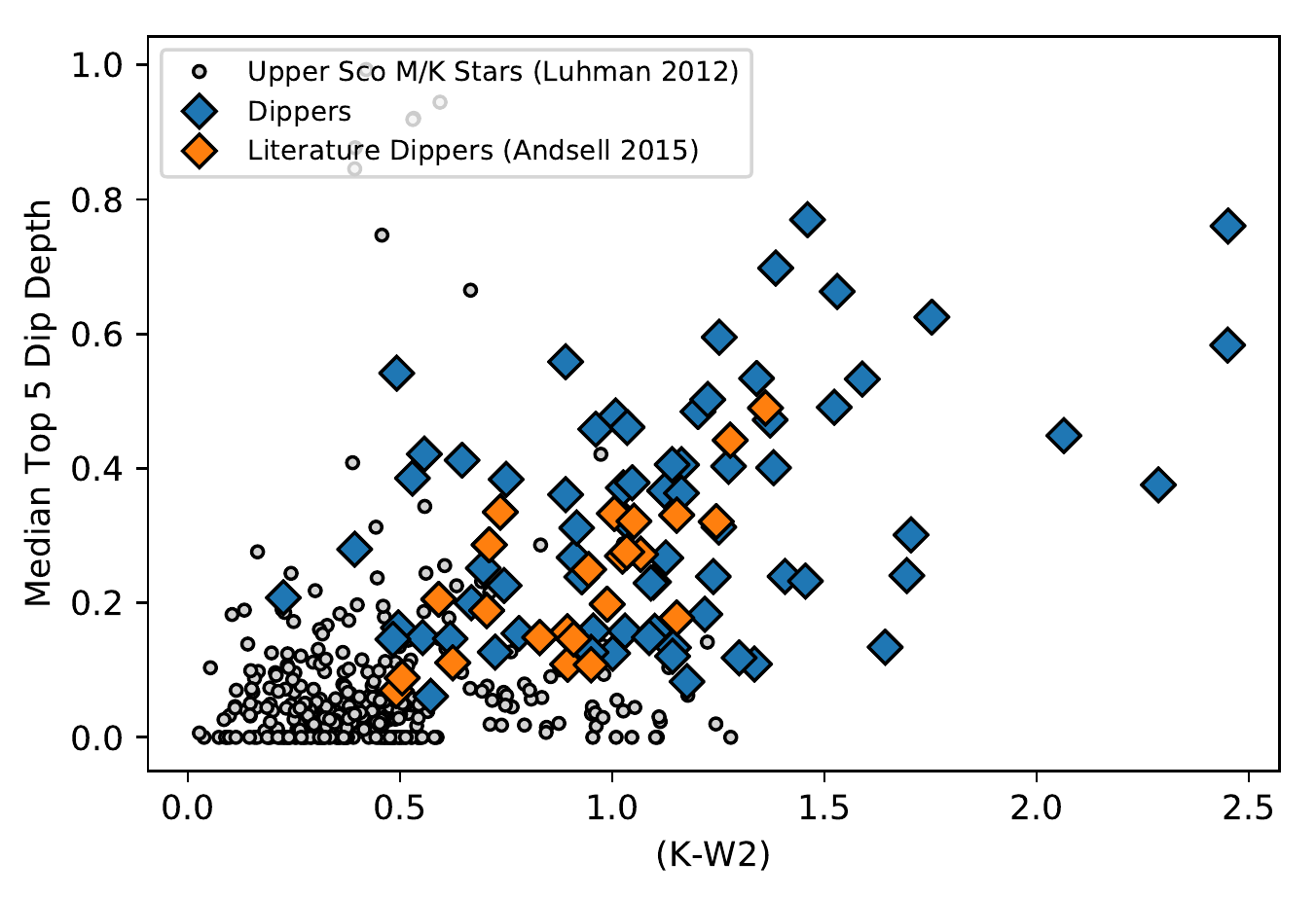}
  \caption{Average depth of the five deepest dips as a function of K$_s$-W2 excess. Here blue diamond points indicate dippers found with the machine learning algorithm from this work and red diamonds indicate dippers found in \protect\cite{Ansdell2015}. Dippers have not been corrected for extinction. We find a correlation between infrared excess and dip depth similar to the correlation found in that work.}
  \label{megancomp}
\end{figure}

In \cite{Ansdell2015} a positive correlation is seen between extinction corrected K$_s$ band - W2 band magnitude, a measure of dust content in the inner disk, and dip depth D$_{Dip}$. In this work we have not obtained spectra for our objects, and so we are not able to uniformly correct for reddening in all of our targets. To investigate the entire sample, we show the relation between dip depth and K$_s$-W2 (Figure~\ref{megancomp}), without correcting any of the sample for extinction. This shows a clear correlation. A Spearman's rank test gives a correlation of 0.424 and a p-value of 1.8$\times$ $10^{-5}$, indicating the correlation is significant. We therefore find strong evidence for a connection between the emitting area of material close to the star and the dip depth.

\subsection{Dipper Fraction by Region}

\begin{figure*}
  \centering
  \includegraphics[width=\textwidth]{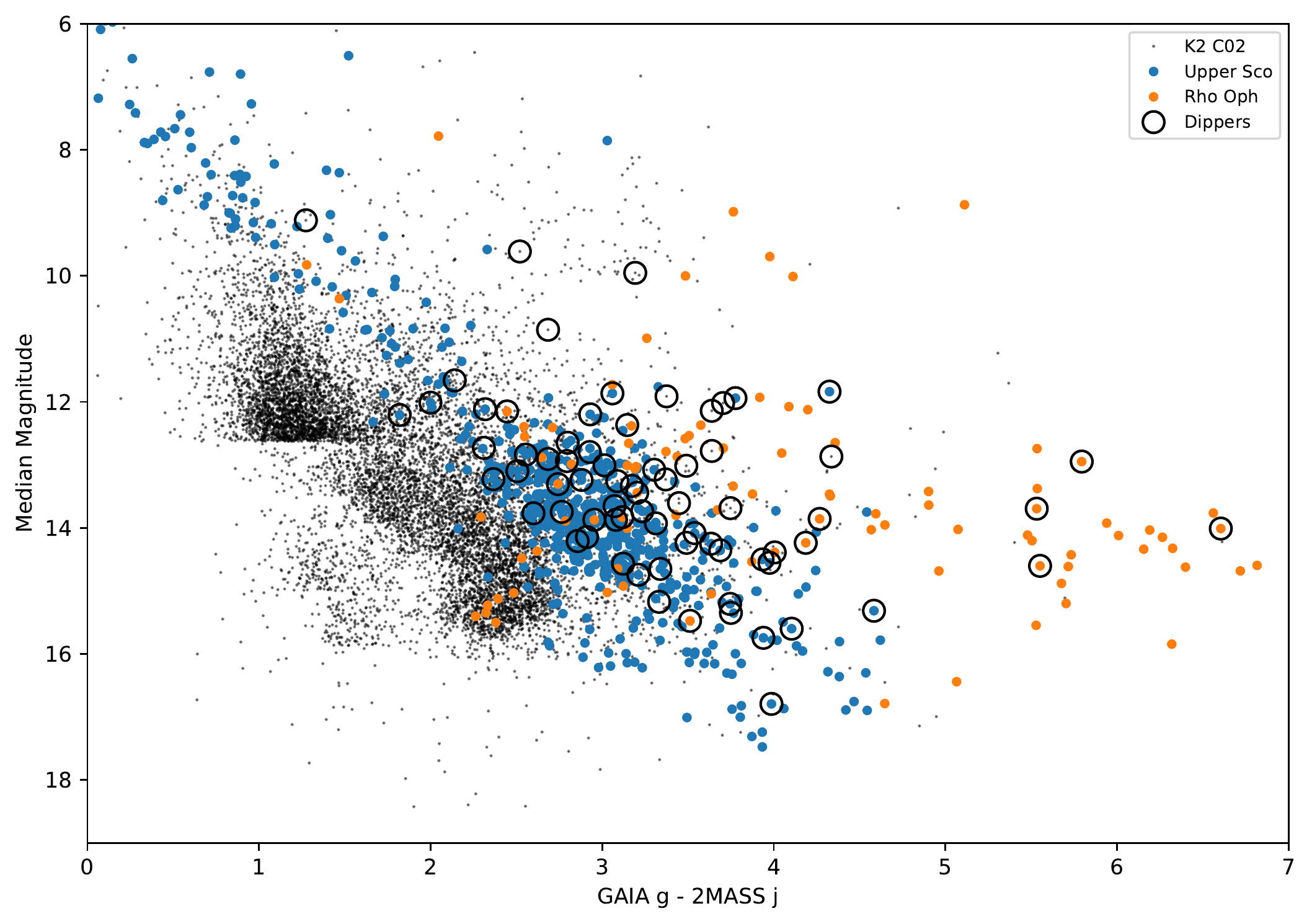}
  \caption{Median magnitude of K2 light curves compared with G-J colour, with magnitudes from Gaia DR1 and 2MASS. The Upper Sco and $\rho$ Oph populations are broadly distinguishable; $\rho$ Oph is seen to be redder as it is embedded in a cloud. At a Kepler magnitude of 18 the SNR of each target is too low to accurately find dippers in the K2 photometry. We find all dippers have an IR excess consistent with disk bearing stars. Membership lists have been built from \protect\cite{luhman,Rizzuto2015,lodieu2013,dezeeuw,slesnick2008,parks,wilking,ratzka}}
  \label{color1}
\end{figure*}

Figure~\ref{color1} shows a colour-magnitude diagram of all of the targets within K2 C02 with members of Upper Scorpius and $\rho$ Ophiuchus highlighted. All dippers are consistent with being members of either Upper Sco or $\rho$ Oph. There is no evidence for dippers in the older, background population.

\begin{table}
\centering
\caption{Members and dippers in each cluster observed by K2. The memberships are likely incomplete and several dippers were not found in any membership. However, they are consistent in colour space with being members of one or both of the young star forming regions.}
\label{clustertab}
\begin{tabular}{lll}
\hline
\textbf{}	&	\textbf{Upper Scorpius }&	\textbf{$\rho$ Ophiuchus }	\\ \hline
Number of Members & 701 & 199  \\ \hline\hline
Number of Dippers & 42 & 22  \\ \hline
Percentage of Dippers & 5.9\%$\pm$0.9\% & 11.0\%$\pm$2.3\%  \\ \hline \hline
Number of Bursters & 8 & 15  \\ \hline
Percentage of Bursters & 1.1\%$\pm$0.4\% & 7.5\%$\pm$1.9\%   \\ \hline
\end{tabular}
\end{table}

We find there is a modest difference between the percentage of dippers in each region out of all members). In Upper Sco, (10 Myr), we find 5.9\%$\pm$0.9\% are dippers and in $\rho$ Oph, (1Myr), a member occurrence rate of 11.0\%$\pm$2.3\%. However, $\rho$ Oph has many more disk bearing stars, and the percentage of disk bearing stars that are dippers is a more useful metric for the prevalence of dippers.

\cite{Cody2014} discuss the fraction of the disk bearing population that are dippers (the `dipper fraction') in NGC 2264. Of their sample of 1266 cluster members 162 were found to have both disks and high quality CoRoT and Spitzer light curves. Of this sample 35 were found to be dippers in the optical.

Using the prescription in \cite{Cody2014} and \cite{Wilking2001} we identify the disk bearing stars as those with $\alpha>-1.6$ where $\alpha=d\log\lambda F_\lambda/d\log\lambda$, using K band photometry from 2MASS and W4 photometry from WISE. We then compare the fraction of dippers among the disk bearing stars for Upper Sco and $\rho$ Oph. All dippers found with our method have disks and meet the $\alpha>-1.6$ criterion. Table \ref{dipfrac} shows the final dipper fraction for each region, including NGC 2264. We find between the three regions the dipper fraction in disk bearing stars is approximately 20\%, regardless of age.

There are 31 dippers with an unknown membership thought to belong to either region, and these are excluded from the statistics in Table~\ref{dipfrac}, but are listed in Table \ref{dippertab} for completeness. With the planned release of parallax data from the Gaia mission in the 2nd Data Release, it will be possible to obtain better membership lists and measure more precisely the dipper fraction and dipper occurrence rate.

\begin{table*}
  \centering
  \caption{Dipper fractions for the three regions with well studied dipper populations. The Upper Scorpius association and $\rho$ Ophiuchus cluster have been added in this work. The dipper fraction quoted here is as a function of disk bearing stars only.}
  \label{dipfrac}
  \begin{tabular}{lllll}
    \hline
    Region & Age & Disk Fraction & Dipper Fraction & Dipper Number \\ \hline\hline
    $\rho$ Ophiuchus & 0.1-1 Myr \citep{Luhman1999}& 40.2\%$\pm$4.3\% & 20.1$\% \pm $4.3$\%$ &  22\\ \hline
    NGC 2264& 1-5 Myr \citep{Rebull2002}\cite{Dahm2008}& 12.8\%$\pm$1.0\% & 21.6$\% \pm $3.7$\%$  \cite{Cody2014} & 35\\ \hline
    Upper Scorpius & 10 Myr \citep{Pecaut2011}& 26.7\%$\pm$2.0\% & 21.8$\% \pm $3.4$\%$ &  42\\ \hline
  \end{tabular}
\end{table*}

\subsubsection{Disk Evolution}
\label{disks}

For some dippers, classifications of their disk-bearing nature can be found using literature. \cite{luhman} use infrared photometry (from 2MASS and WISE) to distinguish the evolutionary stage of disk bearing stars in Upper Scorpius. There is some overlap between their sample and the targets in K2 C02. They define disks as full (optically thick with no evidence in the SED of clearing), evolved (optically thin in the IR with no evidence in the SED of clearing) and transitional (with an SED showing evidence for gaps and holes). They also define debris/evolved transitional disks that are composed of second-generation dust and are considered to be the final stage of disk evolution.

\begin{figure*}
  \centering
  \includegraphics[width=\textwidth]{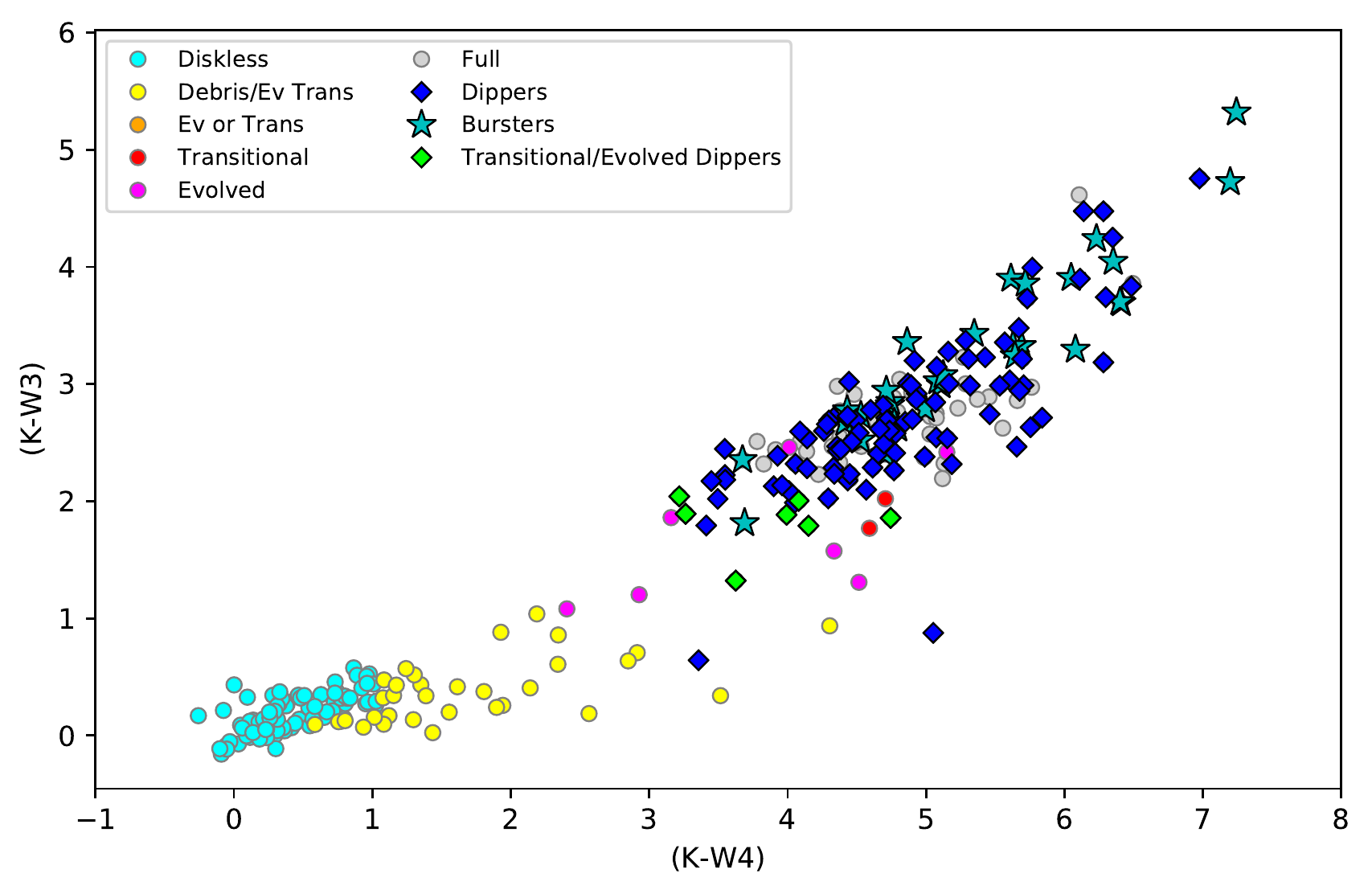}
  \caption{Disk types of dippers and bursters found in this work, where filled circles indicate disk types from \protect\cite{luhman} in Upper Sco that were also observed in K2. Colours are not corrected for reddening. Blue diamonds are all dippers found in this work with any membership, stars indicate bursters also with any membership (which we expect to be an incomplete sample). Green diamonds indicate dippers that were also observed in \protect\cite{luhman} that were found to have evolved or transitional disks given in Table \ref{evolvedtab}.}
  \label{disks2}
\end{figure*}

\begin{figure*}
  \centering
  \includegraphics[width=\textwidth]{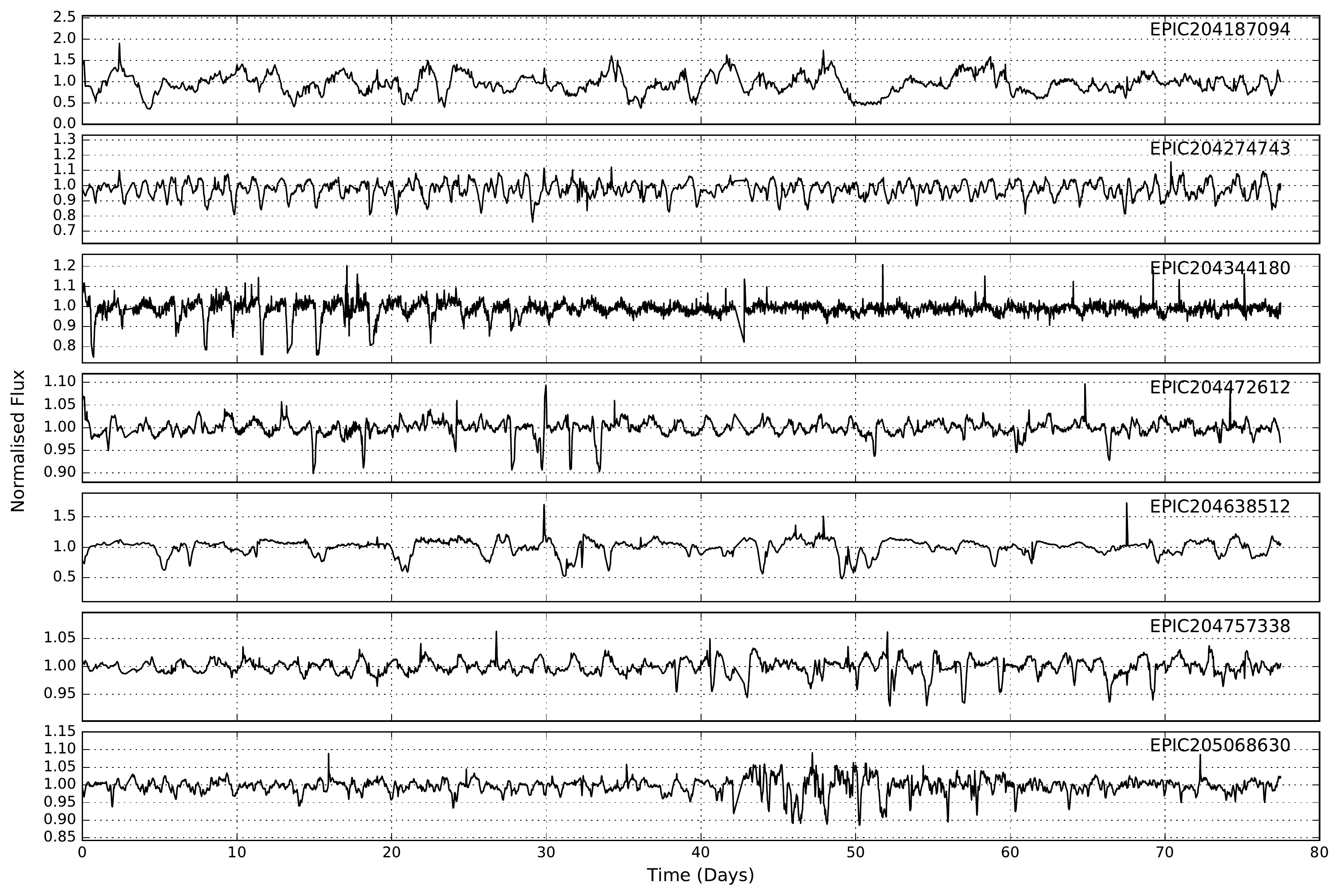}
  \caption{Light curves for evolved or transitional type dippers, as identified by \protect\cite{luhman}. These are indistinguishable from the rest of our dipper sample in any of the features used to train the machine. See also Table \ref{evolvedtab}. }
  \label{evolved}
\end{figure*}

Figure~\ref{disks2} shows a reproduction of Figure 2 from \cite{luhman} where the disk type of each target has been labelled. Dippers and bursters found by the machine learning algorithm and vetted by eye are plotted over the disk bearing stars. For the majority of the dippers we have no spectral types or reddening coefficients, and so the targets in this Figure have not been corrected for reddening. Dippers observed in \cite{luhman} that are labeled as 'evolved'  are given in Table~\ref{evolvedtab} and highlighted in Figure~\ref{disks2} in green.  Light curves for these dippers are also shown in Figure~\ref{evolved}. There are also four other dippers that have IR excesses consistent with being more evolved disks but were not observed in \cite{luhman}.

Of the 11 dippers consistent in colour space with being an evolved or transitional disk, none are distinct from the full disk dipper population in any of the features used to train the machine. They are consistent in period, dip depth and duty cycle (fraction of time in dip compared with out of dip) with the full disk bearing dipper population. This suggests that the dipper phenomenon is not affected by the early stages of disk evolution.

\begin{table}
  \centering
  \caption{Evolved or Transitional Disks as highlighted by \protect\cite{luhman} that also exhibit dipper behaviour.}
  \label{evolvedtab}
  \begin{tabular}{llll}

 \hline
EPIC ID & RA & Dec & Disk Type \\ \hline\hline
204187094 & 242.82956 & -23.322251 & evolved \\ \hline
204274743 & 239.37443 & -22.978846 & evolved \\ \hline
204344180 & 243.63696 & -22.703707 & ev or trans \\ \hline
204472612 & 242.14397 & -22.198866 & evolved \\ \hline
204638512 & 241.09023 & -21.507916 & transitional \\ \hline
204757338 & 241.86443 & -20.995618 & evolved \\ \hline
205068630 & 242.79566 & -19.558912 & evolved \\ \hline
  \end{tabular}
\end{table}

\subsubsection{Spectral Types of Dippers}
\label{spectypes}
\cite{Stauffer2015} and \cite{Ansdell2016} find that the dipper behaviour is limited to late-type KM stars. We similarly find only late-type stars among our dipper sample. Our spectral types are based on \cite{luhman} (who provide spectral types for their sample of 485 members of Upper Scorpius) and \cite{wilking} (who provide spectral types for 124 members of $\rho$ Ophiuchus). Table \ref{spt} shows the expected number of dippers in each spectral type (assuming the dipper behaviour was not spectral type dependent) based on these works. We make the assumption that all stellar types were observed uniformly. Table~\ref{spt} shows we would expect low numbers of stars larger than K and M type, and so we are not able to rule out dippers among more massive stars.

Additionally, young circumstellar disks are shown to disperse faster around more massive stars \citep{Carpenter2006,Kennedy2009}, which limits the number of high mass stars with disks in clusters older than a few Myr. We may also be insensitive to dippers around F and G stars due to their lack of disks in these young star forming regions.

\begin{table}
  \centering
  \caption{Expected and observed number of dippers in Upper Sco and $\rho$ Oph where spectral types have been determined in the works of \protect\cite{luhman} and \protect\cite{wilking}}
  \label{spt}
  \begin{tabular}{llll}
 \hline
Spectral Type & Expected Dippers & Observed Dippers \\ \hline \hline
B& 2.2$\pm$ 0.9& 0 \\ \hline
A& 1.7$\pm$ 0.7& 0 \\ \hline
F& 0.9$\pm$ 0.4& 0 \\ \hline
G& 1.7$\pm$ 0.7& 0 \\ \hline
K& 5.2$\pm$ 1.9& 14 \\ \hline
M& 41.2$\pm$ 14.0& 39  \\ \hline
L& 0.1$\pm$ 0.1& 0 \\ \hline
  \end{tabular}
\end{table}

\subsubsection{Orbital Distance to Inner Disk Edge}

Direct measurements of the inner disk location have been made for Herbig Ae/Be and T Tauri stars using interferometric techniques, and are used to infer the structure of the inner disk. (For examples see \cite{Kraus2015} and \cite{Akeson2005, Anthonioz2014}). The heating mechanism may be inferred by comparing the observed inner edge location with the expected dust sublimation radius for different assumptions (e.g. \cite{Millan-Gabet2006}). Extending such interferometric measurements to masses lower than solar is difficult. Here we suggest that dippers provide an indirect estimate of the location of the inner disk edge. Using dippers to measure the inner disk edge could extend the range of stellar luminosities over which inner edges can be estimated by approximately three orders of magnitude.

The dipper phenomenon is attributed to occultation of the star by opaque material (dust) at, or close to, the inner edge of the disk (e.g. \cite{Bouvier1999, Bodman2016}). The specific mechanism is unclear, and could be due to azimuthal variations in the inner disk edge vertical scale, and/or dusty accretion streams flowing from the disk down to the star. Thus the dipping phenomena is intrinsically tied into the inner edge of the disk, and can be used to study the region.

A period can be estimated, either by using a Lomb-Scargle periodogram across the entire time series or localised where dips repeat. Assuming Keplarian velocity and using the spectral types from \cite{luhman}, \cite{wilking} and \cite{Ansdell2015} we can estimate an orbital separation from the host star.

Even in the event that the occulting material resides in accretion streams that are well interior to the disk inner edge, inner radius estimates may still be robust if we assume these streams are locked to the inner disk edge (which \cite{Bodman2016} argue is set by magnetospheric disk truncation).

The sublimation temperature of dust in the disk is expected to be approximately 1300-1600K, depending on the species of dust \citep{2011EP&S...63.1067K} . This places a simple limit on the distance to the inner edge of the disk, since material closer than this radius will sublimate. As stellar luminosity increases this radius will increase. The period corresponding to this orbital distance, assuming the material orbits at a Keplerian velocity, is given by
\begin{equation}
\label{keplereqn}
P \approx \big(\frac{4\pi^2r^3}{GM}\big)^{\frac{1}{2}}.
\end{equation}

where r is the orbital distance, M is the mass of the star and G is the gravitational constant. The distance to the inner edge of the disk can be estimated using the effective temperature at a distance from a blackbody
\begin{equation}
\label{subl}
r=\big(\frac{L}{16\pi \sigma T_{sub}^4}\big)^{\frac{1}{2}}
\end{equation}
where L is the luminosity of the star and $\sigma$ is the Stephan-Boltzman constant. By measuring the periodicity of the dippers it is possible to use Equations \ref{keplereqn} and \ref{subl} to estimate the location of the material causing the dips and compare it with the sublimation distance.

\begin{figure}
  \centering
  \includegraphics[width=\columnwidth]{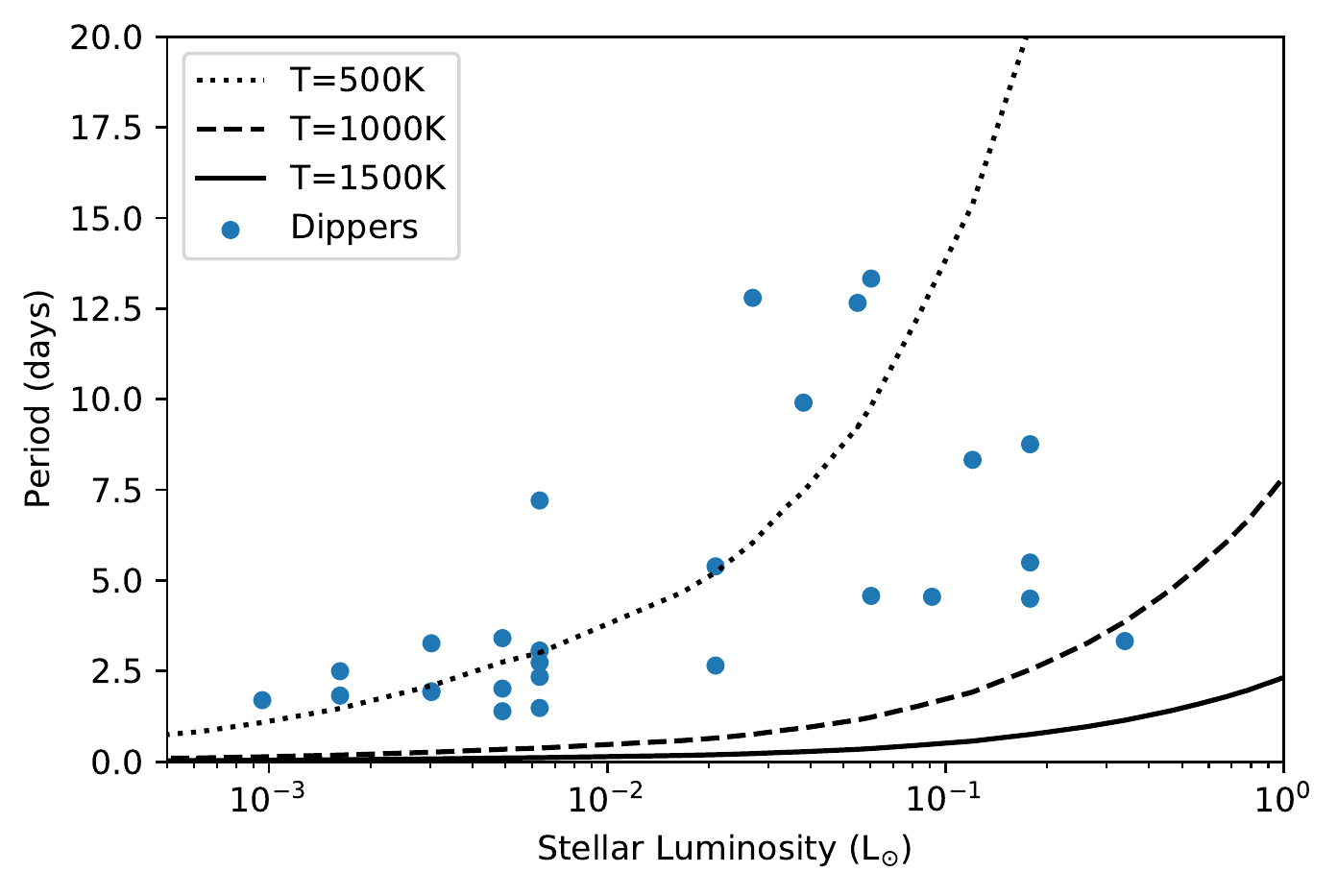}
  \includegraphics[width=\columnwidth]{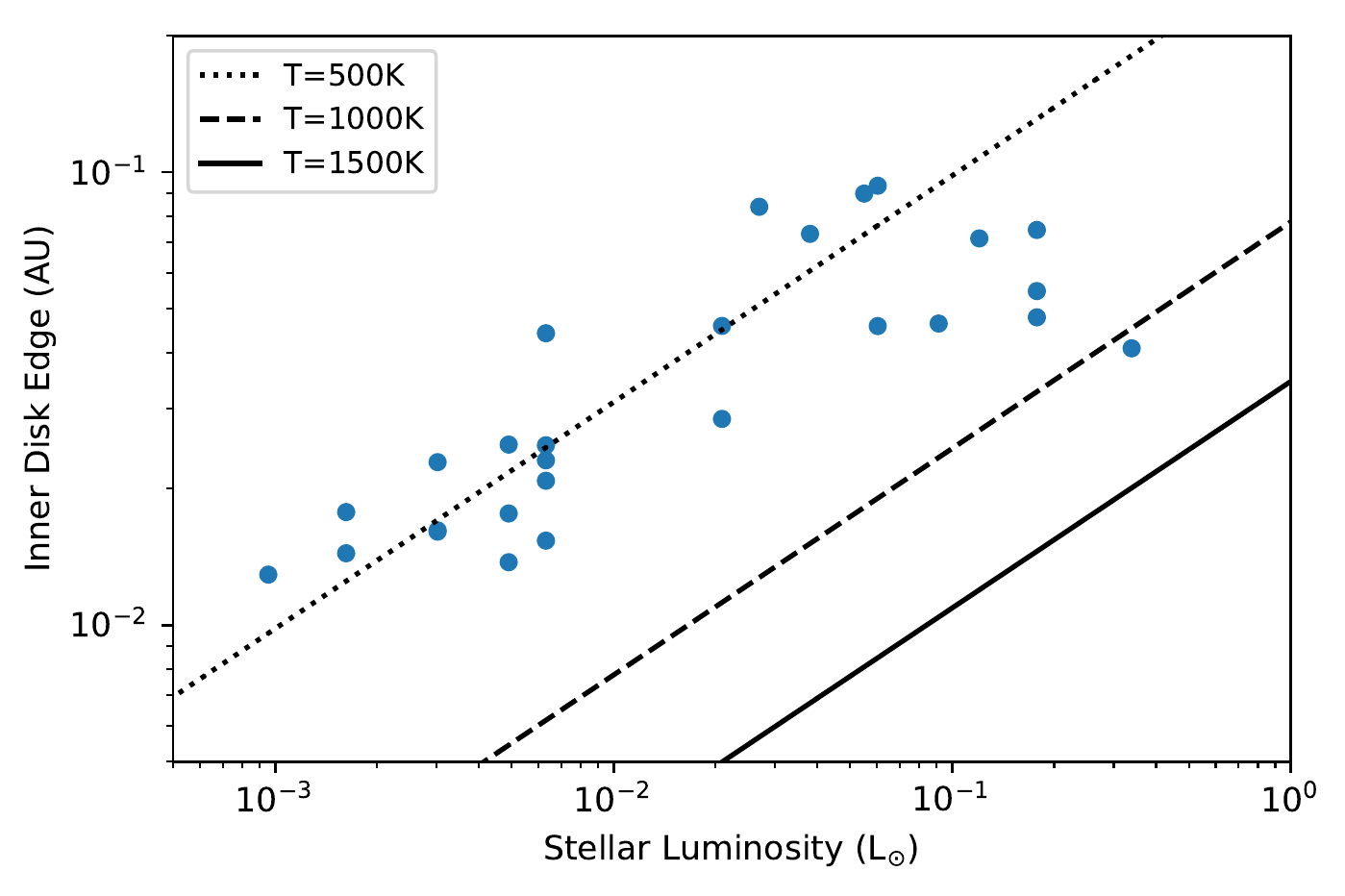}
  \caption{Inner disk edge in AU as a function of the luminosity of the star. We see dippers have systematically larger inner disk edge radii than predicted from a simple blackbody. Invoking backwarming or increased luminosity from accretion may account for this additional heating. We find a median temperature of 580$\pm ^{300}_{120}K$ for all dippers.}
  \label{RvsL}
\end{figure}

Figure~\ref{RvsL} shows the period and the inferred orbital distance in AU (which we use as a proxy for the inner edge radius) for periodic dippers as a function of stellar luminosity (aperiodic dippers are not included). Solid and dashed lines indicate a simple blackbody heating relation given in Eq \ref{subl}. We find the best fitting sublimation temperature for all dippers to be 580$\pm ^{300}_{120}K$. While the dipper population lies along a line of approximately constant temperature, they are notably cooler than the blackbody sublimation temperature.

Other works similarly find that the inner disk edge of T Tauri stars lies further out than predicted by a simple blackbody model and dust sublimation. \cite{Eisner2007} shows using interferometry that low luminosity ($\sim$ 1-5 L$_{\odot}$) T Tauri stars have more distant inner disk edges than predicted by simple models. \cite{Millan-Gabet2006} also show the same result, that low luminosity T Tauri stars have larger sublimation radii compared with that estimated using a blackbody assumption and the sublimation temperature T$_{sub}$. They suggest backwarming in the disk could increase the sublimation radius by a factor of two. They also note that smaller dust grains could cause the sublimation radius to increase. \cite{Muzerolle2003} suggest that the inner edge of the disk could be increased in T Tauri stars due to extra luminosity from accretion. If the dipper phenomenon can be attributed to the inner edge of the disk, some heating mechanism is also required to push the inner edge of the disk further out for K and M type stars.

In this work we extend the range of luminosities for T Tauri stars down to 10$^{-3}$L$_{\odot}$, and similarly find that the material causing the dips exists at a radius much cooler than than blackbody sublimation temperature. In summary, we have shown how dippers provide an estimate of the location of the inner disk edge around low-mass stars. Our measurements have systematic uncertainties because the specific location of the occulting material relative to the disk inner edge is uncertain. Nevertheless, our estimates appear consistent with results for more massive T Tauri and Herbig Ae/Be stars, and extend the stellar luminosity range over which such measurements can be made by several orders of magnitude.

\subsubsection{Insensitivity to Long Period Dippers}
\label{starsize}

If we assume the occulting material is at the sublimation radius then the period of the dips should increase as a function of spectral type as stellar luminosities (and so stellar masses) increase (see Figure~\ref{RvsL}). Using the stellar models from \cite{Pecaut2013} it is possible to calculate the expected period for early-type (BAFG) stars if the same dipper phenomenon were to occur in these spectral types.  In this case we use a sublimation temperature of T$_{sub}$=580$\pm ^{300}_{120}K$ (which is found to be the median sublimation temperature for the dippers in this sample as discussed above).

For a solar-like G star we would expect a period of 40 $\pm_{28}^{41}$ days, and for an F type star we would expect a period of 98 $\pm_{70}^{98}$ days. As K2 C02 is an 80 day long campaign (and we require at least 3-5 events for a detection) we would expect F- and G-type dippers to be barely detectable, if they were physically possible and present (see \cite{Bodman2016,Kennedy2017} for discussions of physical reasons dippers may be restricted to late-type stars). Also, as shown in Table \ref{spt}, there are low numbers of F- and G-type stars expected in this sample. For earlier B- and A-type stars we expect the rotation period at the sublimation temperature to be $>$ 80 days and too long to be detectable with K2. (As discussed in Section~\ref{spectypes} there are also few massive stars in Upper Sco and $\rho$ Oph.)

\section{Bursters}
\label{bursters}
\begin{figure*}
  \centering
  \includegraphics[width=\textwidth]{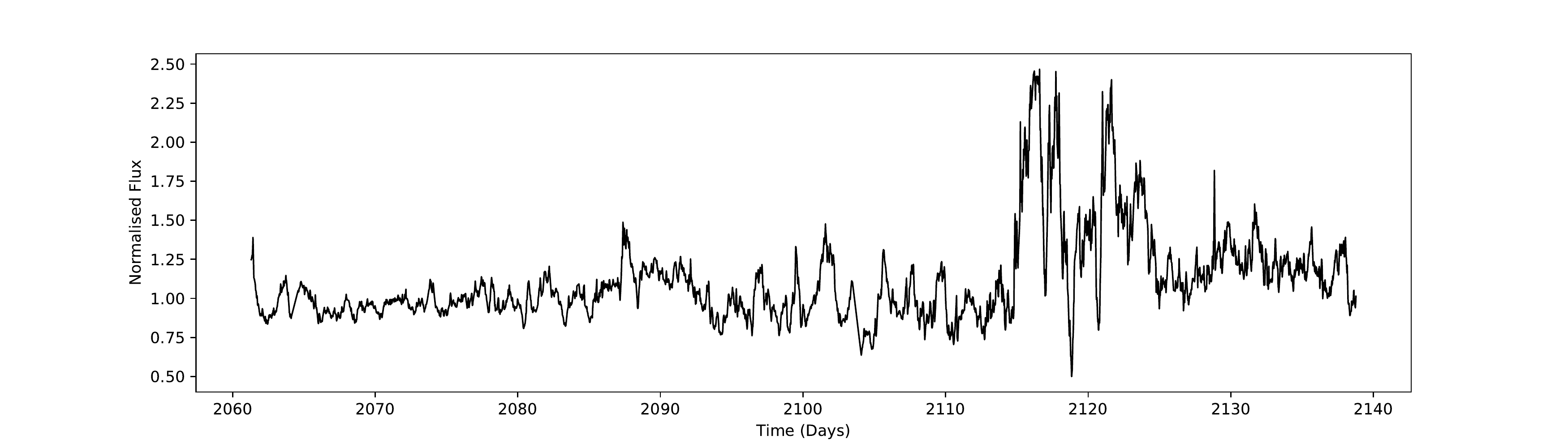}
  \caption{Example of a burster from K2 C02. Here we see the flux of the star doubles over a few days before returning to normal. The shape and duration of the burst event is inconsistent with stellar flares. 30 such object have been found in this work.}
  \label{flareexample}
\end{figure*}

A subset of the candidates flagged by the RF classifier were found to become brighter in bursting events similar to those found in \cite{Cody2014} and \cite{Stauffer2014}. These were distinguished by eye and flagged as bursters. This subset also contained targets that both dipped and bursted, referred to in \cite{Cody2014} as a symmetric morphology. All bursters were noted to have a significant infrared excess and \emph{M-statistic} of $\lesssim$ 0 (see Figure~/ref{Mstat}), and are otherwise indistinguishable from the dipper sample, they cluster in the same spaces of all other features. An example of a burster is given in Figure~\ref{flareexample}.

The burst events found in this sample do not resemble stellar flares, which are characterised as a sharp increase in flux with an exponential drop off. These events much more closely resemble the dipping behaviour in shape. \cite{Stauffer2014} suggest these events are accretion bursts, where short duration mass in-falls onto the stellar photosphere cause brightening events, potentially also causing a hot spot. They show other candidates that are not bursting but may display a rotating hot spot from such accretion mechanics.

It would have been possible to iterate further and build a new burster class to find this type of object. We chose to have one broad classification rather than split into two classes, dipper and burster, for two reasons: firstly the number of bursters is small which limits our ability to produce a reliable training set. Secondly there are some targets that are difficult to classify that both dip and flare. To ensure the dipper population is as clean as possible any cases where there was an equal amount of bursting events compared with dipping events were added to the bursting class.

The occurrence rate of bursters is 1.1\%$\pm$0.4\% for Upper Sco (10 Myr) and 7.5\%$\pm$1.9\% for $\rho$ Oph (1 Myr) as shown in Table \ref{clustertab}, though this may be an underestimate. If our incompleteness is the same for each region (which we expect) we find more bursters in the younger $\rho$ Oph cluster. The fraction of disk bearing stars that are bursters is 4.3\%$\pm$1.5\% for Upper Sco and 14.3\%$\pm$4.0\% for $\rho$ Oph. Thus, there is weak (2$\sigma$) evidence that, unlike the dipper fraction, there is a decrease in burster fraction (and hence the rate or magnitude of accretion bursts) with age.

A full description of all targets, including whether they were classified by eye as bursting or dipping, is given in Table \ref{dippertab} in the Appendix. The light curves of all bursters found in this work are shown in Figure~\ref{fig:bursters} in the Appendix.

\section{Summary}
\label{summary}

Dippers are a subclass of Classical T Tauri stars, so called because of their frequent drops in flux of 10-50\%. In this work we have used a Random Forest (RF) Machine Learning (ML) classification algorithm to find new dipper candidates within K2 Campaign 2 (K2 C02). We have employed 25 known dippers as a training set, and find 70 new dippers in this dataset. Our final sample of 95 dippers represents almost a factor 4 increase in the known population in the two clusters. Machine Learning is an efficient approach for the identification of this class of variable star.

The ML algorithm used 25 different, uncorrelated features based solely on light curve metrics. We included standard light curve metrics, Gaussian Mixture Models and wavelet analysis to characterise the light curves. As our training sample was small the algorithm was iterated 200 times to improve robustness. We found from cross-validation that 86\% of the dippers from the training sample were retrieved successfully.

Based on IR excesses from WISE, all dippers are consistent with being disk bearing members of Upper Scorpius and $\rho$ Ophiuchus. We measure the dipper occurrence rate, and find 5.9\%$\pm$0.9\% of members are dippers for Upper Scorpius and 11.0\%$\pm$2.3\% of members are dippers for $\rho$ Ophiuchus.

We determine a dipper fraction (defined as the fraction of disk bearing stars that are also dippers) of 20.1\%$\pm$4.3\% for $\rho$ Ophiuchus (age $\sim1$ Myr) and 21.8\%$\pm$3.4\% for Upper Scorpius (age $\sim10$ Myr). This is similar to the findings of \cite{Cody2014}, where a dipper fraction of 21.6 $\pm$ 3.7\% was found in the young cluster NGC 2264 (age $\sim1-5$ Myr).

Of our sample, 31 dippers are not catalogued members of either cluster, but are consistent in colour-magnitude space with being a member of either Upper Sco or $\rho$ Oph. We note that with the release of parallax information from Gaia in 2018 the two regions should become much easier to distinguish. This will not only allow us to establish the membership of these unknown dippers but calculate more accurate dipper fractions based on more complete membership lists.

We find that 35\% of our dipper sample do not meet the criteria set out in \cite{Ansdell2015} but are still clustered in the same feature space. Most of these cases were close to the cuts suggested by \cite{Ansdell2015}, but have slightly longer periods or slightly smaller measurements of R$_{Dip}$; we suggest their metric may be too strict. We find further evidence to support a correlation between K$_s$-W2 and dip depth, suggesting that stars with more material close to the star show deeper dips.

Dippers present an opportunity to study the inner edge of disks around small stars without the use of interferometry. Using our sample of periodic dippers we are able to estimate the orbital separation of the occulting material as a function of stellar luminosity. If the material is at the inner edge of the disk we would expect it to have a temperature similar to the sublimation temperature of dust (T$\sim$1300K). We find the material to be much cooler, with a blackbody temperature of 580$\pm ^{300}_{120}K$. This is consistent with interferometric measurements of T Tauri stars which show the inner edge to be further out than the equilibrium blackbody sublimation radius.

Similarly to \cite{Ansdell2015} and \cite{Stauffer2015} we find the dipper phenomenon only in late-type (K--M) stars. Based on spectral types from \cite{luhman} and \cite{wilking}, we would expect to observe mostly late-type stars in our sample (see Table \ref{spt}) and few, if any, OBAFG types. This limits our sensitivity to dippers among these more massive stars. Additionally, if we assume a rotation period consistent with the average temperature of the dippers as a function of stellar luminosity, the 80 day K2 campaign is too short to observe the dipper phenomenon around early-type stars. F- and G-type dippers may be barely detectable, and B- and A-type dippers will have orbital periods of months or more. Based on this we cannot rule out the possibility of early-type dippers.

According to \cite{luhman}, several of our dippers have evolved disk types (see Table \ref{evolvedtab} and Figure~\ref{evolved}), with smaller IR excesses than the rest of the sample. These dippers are otherwise identical to the rest of the sample and are not distinguishable in any of the features used in this work. We find no dippers with debris disks or diskless dippers, consistent with the theory that the dipper phenomenon is caused by dust in primordial gas-rich disks occulting the star. Similarly, \cite{Alencar2010} find some dippers to be evolved in NGC 2264. As the dipper fraction (of stars with disks) is consistent across three clusters spanning 1--10 Myr in age, and as evolved disk types show no difference in the dipper phenomenon, we suggest that the dipper phenomenon is unaffected by the early stages of disk evaporation.

We find 30 bursters which exhibit brightening events on time scales of more than a day which are inconsistent with stellar flares, due to shape, frequency and duration. This is similar to findings in the NGC 2264 cluster discussed in \cite{Cody2014} and \cite{Stauffer2014}, who suggest that bursters are young disk bearing objects undergoing accretion events. The census of bursters is likely incomplete in our survey as there is no specific training set for them. Their identification arose because they cluster similarly in feature space to dippers. The burster fraction (of stars with disks) is 4.3\%$\pm$1.5\% for Upper Sco (10 Myr) and 14.3\%$\pm$4.0\% for $\rho$ Oph (1 Myr). This suggests that the younger $\rho$ Oph cluster has a higher rate of accretion bursts at the $\sim 2 \sigma$ level, and that the rate of accretion does reduce as a function of cluster age.

With K2 observations of the Pleiades and Hyades in Campaign 4 and new observations of the Taurus region recently taken in Campaign 13 we will have an opportunity to study more young stars. This will allow us to investigate the dipper fraction over a wider range of ages. Upper Scorpius will also be reobserved in Campaign 15, though coverage will be partial, allowing us to follow up on several of the dippers found in this work. By expanding the sample of dippers we aim to understand more about the inner disk region around young stars.

\section*{Acknowledgements}
We thank Andrea Navas for getting the ball rolling. We thank Sergey Koposov and Benjamin Pope for helpful discussions on ML techniques and data reduction for K2. We thank the reviewer for their thorough and constructive comments. CH and STH acknowledge support from the Science and Technology Facilities Council (STFC) (CH for PhD funding). GMK is supported by the Royal Society as a Royal Society University Research Fellow. This work has made used of data from the 2MASS, Gaia and WISE surveys. This paper includes data collected by the Kepler/K2 mission. Funding for the Kepler/K2 mission is provided by the NASA Science Mission directorate.

\appendix

\section{Appendix: All Dipper and Burster Targets}
\begin{figure*}
  \centering
  \includegraphics[width=\linewidth]{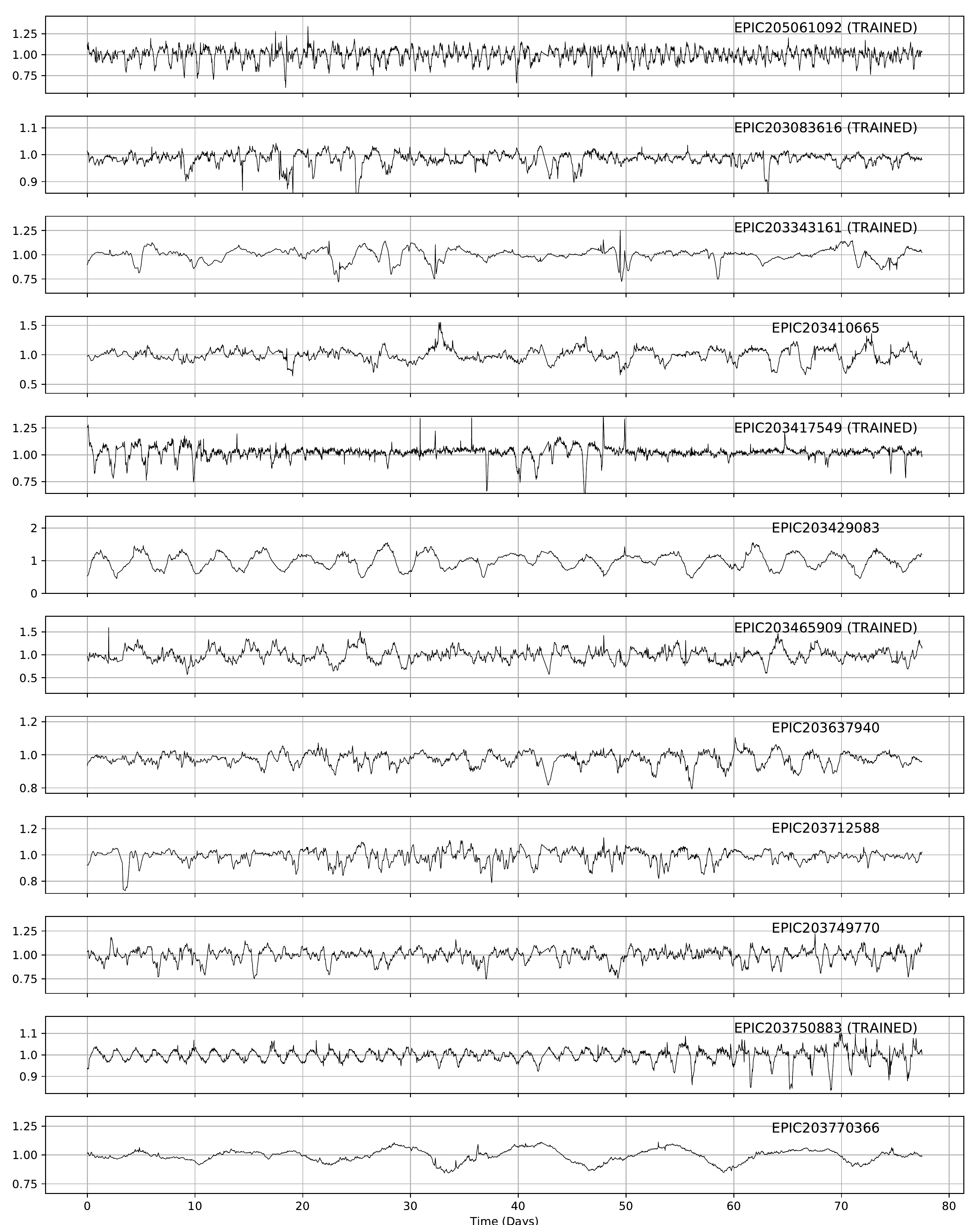}
  \caption{Full dipper sample from this work. Dippers that were used to train the algorithm are labelled. 22 of these training candidates are taken from \protect\cite{Ansdell2015} and \protect\cite{Bodman2016}. 18 are taken from the first iteration of this machine, see Section \ref{iterations} for more detail.}
    \label{knowndip}
\end{figure*}

\addtocounter{figure}{-1}
\begin{figure*}
  \centering
  \includegraphics[width=\linewidth]{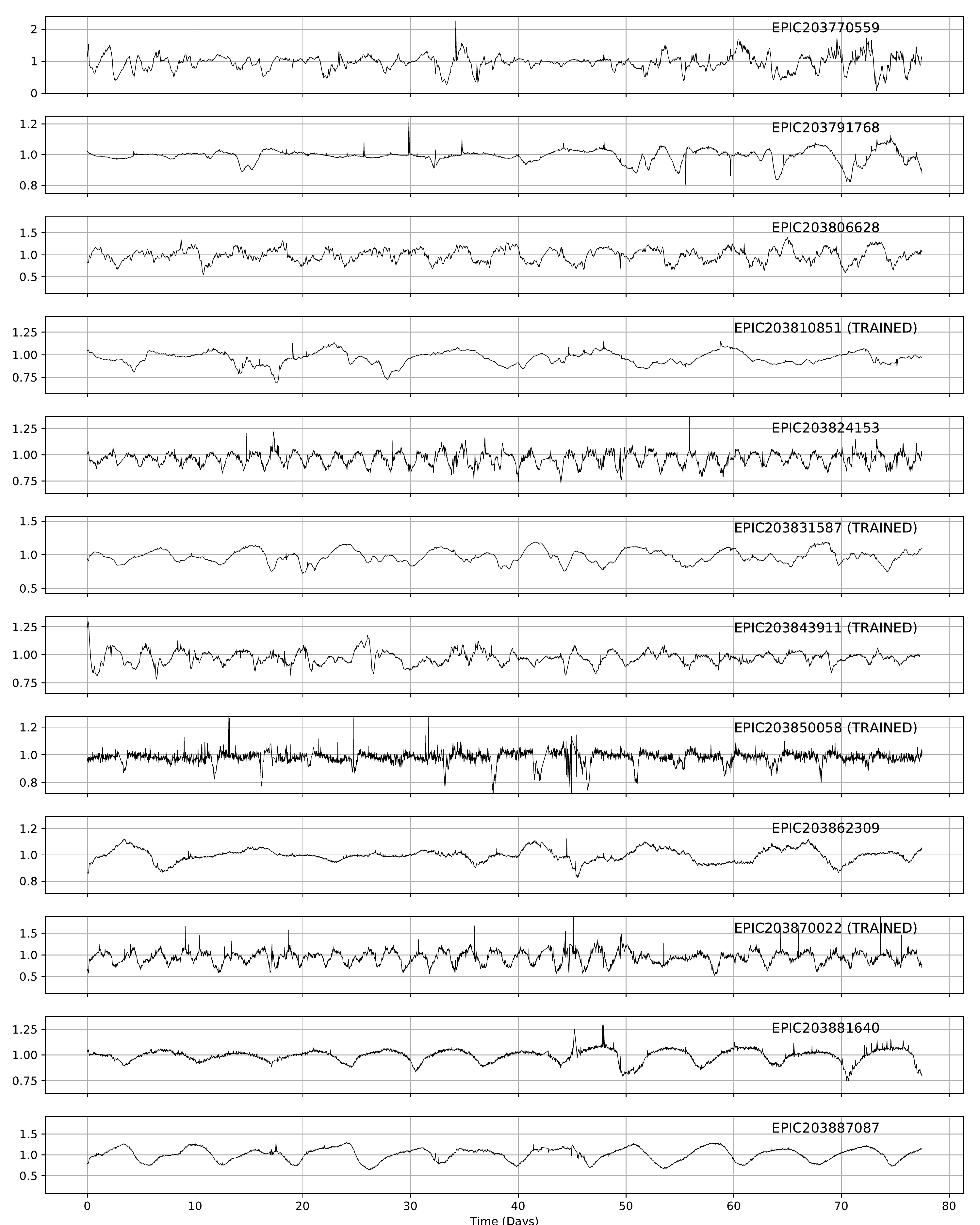}
  \caption{continued}
\end{figure*}

\addtocounter{figure}{-1}
\begin{figure*}
  \centering
  \includegraphics[width=\linewidth]{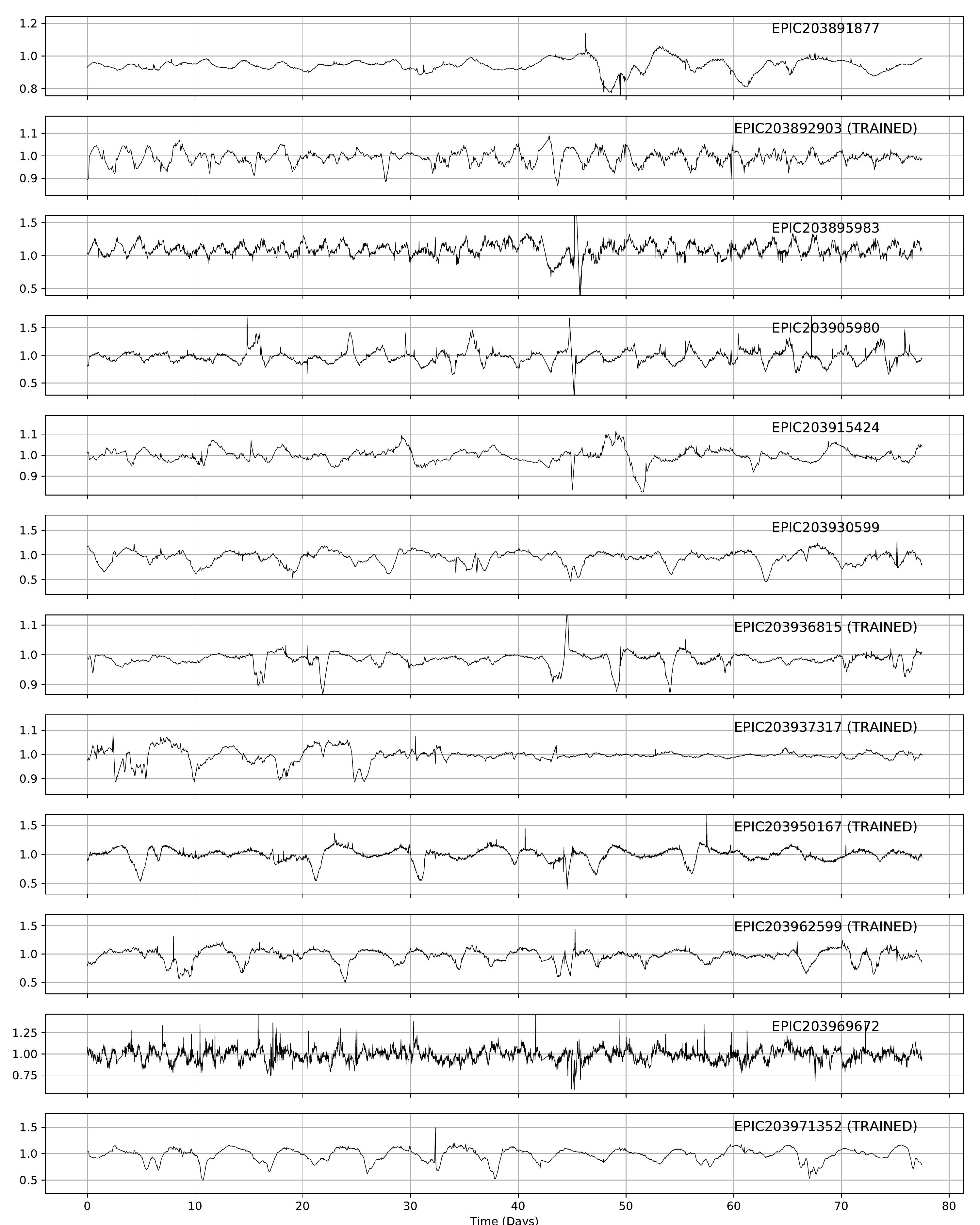}
  \caption{continued}
\end{figure*}

\addtocounter{figure}{-1}
\begin{figure*}
  \centering
  \includegraphics[width=\linewidth]{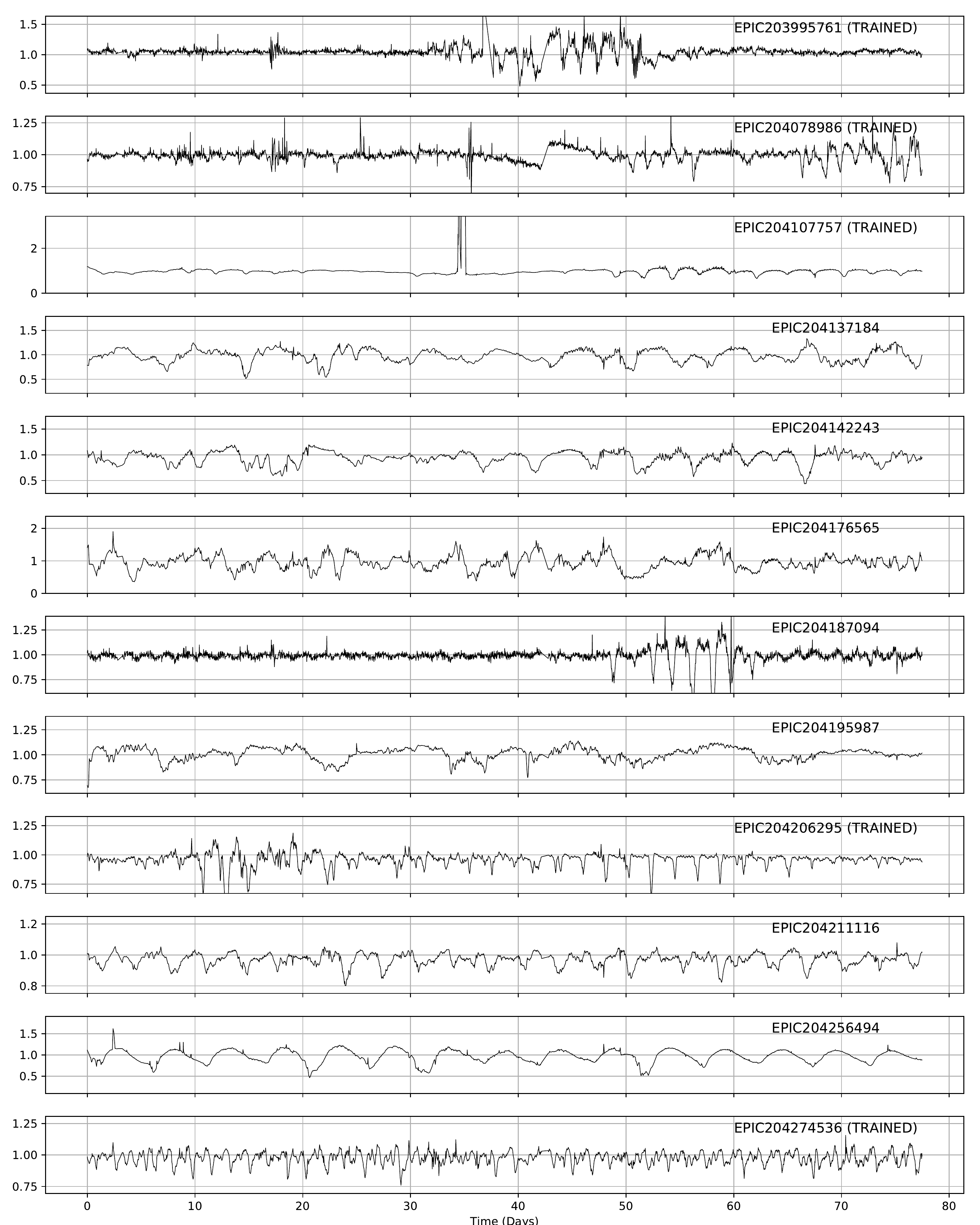}
  \caption{continued}
\end{figure*}
\addtocounter{figure}{-1}
\begin{figure*}
  \centering
  \includegraphics[width=\linewidth]{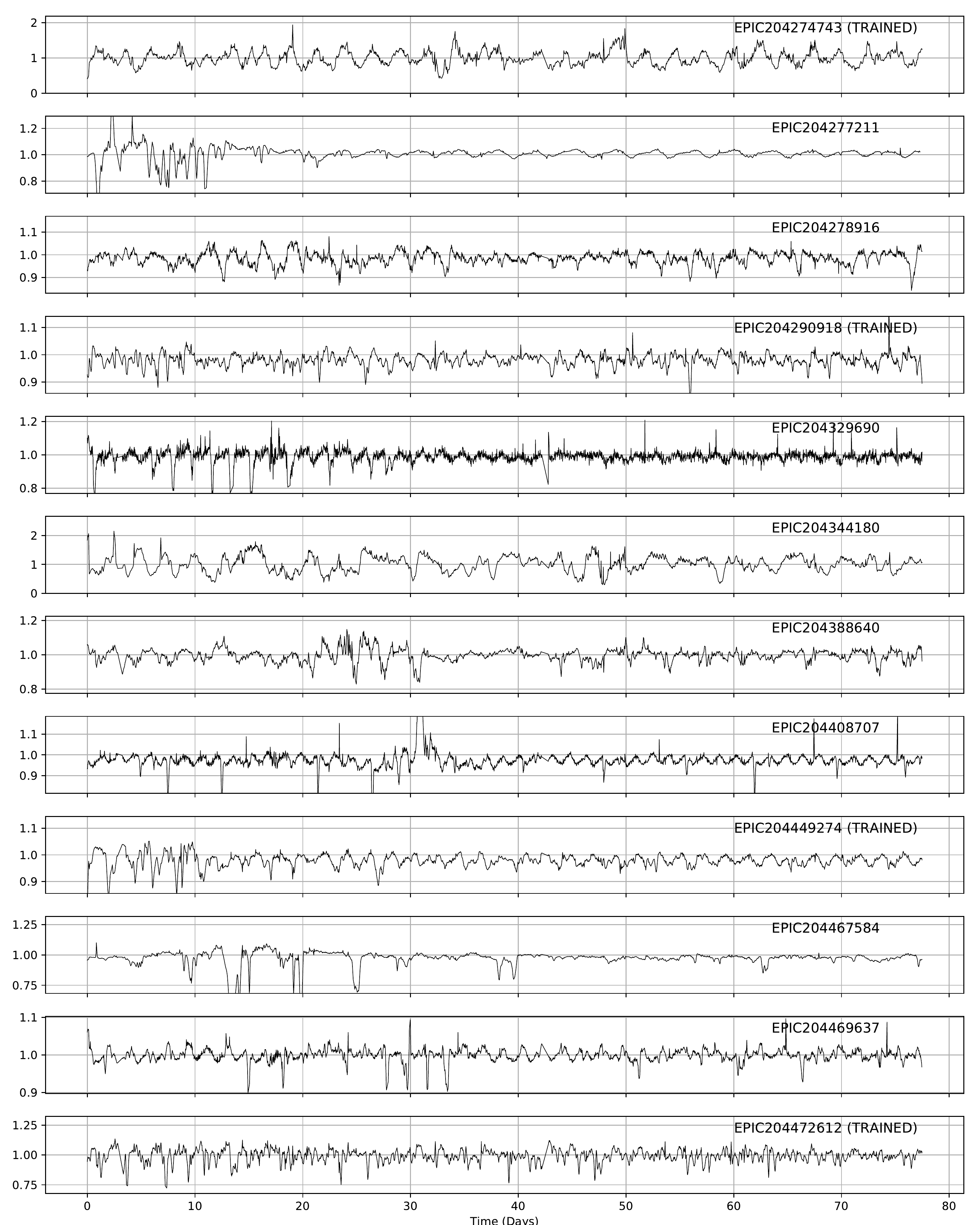}
  \caption{continued}
\end{figure*}
\addtocounter{figure}{-1}
\begin{figure*}
  \centering
  \includegraphics[width=\linewidth]{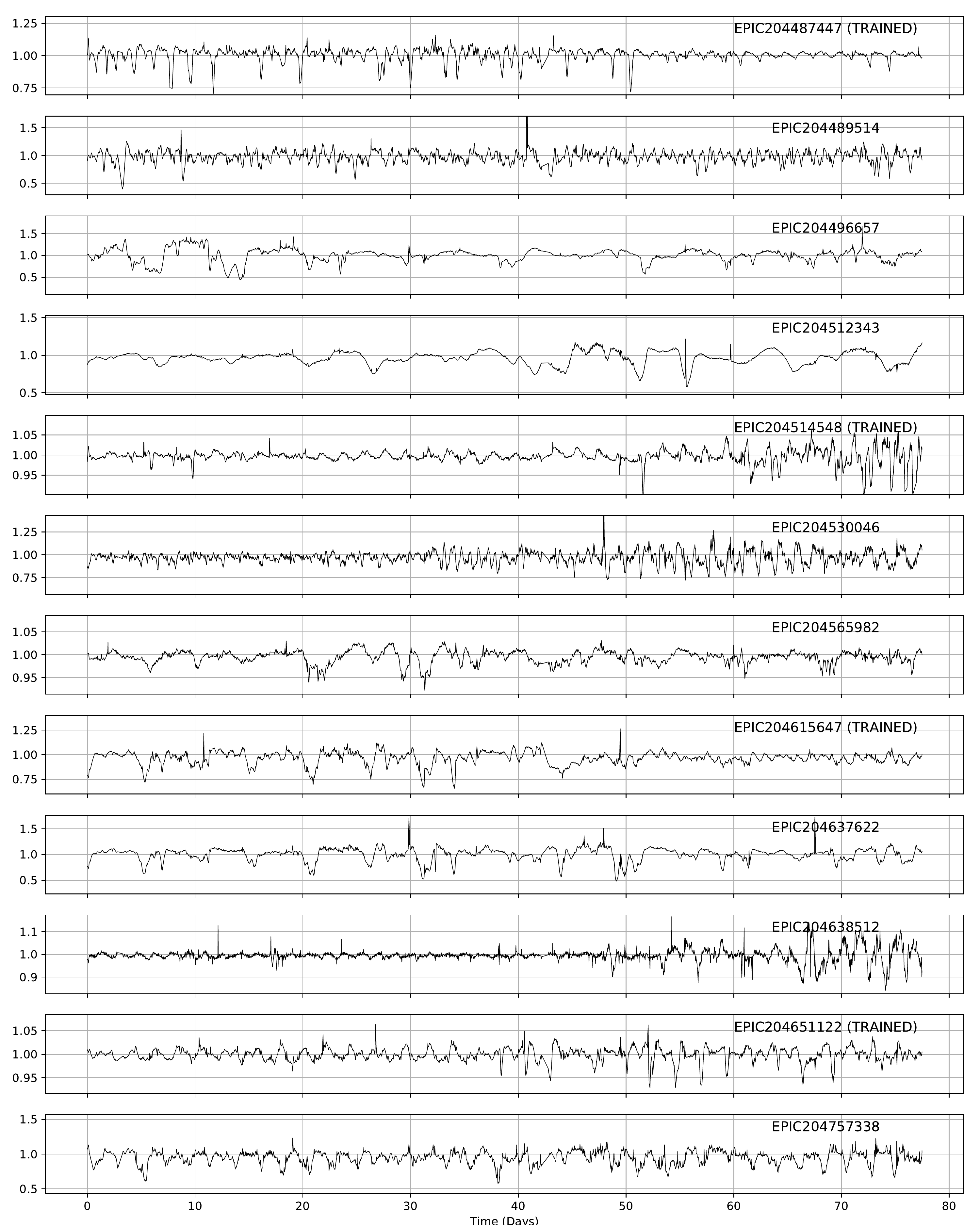}
  \caption{continued}
\end{figure*}
\addtocounter{figure}{-1}
\begin{figure*}
  \centering
  \includegraphics[width=\linewidth]{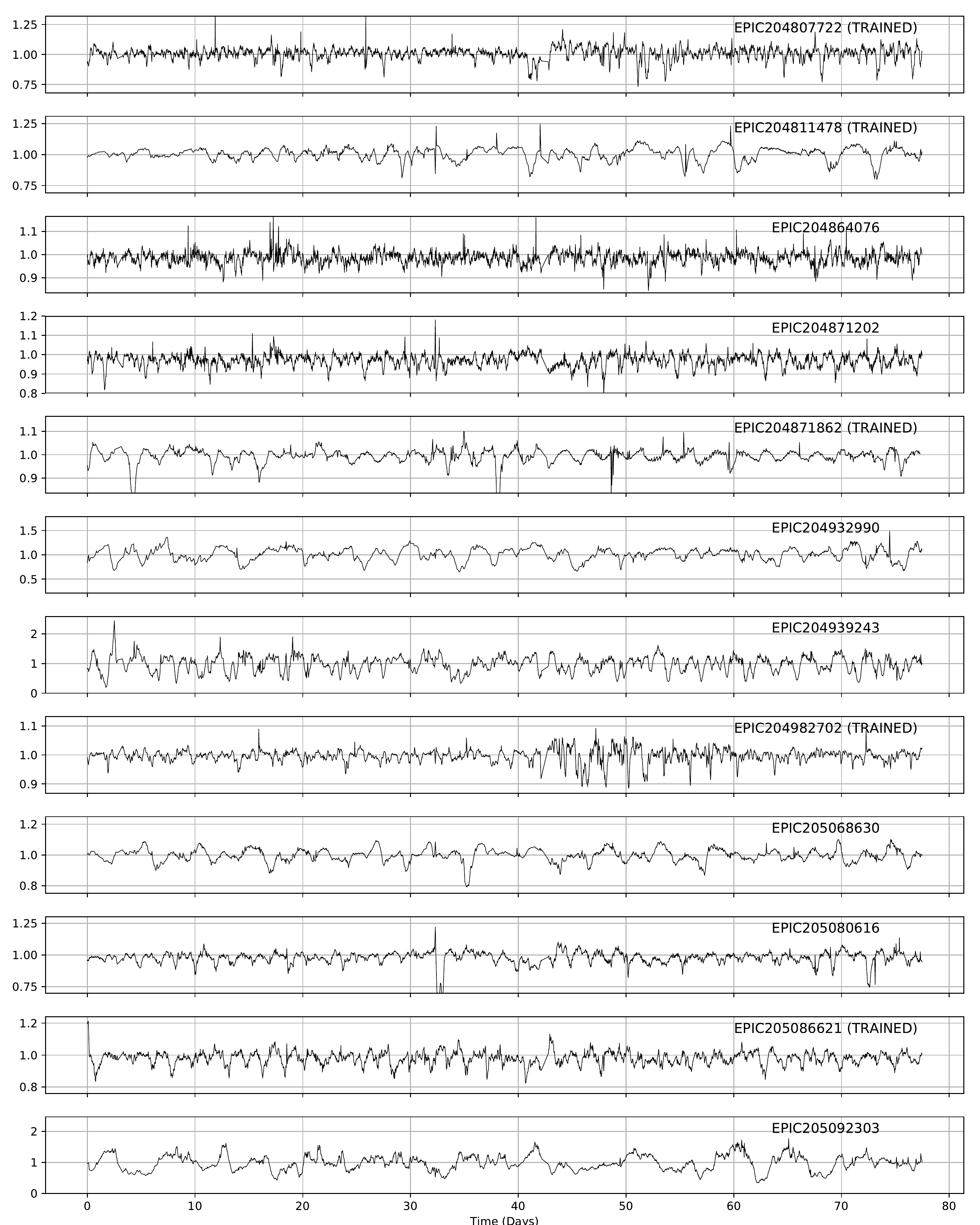}
  \caption{continued}
\end{figure*}
\addtocounter{figure}{-1}
\begin{figure*}
  \centering
  \includegraphics[width=\linewidth]{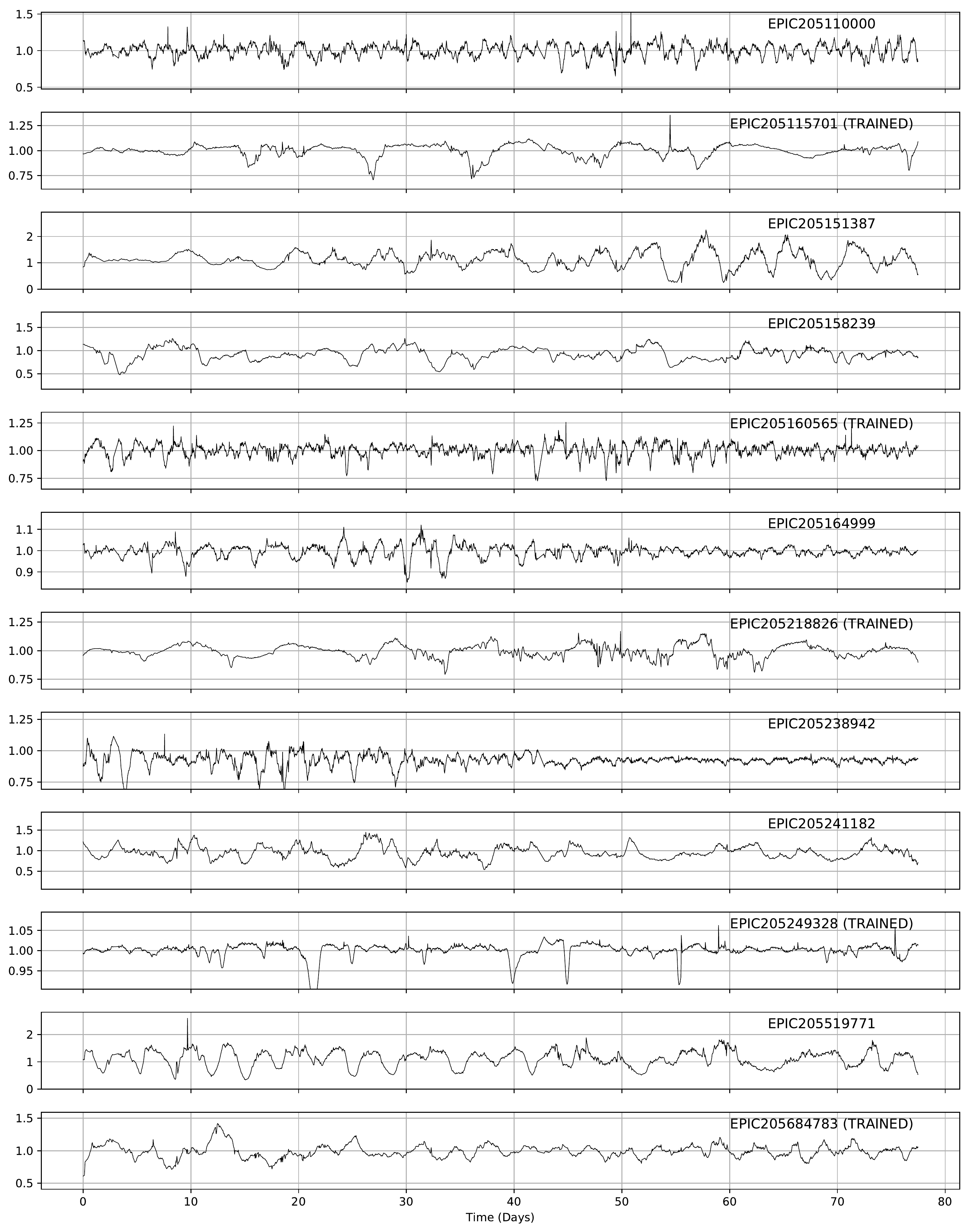}
  \caption{continued}
\end{figure*}

\begin{figure*}
  \centering
  \includegraphics[width=\linewidth]{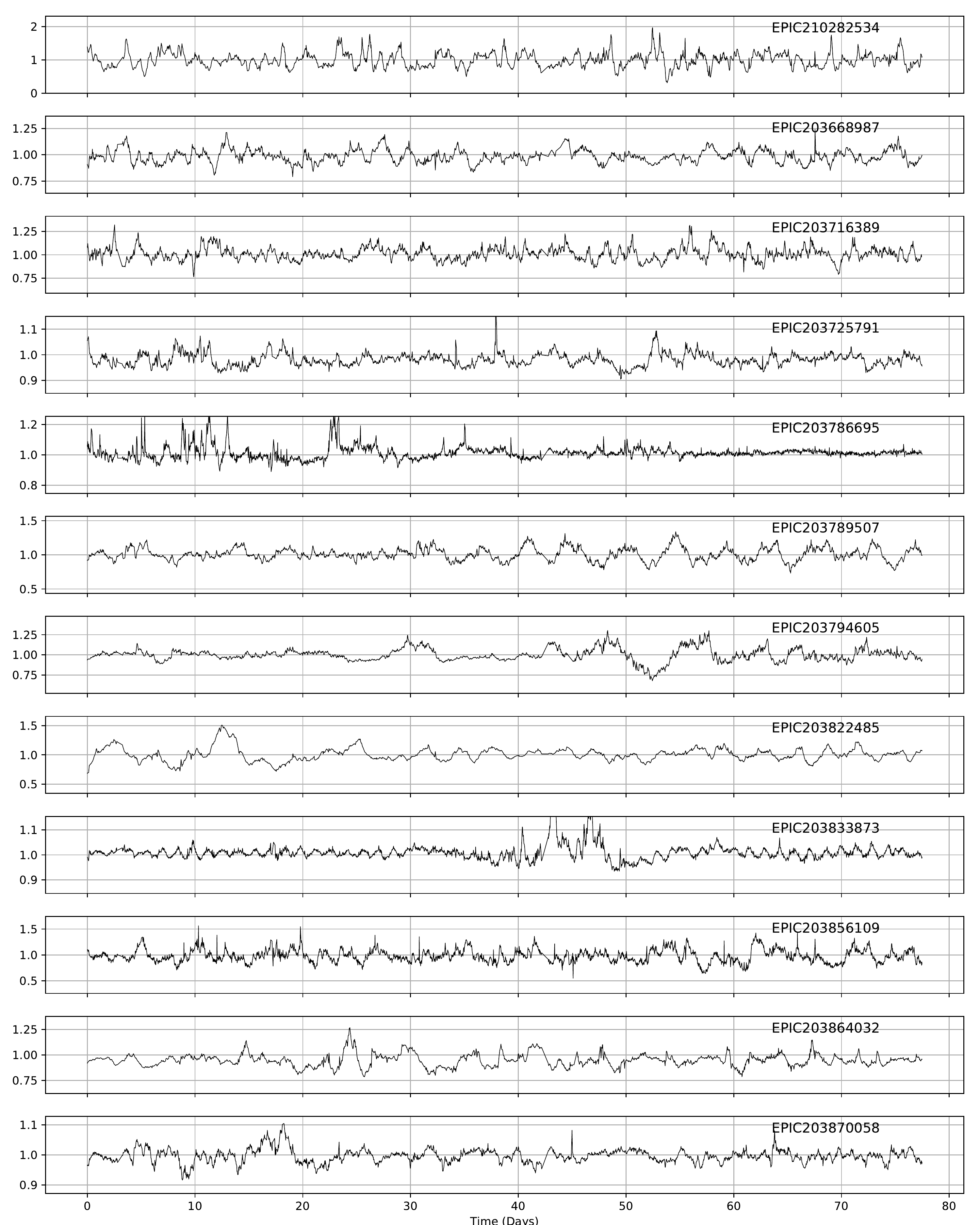}
  \caption{Bursters identified in this work. These are targets that were identified by the machine but have been separated by eye as bursting targets rather than dipping. See Section \ref{bursters} for details.}
    \label{fig:bursters}
\end{figure*}
\addtocounter{figure}{-1}
\begin{figure*}
  \centering
  \includegraphics[width=\linewidth]{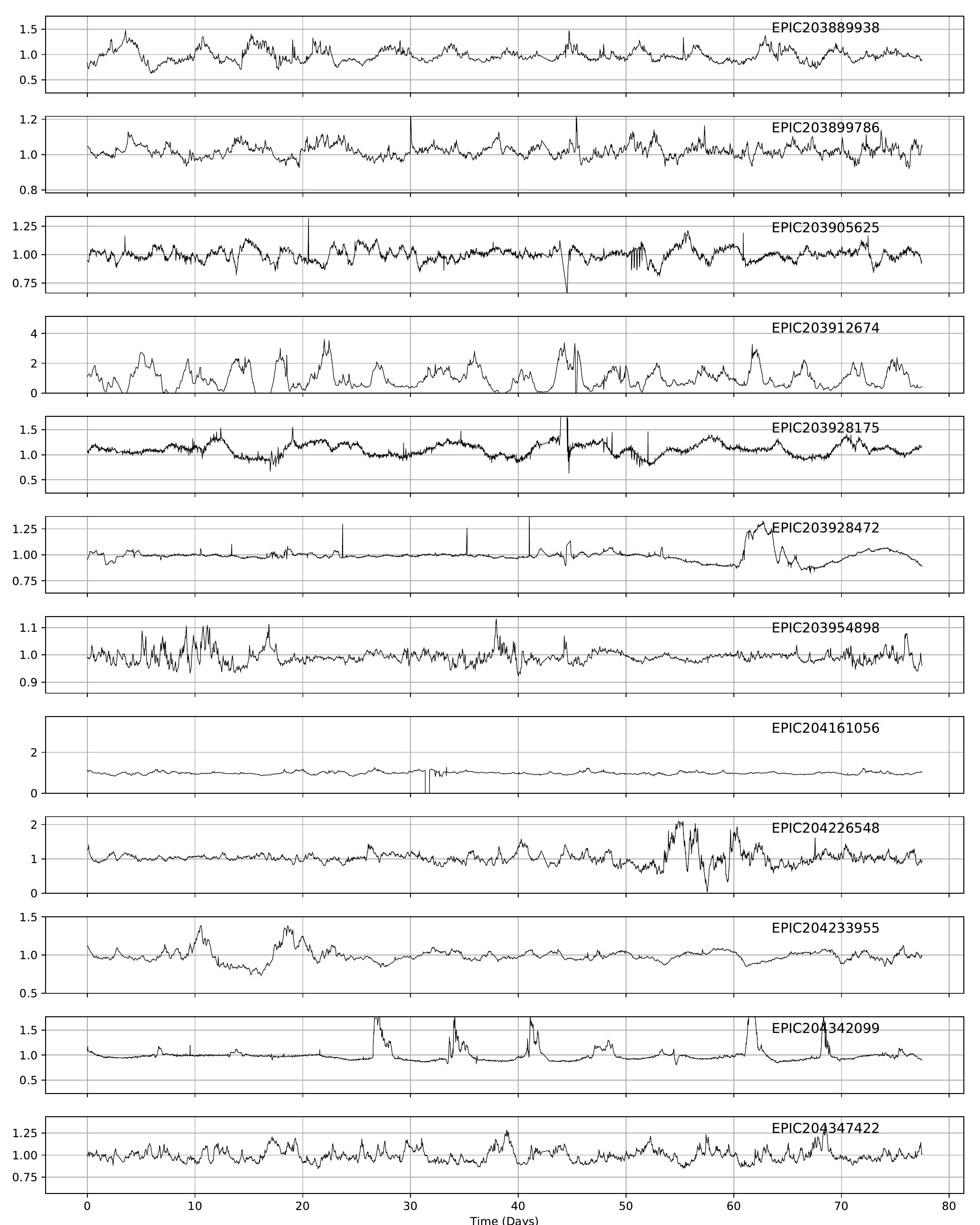}
  \caption{continued}
\end{figure*}
\addtocounter{figure}{-1}
\begin{figure*}
  \centering
  \includegraphics[width=\linewidth]{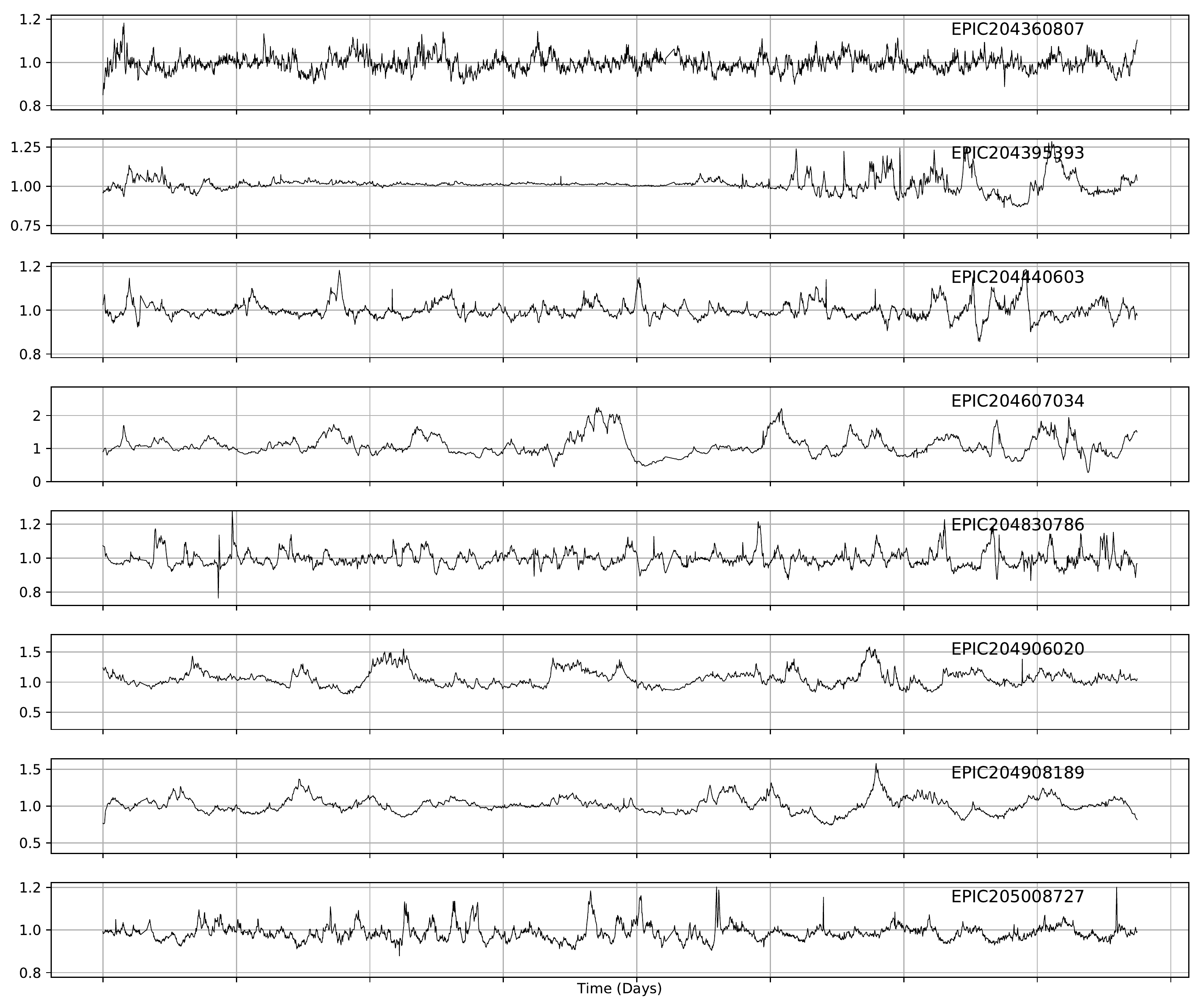}
  \caption{continued}
\end{figure*}

\begin{table}
\centering
\caption{Dippers identified in this work}
\label{dippertab}
\begin{tabular}{l l l l}
\hline
EPIC ID & RA & Dec & Membership\\ \hline\hline
203083616 & 243.71889 & -27.315475 & -\\ \hline
203343161 & 246.23282 & -26.455029 & -\\ \hline
203410665 & 246.41035 & -26.231661 & -\\ \hline
203417549 & 245.39461 & -26.207482 & -\\ \hline
203429083 & 239.26462 & -26.168919 & -\\ \hline
203465909 & 246.95435 & -26.045494 & -\\ \hline
203637940 & 246.61564 & -25.456825 & RO\\ \hline
203712588 & 246.31341 & -25.198359 & RO\\ \hline
203749770 & 246.80307 & -25.067169 & US\\ \hline
203750883 & 243.40213 & -25.063164 & US\\ \hline
203770366 & 243.77187 & -24.99309 & US\\ \hline
203770559 & 246.25882 & -24.992335 & RO\\ \hline
203791768 & 246.82646 & -24.914865 & RO\\ \hline
203806628 & 246.81297 & -24.860783 & RO\\ \hline
203810851 & 239.47685 & -24.845107 & -\\ \hline
203824153 & 247.22528 & -24.795612 & RO\\ \hline
203831587 & 246.29556 & -24.767736 & RO\\ \hline
203843911 & 246.59875 & -24.720513 & RO\\ \hline
203850058 & 246.77748 & -24.696901 & RO\\ \hline
203862309 & 246.92793 & -24.647396 & RO\\ \hline
203870022 & 246.90968 & -24.616274 & RO\\ \hline
203881640 & 246.78794 & -24.568928 & RO\\ \hline
203887087 & 247.05747 & -24.547056 & RO\\ \hline
203891877 & 247.05712 & -24.527538 & RO\\ \hline
203892903 & 245.68921 & -24.52336 & RO\\ \hline
203895983 & 241.0789 & -24.510927 & US\\ \hline
203915424 & 246.86076 & -24.431774 & RO\\ \hline
203930599 & 246.9178 & -24.367814 & RO\\ \hline
203936815 & 246.67857 & -24.341642 & RO\\ \hline
203937317 & 246.57117 & -24.339449 & RO\\ \hline
203950167 & 245.78846 & -24.284773 & RO\\ \hline
203962599 & 246.73655 & -24.230991 & RO\\ \hline
203969672 & 246.78781 & -24.200216 & RO\\ \hline
203971352 & 247.05298 & -24.193214 & RO\\ \hline
203995761 & 247.06973 & -24.087296 & RO\\ \hline
204078986 & 242.03105 & -23.751549 & US\\ \hline
204107757 & 239.00434 & -23.635588 & US\\ \hline
204137184 & 240.52158 & -23.518595 & -\\ \hline
\end{tabular}
\end{table}
\begin{table}
\centering
\begin{tabular}{l l l l}
\hline
EPIC ID & RA & Dec & Membership\\ \hline\hline
204142243 & 245.60399 & -23.49863 & RO\\ \hline
204176565 & 245.57722 & -23.36339 & RO\\ \hline
204187094 & 242.82956 & -23.322251 & US\\ \hline
204195987 & 245.68725 & -23.287067 & RO\\ \hline
204206295 & 246.69764 & -23.247469 & -\\ \hline
204211116 & 245.42498 & -23.228676 & RO\\ \hline
204256494 & 246.15225 & -23.050051 & -\\ \hline
204274536 & 245.88681 & -22.979681 & RO\\ \hline
204274743 & 239.37443 & -22.978846 & US\\ \hline
204277211 & 240.42029 & -22.969607 & US\\ \hline
204278916 & 240.53157 & -22.963025 & US\\ \hline
204290918 & 245.32701 & -22.916067 & RO\\ \hline
204329690 & 245.50811 & -22.761395 & -\\ \hline
204344180 & 243.63696 & -22.703707 & US\\ \hline
204388640 & 240.51791 & -22.529692 & US\\ \hline
204408707 & 245.09545 & -22.45115 & -\\ \hline
204449274 & 245.59001 & -22.291862 & US\\ \hline
204467584 & 242.82111 & -22.219075 & US\\ \hline
204469637 & 245.02565 & -22.210701 & US\\ \hline
204472612 & 242.14397 & -22.198866 & US\\ \hline
204487447 & 242.62791 & -22.139708 & -\\ \hline
204489514 & 240.75673 & -22.131222 & US\\ \hline
204496657 & 239.27675 & -22.101694 & US\\ \hline
204512343 & 239.3379 & -22.036949 & -\\ \hline
204514548 & 239.16676 & -22.027779 & -\\ \hline
204530046 & 242.70879 & -21.963379 & -\\ \hline
204565982 & 246.78925 & -21.812695 & US\\ \hline
204615647 & 243.3413 & -21.603808 & US\\ \hline
204637622 & 241.08741 & -21.511577 & US\\ \hline
204638512 & 241.09023 & -21.507916 & US\\ \hline
204651122 & 243.09541 & -21.454407 & -\\ \hline
204757338 & 241.86443 & -20.995618 & US\\ \hline
204807722 & 239.25609 & -20.771812 & -\\ \hline
204811478 & 238.98336 & -20.755199 & US\\ \hline
204864076 & 240.9903 & -20.518229 & US\\ \hline
204871202 & 242.25299 & -20.485743 & US\\ \hline
204871862 & 241.75707 & -20.482767 & US\\ \hline
204932990 & 242.9622 & -20.202746 & US\\ \hline
\end{tabular}
\end{table}
\begin{table}
\centering
\begin{tabular}{l l l l}
\hline
EPIC ID & RA & Dec & Membership\\ \hline\hline
204939243 & 243.88417 & -20.173256 & US\\ \hline
204982702 & 242.46694 & -19.968491 & -\\ \hline
205068630 & 242.79566 & -19.558912 & US\\ \hline
205080616 & 242.09685 & -19.500266 & US\\ \hline
205086621 & 242.93893 & -19.470343 & US\\ \hline
205092303 & 242.33561 & -19.442186 & US\\ \hline
205110000 & 243.93404 & -19.354803 & US\\ \hline
205115701 & 242.52257 & -19.32674 & US\\ \hline
205151387 & 242.25317 & -19.147971 & US\\ \hline
205158239 & 243.5846 & -19.113386 & US\\ \hline
205160565 & 243.58712 & -19.101454 & -\\ \hline
205164999 & 243.25979 & -19.079185 & US\\ \hline
205218826 & 242.40225 & -18.800259 & US\\ \hline
205238942 & 241.69976 & -18.695508 & US\\ \hline
205241182 & 242.69317 & -18.683298 & US\\ \hline
205249328 & 242.88061 & -18.640545 & US\\ \hline
205519771 & 241.80841 & -17.045187 & US\\ \hline
205684783 & 248.53822 & -15.804683 & -\\ \hline
210282534 & 246.74378 & -24.76024 & RO\\ \hline
\end{tabular}
\end{table}

\begin{table}
\centering
\label{bursttab}
\caption{Bursters identified in this work. Note bursters were not a trained class and this sample is likely incomplete.}
\begin{tabular}{l l l l}

\hline
EPIC ID & RA & Dec & Membership\\ \hline
203668987 & 246.57672 & -25.348854 & RO\\ \hline
203716389 & 246.32199 & -25.184946 & RO\\ \hline
203725791 & 240.37096 & -25.151953 & US\\ \hline
203786695 & 246.24892 & -24.933565 & RO\\ \hline
203789507 & 239.27043 & -24.922976 & -\\ \hline
203794605 & 247.59752 & -24.90448 & RO\\ \hline
203822485 & 246.84569 & -24.801971 & RO\\ \hline
203833873 & 246.74345 & -24.758829 & RO\\ \hline
203856109 & 242.46659 & -24.67214 & -\\ \hline
203864032 & 246.70408 & -24.640345 & RO\\ \hline
203870058 & 247.06878 & -24.616106 & RO\\ \hline
203889938 & 241.85937 & -24.535546 & US\\ \hline
203899786 & 246.35149 & -24.49577 & RO\\ \hline
203905576 & 246.57862 & -24.472137 & RO\\ \hline
203905625 & 247.18861 & -24.471941 & RO\\ \hline
203912674 & 246.41492 & -24.443031 & RO\\ \hline
203928175 & 247.09718 & -24.378053 & RO\\ \hline
203954898 & 246.65344 & -24.264414 & RO\\ \hline
204161056 & 246.42873 & -23.424052 & RO\\ \hline
204233955 & 241.87311 & -23.139515 & US\\ \hline
204342099 & 243.89405 & -22.711803 & US\\ \hline
204347422 & 244.96421 & -22.690742 & -\\ \hline
204360807 & 245.48918 & -22.638388 & -\\ \hline
204395393 & 240.07679 & -22.50318 & US\\ \hline
204440603 & 243.59637 & -22.326067 & US\\ \hline
204607034 & 240.67302 & -21.640156 & US\\ \hline
204830786 & 241.99149 & -20.669079 & US\\ \hline
204906020 & 241.75883 & -20.327436 & US\\ \hline
205008727 & 244.89878 & -19.845192 & -\\ \hline
205061092 & 243.71577 & -19.594519 & -\\ \hline

\end{tabular}
\end{table}
\bibliographystyle{mnras}
\bibliography{simon}

\begin{thebibliography}{}
\makeatletter
\relax
\def\mn@urlcharsother{\let\do\@makeother \do\$\do\&\do\#\do\^\do\_\do\%\do\~}
\def\mn@doi{\begingroup\mn@urlcharsother \@ifnextchar [ {\mn@doi@}
  {\mn@doi@[]}}
\def\mn@doi@[#1]#2{\def\@tempa{#1}\ifx\@tempa\@empty \href
  {http://dx.doi.org/#2} {doi:#2}\else \href {http://dx.doi.org/#2} {#1}\fi
  \endgroup}
\def\mn@eprint#1#2{\mn@eprint@#1:#2::\@nil}
\def\mn@eprint@arXiv#1{\href {http://arxiv.org/abs/#1} {{\tt arXiv:#1}}}
\def\mn@eprint@dblp#1{\href {http://dblp.uni-trier.de/rec/bibtex/#1.xml}
  {dblp:#1}}
\def\mn@eprint@#1:#2:#3:#4\@nil{\def\@tempa {#1}\def\@tempb {#2}\def\@tempc
  {#3}\ifx \@tempc \@empty \let \@tempc \@tempb \let \@tempb \@tempa \fi \ifx
  \@tempb \@empty \def\@tempb {arXiv}\fi \@ifundefined
  {mn@eprint@\@tempb}{\@tempb:\@tempc}{\expandafter \expandafter \csname
  mn@eprint@\@tempb\endcsname \expandafter{\@tempc}}}

\bibitem[\protect\citeauthoryear{{Aigrain}, {Parviainen}  \& {Pope}}{{Aigrain}
  et~al.}{2016}]{Aigrain2016}
{Aigrain} S.,  {Parviainen} H.,   {Pope} B.~J.~S.,  2016, \mn@doi [\mnras]
  {10.1093/mnras/stw706}, \href
  {http://adsabs.harvard.edu/abs/2016MNRAS.459.2408A} {459, 2408}

\bibitem[\protect\citeauthoryear{{Akeson} et~al.,}{{Akeson}
  et~al.}{2005}]{Akeson2005}
{Akeson} R.~L.,  et~al., 2005, \mn@doi [\apj] {10.1086/427770}, \href
  {http://adsabs.harvard.edu/abs/2005ApJ...622..440A} {622, 440}

\bibitem[\protect\citeauthoryear{{Alencar} et~al.,}{{Alencar}
  et~al.}{2010}]{Alencar2010}
{Alencar} S.~H.~P.,  et~al., 2010, \mn@doi [\aap]
  {10.1051/0004-6361/201014184}, \href
  {http://adsabs.harvard.edu/abs/2010A%26A...519A..88A} {519, A88}

\bibitem[\protect\citeauthoryear{{Andrews} \& {Williams}}{{Andrews} \&
  {Williams}}{2007}]{Andrews2007}
{Andrews} S.~M.,  {Williams} J.~P.,  2007, \mn@doi [\apj] {10.1086/522885},
  \href {http://adsabs.harvard.edu/abs/2007ApJ...671.1800A} {671, 1800}

\bibitem[\protect\citeauthoryear{{Ansdell}, {Gaidos}, {Williams}, {Kennedy},
  {Wyatt}, {LaCourse}, {Jacobs}  \& {Mann}}{{Ansdell}
  et~al.}{2016a}]{Ansdell2016}
{Ansdell} M.,  {Gaidos} E.,  {Williams} J.~P.,  {Kennedy} G.,  {Wyatt} M.~C.,
  {LaCourse} D.~M.,  {Jacobs} T.~L.,   {Mann} A.~W.,  2016a, \mn@doi [\mnras]
  {10.1093/mnrasl/slw140}, \href
  {http://adsabs.harvard.edu/abs/2016MNRAS.462L.101A} {462, L101}

\bibitem[\protect\citeauthoryear{{Ansdell} et~al.,}{{Ansdell}
  et~al.}{2016b}]{Ansdell2015}
{Ansdell} M.,  et~al., 2016b, \mn@doi [\apj] {10.3847/0004-637X/816/2/69},
  \href {http://adsabs.harvard.edu/abs/2016ApJ...816...69A} {816, 69}

\bibitem[\protect\citeauthoryear{{Anthonioz} et~al.,}{{Anthonioz}
  et~al.}{2015}]{Anthonioz2014}
{Anthonioz} F.,  et~al., 2015, \mn@doi [\aap] {10.1051/0004-6361/201424520},
  \href {http://adsabs.harvard.edu/abs/2015A%26A...574A..41A} {574, A41}

\bibitem[\protect\citeauthoryear{{Armstrong} et~al.,}{{Armstrong}
  et~al.}{2015}]{Armstrong2015}
{Armstrong} D.~J.,  et~al., 2015, \mn@doi [\aap] {10.1051/0004-6361/201525889},
  \href {http://ukads.nottingham.ac.uk/abs/2015A%26A...579A..19A} {579, A19}

\bibitem[\protect\citeauthoryear{{Armstrong} et~al.,}{{Armstrong}
  et~al.}{2016}]{Armstrong2016}
{Armstrong} D.~J.,  et~al., 2016, \mn@doi [\mnras] {10.1093/mnras/stv2836},
  \href {http://adsabs.harvard.edu/abs/2016MNRAS.456.2260A} {456, 2260}

\bibitem[\protect\citeauthoryear{{Bloom} et~al.,}{{Bloom}
  et~al.}{2012}]{Bloom2011}
{Bloom} J.~S.,  et~al., 2012, \mn@doi [\pasp] {10.1086/668468}, \href
  {http://adsabs.harvard.edu/abs/2012PASP..124.1175B} {124, 1175}

\bibitem[\protect\citeauthoryear{{Bodman} et~al.,}{{Bodman}
  et~al.}{2017}]{Bodman2016}
{Bodman} E.~H.~L.,  et~al., 2017, \mn@doi [\mnras] {10.1093/mnras/stx1034},
  \href {http://adsabs.harvard.edu/abs/2017MNRAS.470..202B} {470, 202}

\bibitem[\protect\citeauthoryear{{Bouvier} et~al.,}{{Bouvier}
  et~al.}{1999}]{Bouvier1999}
{Bouvier} J.,  et~al., 1999, \aap, \href
  {http://adsabs.harvard.edu/abs/1999A%26A...349..619B} {349, 619}

\bibitem[\protect\citeauthoryear{{Bouvier} et~al.,}{{Bouvier}
  et~al.}{2003}]{Bouvier2003}
{Bouvier} J.,  et~al., 2003, \mn@doi [\aap] {10.1051/0004-6361:20030938}, \href
  {http://adsabs.harvard.edu/abs/2003A%26A...409..169B} {409, 169}

\bibitem[\protect\citeauthoryear{{Bravo}, {Roque}, {Estrela}, {Le{\~a}o}  \&
  {De Medeiros}}{{Bravo} et~al.}{2014}]{Bravo2014}
{Bravo} J.~P.,  {Roque} S.,  {Estrela} R.,  {Le{\~a}o} I.~C.,   {De Medeiros}
  J.~R.,  2014, \mn@doi [\aap] {10.1051/0004-6361/201323032}, \href
  {http://adsabs.harvard.edu/abs/2014A%26A...568A..34B} {568, A34}

\bibitem[\protect\citeauthoryear{{Carpenter}, {Mamajek}, {Hillenbrand}  \&
  {Meyer}}{{Carpenter} et~al.}{2006}]{Carpenter2006}
{Carpenter} J.~M.,  {Mamajek} E.~E.,  {Hillenbrand} L.~A.,   {Meyer} M.~R.,
  2006, \mn@doi [\apjl] {10.1086/509121}, \href
  {http://adsabs.harvard.edu/abs/2006ApJ...651L..49C} {651, L49}

\bibitem[\protect\citeauthoryear{{Cody} et~al.,}{{Cody}
  et~al.}{2014}]{Cody2014}
{Cody} A.~M.,  et~al., 2014, \mn@doi [\aj] {10.1088/0004-6256/147/4/82}, \href
  {http://adsabs.harvard.edu/abs/2014AJ....147...82C} {147, 82}

\bibitem[\protect\citeauthoryear{{Dahm}}{{Dahm}}{2008}]{Dahm2008}
{Dahm} S.~E.,  2008, {The Young Cluster and Star Forming Region NGC 2264}.
The Southern Sky ASP Monograph Publications, p.~966

\bibitem[\protect\citeauthoryear{{Dubath} et~al.,}{{Dubath}
  et~al.}{2011}]{Dubath2011}
{Dubath} P.,  et~al., 2011, \mn@doi [\mnras]
  {10.1111/j.1365-2966.2011.18575.x}, \href
  {http://adsabs.harvard.edu/abs/2011MNRAS.414.2602D} {414, 2602}

\bibitem[\protect\citeauthoryear{{Eisner}, {Hillenbrand}, {White}, {Bloom},
  {Akeson}  \& {Blake}}{{Eisner} et~al.}{2007}]{Eisner2007}
{Eisner} J.~A.,  {Hillenbrand} L.~A.,  {White} R.~J.,  {Bloom} J.~S.,  {Akeson}
  R.~L.,   {Blake} C.~H.,  2007, \mn@doi [\apj] {10.1086/521874}, \href
  {http://adsabs.harvard.edu/abs/2007ApJ...669.1072E} {669, 1072}

\bibitem[\protect\citeauthoryear{{Herbst}}{{Herbst}}{2012}]{Herbst2012}
{Herbst} W.,  2012, Journal of the American Association of Variable Star
  Observers (JAAVSO), \href {http://adsabs.harvard.edu/abs/2012JAVSO..40..448H}
  {40, 448}

\bibitem[\protect\citeauthoryear{{Kennedy} \& {Kenyon}}{{Kennedy} \&
  {Kenyon}}{2009}]{Kennedy2009}
{Kennedy} G.~M.,  {Kenyon} S.~J.,  2009, \mn@doi [\apj]
  {10.1088/0004-637X/695/2/1210}, \href
  {http://adsabs.harvard.edu/abs/2009ApJ...695.1210K} {695, 1210}

\bibitem[\protect\citeauthoryear{Kennedy, Kenworthy, Pepper, Rodriguez, Siverd,
  Stassun  \& Wyatt}{Kennedy et~al.}{2017}]{Kennedy2017}
Kennedy G.~M.,  Kenworthy M.~A.,  Pepper J.,  Rodriguez J.~E.,  Siverd R.~J.,
  Stassun K.~G.,   Wyatt M.~C.,  2017, \mn@doi [Royal Society Open Science]
  {10.1098/rsos.160652}, 4

\bibitem[\protect\citeauthoryear{{Kobayashi}, {Kimura}, {Watanabe}, {Yamamoto}
  \& {M{\"u}ller}}{{Kobayashi} et~al.}{2011}]{2011EP&S...63.1067K}
{Kobayashi} H.,  {Kimura} H.,  {Watanabe} S.-i.,  {Yamamoto} T.,   {M{\"u}ller}
  S.,  2011, \mn@doi [Earth, Planets, and Space] {10.5047/eps.2011.03.012},
  \href {http://adsabs.harvard.edu/abs/2011EP%26S...63.1067K} {63, 1067}

\bibitem[\protect\citeauthoryear{{Kraus}}{{Kraus}}{2015}]{Kraus2015}
{Kraus} S.,  2015, \mn@doi [\apss] {10.1007/s10509-015-2226-6}, \href
  {http://adsabs.harvard.edu/abs/2015Ap%26SS.357...97K} {357, 97}

\bibitem[\protect\citeauthoryear{{Lodieu}}{{Lodieu}}{2013}]{lodieu2013}
{Lodieu} N.,  2013, \mn@doi [\mnras] {10.1093/mnras/stt402}, \href
  {http://ukads.nottingham.ac.uk/abs/2013MNRAS.431.3222L} {431, 3222}

\bibitem[\protect\citeauthoryear{{Loinard}, {Torres}, {Mioduszewski}  \&
  {Rodr{\'{\i}}guez}}{{Loinard} et~al.}{2008}]{Loinard2008}
{Loinard} L.,  {Torres} R.~M.,  {Mioduszewski} A.~J.,   {Rodr{\'{\i}}guez}
  L.~F.,  2008, \mn@doi [\apjl] {10.1086/529548}, \href
  {http://adsabs.harvard.edu/abs/2008ApJ...675L..29L} {675, L29}

\bibitem[\protect\citeauthoryear{{Lomb}}{{Lomb}}{1976}]{Lomb}
{Lomb} N.~R.,  1976, \mn@doi [\apss] {10.1007/BF00648343}, \href
  {http://adsabs.harvard.edu/abs/1976Ap%26SS..39..447L} {39, 447}

\bibitem[\protect\citeauthoryear{{Loomis}, {{\"O}berg}, {Andrews}  \&
  {MacGregor}}{{Loomis} et~al.}{2017}]{Loomis2017}
{Loomis} R.~A.,  {{\"O}berg} K.~I.,  {Andrews} S.~M.,   {MacGregor} M.~A.,
  2017, \mn@doi [\apj] {10.3847/1538-4357/aa6c63}, \href
  {http://adsabs.harvard.edu/abs/2017ApJ...840...23L} {840, 23}

\bibitem[\protect\citeauthoryear{{Luger}, {Agol}, {Kruse}, {Barnes}, {Becker},
  {Foreman-Mackey}  \& {Deming}}{{Luger} et~al.}{2016}]{Luger2016}
{Luger} R.,  {Agol} E.,  {Kruse} E.,  {Barnes} R.,  {Becker} A.,
  {Foreman-Mackey} D.,   {Deming} D.,  2016, \mn@doi [\aj]
  {10.3847/0004-6256/152/4/100}, \href
  {http://adsabs.harvard.edu/abs/2016AJ....152..100L} {152, 100}

\bibitem[\protect\citeauthoryear{{Luhman} \& {Mamajek}}{{Luhman} \&
  {Mamajek}}{2012}]{luhman}
{Luhman} K.~L.,  {Mamajek} E.~E.,  2012, \mn@doi [\apj]
  {10.1088/0004-637X/758/1/31}, \href
  {http://adsabs.harvard.edu/abs/2012ApJ...758...31L} {758, 31}

\bibitem[\protect\citeauthoryear{{Luhman} \& {Rieke}}{{Luhman} \&
  {Rieke}}{1999}]{Luhman1999}
{Luhman} K.~L.,  {Rieke} G.~H.,  1999, \mn@doi [\apj] {10.1086/307891}, \href
  {http://adsabs.harvard.edu/abs/1999ApJ...525..440L} {525, 440}

\bibitem[\protect\citeauthoryear{{Marino}, {Perez}  \& {Casassus}}{{Marino}
  et~al.}{2015}]{Marino2015}
{Marino} S.,  {Perez} S.,   {Casassus} S.,  2015, \mn@doi [\apjl]
  {10.1088/2041-8205/798/2/L44}, \href
  {http://adsabs.harvard.edu/abs/2015ApJ...798L..44M} {798, L44}

\bibitem[\protect\citeauthoryear{{McGinnis} et~al.,}{{McGinnis}
  et~al.}{2015}]{Mcginnis2015}
{McGinnis} P.~T.,  et~al., 2015, \mn@doi [\aap] {10.1051/0004-6361/201425475},
  577, A11

\bibitem[\protect\citeauthoryear{{Millan-Gabet}, {Malbet}, {Akeson}, {Leinert},
  {Monnier}  \& {Waters}}{{Millan-Gabet} et~al.}{2007}]{Millan-Gabet2006}
{Millan-Gabet} R.,  {Malbet} F.,  {Akeson} R.,  {Leinert} C.,  {Monnier} J.,
  {Waters} R.,  2007, in Protostars and Planets V. pp 539--554

\bibitem[\protect\citeauthoryear{Morales-CalderÃ³n
  et~al.,}{Morales-CalderÃ³n et~al.}{2011}]{Morales-Calderon2011}
Morales-CalderÃ³n M.,  et~al., 2011, \aj, 733, 50

\bibitem[\protect\citeauthoryear{{Muzerolle}, {Calvet}, {Hartmann}  \&
  {D'Alessio}}{{Muzerolle} et~al.}{2003}]{Muzerolle2003}
{Muzerolle} J.,  {Calvet} N.,  {Hartmann} L.,   {D'Alessio} P.,  2003, \mn@doi
  [\apjl] {10.1086/379921}, \href
  {http://adsabs.harvard.edu/abs/2003ApJ...597L.149M} {597, L149}

\bibitem[\protect\citeauthoryear{{Parks}, {Plavchan}, {White}  \&
  {Gee}}{{Parks} et~al.}{2014}]{parks}
{Parks} J.~R.,  {Plavchan} P.,  {White} R.~J.,   {Gee} A.~H.,  2014, \mn@doi
  [\apjs] {10.1088/0067-0049/211/1/3}, \href
  {http://adsabs.harvard.edu/abs/2014ApJS..211....3P} {211, 3}

\bibitem[\protect\citeauthoryear{{Pecaut} \& {Mamajek}}{{Pecaut} \&
  {Mamajek}}{2013}]{Pecaut2013}
{Pecaut} M.~J.,  {Mamajek} E.~E.,  2013, \mn@doi [\apjs]
  {10.1088/0067-0049/208/1/9}, \href
  {http://adsabs.harvard.edu/abs/2013ApJS..208....9P} {208, 9}

\bibitem[\protect\citeauthoryear{{Pecaut}, {Mamajek}  \& {Bubar}}{{Pecaut}
  et~al.}{2012}]{Pecaut2011}
{Pecaut} M.~J.,  {Mamajek} E.~E.,   {Bubar} E.~J.,  2012, \mn@doi [\apj]
  {10.1088/0004-637X/746/2/154}, \href
  {http://adsabs.harvard.edu/abs/2012ApJ...746..154P} {746, 154}

\bibitem[\protect\citeauthoryear{Pedregosa et~al.,}{Pedregosa
  et~al.}{2011}]{scikit-learn}
Pedregosa F.,  et~al., 2011, Journal of Machine Learning Research, 12, 2825

\bibitem[\protect\citeauthoryear{{Ratzka}, {K{\"o}hler}  \& {Leinert}}{{Ratzka}
  et~al.}{2005}]{ratzka}
{Ratzka} T.,  {K{\"o}hler} R.,   {Leinert} C.,  2005, \mn@doi [\aap]
  {10.1051/0004-6361:20042107}, \href
  {http://adsabs.harvard.edu/abs/2005A%26A...437..611R} {437, 611}

\bibitem[\protect\citeauthoryear{Rebull et~al.,}{Rebull
  et~al.}{2002}]{Rebull2002}
Rebull L.~M.,  et~al., 2002, \mn@doi [\aj] {10.1086/338904}, 123, 1528

\bibitem[\protect\citeauthoryear{{Richards} et~al.,}{{Richards}
  et~al.}{2011}]{Richards2011}
{Richards} J.~W.,  et~al., 2011, \mn@doi [\apj] {10.1088/0004-637X/733/1/10},
  \href {http://adsabs.harvard.edu/abs/2011ApJ...733...10R} {733, 10}

\bibitem[\protect\citeauthoryear{{Richards}, {Starr}, {Miller}, {Bloom},
  {Butler}, {Brink}  \& {Crellin-Quick}}{{Richards}
  et~al.}{2012}]{Richards2012}
{Richards} J.~W.,  {Starr} D.~L.,  {Miller} A.~A.,  {Bloom} J.~S.,  {Butler}
  N.~R.,  {Brink} H.,   {Crellin-Quick} A.,  2012, \mn@doi [\apjs]
  {10.1088/0067-0049/203/2/32}, \href
  {http://adsabs.harvard.edu/abs/2012ApJS..203...32R} {203, 32}

\bibitem[\protect\citeauthoryear{{Rizzuto}, {Ireland}  \& {Kraus}}{{Rizzuto}
  et~al.}{2015}]{Rizzuto2015}
{Rizzuto} A.~C.,  {Ireland} M.~J.,   {Kraus} A.~L.,  2015, \mn@doi [\mnras]
  {10.1093/mnras/stv207}, \href
  {http://adsabs.harvard.edu/abs/2015MNRAS.448.2737R} {448, 2737}

\bibitem[\protect\citeauthoryear{{Scargle}}{{Scargle}}{1982}]{Scargle}
{Scargle} J.~D.,  1982, \mn@doi [\apj] {10.1086/160554}, \href
  {http://adsabs.harvard.edu/abs/1982ApJ...263..835S} {263, 835}

\bibitem[\protect\citeauthoryear{{Skrutskie} et~al.,}{{Skrutskie}
  et~al.}{2006}]{twomass}
{Skrutskie} M.~F.,  et~al., 2006, \mn@doi [\aj] {10.1086/498708}, \href
  {http://adsabs.harvard.edu/abs/2006AJ....131.1163S} {131, 1163}

\bibitem[\protect\citeauthoryear{{Slesnick}, {Hillenbrand}  \&
  {Carpenter}}{{Slesnick} et~al.}{2008}]{slesnick2008}
{Slesnick} C.~L.,  {Hillenbrand} L.~A.,   {Carpenter} J.~M.,  2008, \mn@doi
  [\apj] {10.1086/592265}, \href
  {http://adsabs.harvard.edu/abs/2008ApJ...688..377S} {688, 377}

\bibitem[\protect\citeauthoryear{Stauffer et~al.,}{Stauffer
  et~al.}{2014}]{Stauffer2014}
Stauffer J.,  et~al., 2014, \aj, 147, 83

\bibitem[\protect\citeauthoryear{{Stauffer} et~al.,}{{Stauffer}
  et~al.}{2015}]{Stauffer2015}
{Stauffer} J.,  et~al., 2015, \mn@doi [\aj] {10.1088/0004-6256/149/4/130},
  \href {http://adsabs.harvard.edu/abs/2015AJ....149..130S} {149, 130}

\bibitem[\protect\citeauthoryear{{Thalmann} et~al.,}{{Thalmann}
  et~al.}{2015}]{Thalmann2015}
{Thalmann} C.,  et~al., 2015, \mn@doi [\apjl] {10.1088/2041-8205/808/2/L41},
  \href {http://adsabs.harvard.edu/abs/2015ApJ...808L..41T} {808, L41}

\bibitem[\protect\citeauthoryear{{Torrence} \& {Compo}}{{Torrence} \&
  {Compo}}{1998}]{Torrence98apractical}
{Torrence} C.,  {Compo} G.~P.,  1998, \mn@doi [Bulletin of the American
  Meteorological Society] {10.1175/1520-0477(1998)079<0061:APGTWA>2.0.CO;2},
  \href {http://adsabs.harvard.edu/abs/1998BAMS...79...61T} {79, 61}

\bibitem[\protect\citeauthoryear{{Vanderburg} \& {Johnson}}{{Vanderburg} \&
  {Johnson}}{2014}]{Vanderburg2014}
{Vanderburg} A.,  {Johnson} J.~A.,  2014, \mn@doi [\pasp] {10.1086/678764},
  \href {http://adsabs.harvard.edu/abs/2014PASP..126..948V} {126, 948}

\bibitem[\protect\citeauthoryear{{Wenger} et~al.,}{{Wenger}
  et~al.}{2000}]{Wegner2000}
{Wenger} M.,  et~al., 2000, \mn@doi [\aaps] {10.1051/aas:2000332}, \href
  {http://adsabs.harvard.edu/abs/2000A%26AS..143....9W} {143, 9}

\bibitem[\protect\citeauthoryear{{Wilking}, {Meyer}, {Greene}  \&
  {Robinson}}{{Wilking} et~al.}{2001}]{Wilking2001}
{Wilking} B.~A.,  {Meyer} M.~R.,  {Greene} T.~P.,   {Robinson} J.~G.,  2001, in
  American Astronomical Society Meeting Abstracts. p.~563

\bibitem[\protect\citeauthoryear{{Wilking}, {Gagn{\'e}}  \& {Allen}}{{Wilking}
  et~al.}{2008}]{wilking}
{Wilking} B.~A.,  {Gagn{\'e}} M.,   {Allen} L.~E.,  2008, {Star Formation in
  the {$\rho$} Ophiuchi Molecular Cloud}.
The Southern Sky ASP Monograph Publications, p.~351

\bibitem[\protect\citeauthoryear{{Wright} et~al.,}{{Wright}
  et~al.}{2010}]{WISE}
{Wright} E.~L.,  et~al., 2010, \mn@doi [\aj] {10.1088/0004-6256/140/6/1868},
  \href {http://adsabs.harvard.edu/abs/2010AJ....140.1868W} {140, 1868}

\bibitem[\protect\citeauthoryear{{de Zeeuw}, {Hoogerwerf}, {de Bruijne},
  {Brown}  \& {Blaauw}}{{de Zeeuw} et~al.}{1999}]{dezeeuw}
{de Zeeuw} P.~T.,  {Hoogerwerf} R.,  {de Bruijne} J.~H.~J.,  {Brown} A.~G.~A.,
   {Blaauw} A.,  1999, \mn@doi [\aj] {10.1086/300682}, \href
  {http://adsabs.harvard.edu/abs/1999AJ....117..354D} {117, 354}

\makeatother
\end{thebibliography}

\end{document}